\documentclass[twocolumn,aps,prl,superscriptaddress]{revtex4-2}

\usepackage{color,amsmath,amssymb,epsfig,graphicx}
\usepackage{bm}
\usepackage{mathptmx, courier, pifont}
\usepackage[scaled=0.92]{helvet}
\usepackage[T1]{fontenc}
\usepackage{textcomp}
\usepackage[colorlinks=true,linkcolor=blue,filecolor=blue,urlcolor=blue,citecolor=blue]{hyperref}
\providecommand{\boldsymbol}[1]{\mbox{\boldmath $#1$}}

\usepackage{multirow}
\usepackage{longtable}
\usepackage[utf8]{inputenc}
\def\bra{\,<\!} \def\ket{\!>\,} \def\ack{\,|\,}
\begin{document}
\author{S. A. Bhat}
\affiliation{Department of Physics, University of Kashmir, Srinagar,
  Jammu and Kashmir, 190 006, India}
 \author{S. Jehangir}
\email{sheikhahmad.phy@gmail.com}
\affiliation{Department of Physics, Govt. Degree College for Women, Pulwama,
 Jammu and Kashmir, 192301, India}
\author{G. H. Bhat}
\email{gwhr.bhat@gmail.com}
\affiliation{Department of Physics, GDC Shopian, Higher Education, Jammu and Kashmir, 192 303, India}

\author{J. A. Sheikh}
\email{sjaphysics@gmail.com}
\affiliation{Department of Physics, Islamic University of Science and Technology, 
  Jammu and Kashmir, 192 122, India}
\affiliation{Department of Physics, University of Kashmir, Srinagar,
Jammu and Kashmir, 190 006, India}
 
\author{G. B. Vakil}
\affiliation{Department of Physics, University of Kashmir, Srinagar,
  Jammu and Kashmir, 190 006, India}

\title{ Microscopic investigation of wobbling motion in even-even nuclei }

\begin{abstract}

  The possibility of observing wobbling mode in the even-even systems of $^{76}$Ge, $^{112}$Ru, $^{188,192}$Os, $^{192}$Pt and $^{232}$Th
  is explored using the triaxial projected shell model approach. These nuclei are known to have $\gamma$-bands whose odd-spin members are
  lower than the average of the neighbouring even-spin states. It is shown through a detailed analysis of the
  excitation energies and the electromagnetic
  transition probabilities that the observed band structures in these nuclei except for $^{232}$Th can
  be characterised as originating from the wobbling motion. It is further demonstrated that quasiparticle alignment is responsible
  for driving the systems to the wobbling mode.

\end{abstract}

\date{\today}

\maketitle
\section{Introduction}
\label{Sect.01}
The appearance of the rotational spectra in a finite quantum many-body system is a direct manifestation
of the anisotropic density distribution or deformation of the system in the intrinsic frame \cite{BM75}.
  In atomic nuclei, the quantal shell effects are responsible for the deformation in the 
  nuclear equilibrium shape. The nuclear deformation is assumed to have an ellipsoidal shape distribution and is parameterised in terms
  of $\beta$ and $\gamma$, where the former expresses the deviation from the spherical shape and the later signifies the departure
  from the axial symmetry \cite{BM75,RF00,IH83}. The ground-state properties of most of the deformed nuclei are described
  using axial symmetry with
  the non-axial parameter, $\gamma=0$. There are selection rules based on the axial symmetry, for instance, the Alaga rules, which are
  satisfied by most of the deformed nuclei \cite{Alaga1955} 
  in the low-spin region. Nevertheless, there are also a few regions
  in the Segre chart where the nuclei are predicted to have triaxial deformation with non-zero values for $\gamma$ \cite{Aberg1990,IR1989,TB1989,RB04}.

  The studies in the triaxial nuclei have been reinvigorated with the experimental identification of chiral band structures and the
  wobbling mode, the two characteristic phenomena of a triaxial shape. The chiral geometry also called as left-right
  symmetry develops when the total angular-momentum vector is distributed along the three principle axis, and
  the rotation in clock-wise
  and counter clock-wise directions being equivalent in the intrinsic frame, giving rise to two identical band
  structures in the laboratory frame \cite{SF97,TS04,VI00,PO04}.
  The chiral symmetry mandates that all the properties of the doublet bands should be identical. The doublet band structures
  have been identified in several regions of the periodic table \cite{KS01,SU03,EG06}.

  The wobbling mode, the other characteristic rotational mode
  of a triaxial system, was proposed by Bohr and Mottelson in 1975 \cite{BM75}. It was shown that asymmetric rotor Hamiltonian has
  a simple solution for angular-momentum, I $>>$ 1 with the Hamiltonian acquiring harmonic oscillator form. The stationary states
  of the simplified Hamiltonian are labelled by a boson quantum number ``$n_{\omega}$'' apart from the angular-momentum quantum numbers
  of ``I'' and ``M''. The yrast (lowest) configuration has $n_{\omega}=0$ and corresponds to the geometry with the rotational axis
  directed along the principle axis having the largest moment of inertia. For excited configurations with $n_{\omega}$=1,2,3....,
  some angular-momentum is transferred from the axis having largest moment-of-inertia to other axes and for $I >> 1$, gives rise
  to harmonic spectrum near the yrast line. The wobbling motion was originally proposed for even-even systems, but it was odd-proton
  $^{163}$Lu nucleus for which this new collective mode was first experimentally identified \cite{WM1}. For this system,
  triaxial strongly deformed (TSD) bands were known to exist, but the nature of band structures remained obscure \cite{WM41}.
  In the seminal experimental study \cite{WM1}, the inter-band transitions from the first excited TSD (TSD2) band
  to the lowest TSD (TSD1) bands were measured and mixing ratio, $\delta =  90.5 \pm 1.3 \%$ for E2 and $9.4 \pm 1.3 \%$ for
  M1 were extracted. These ratios demonstrate that the transitions from TSD2 to TSD1 bands are dominated by E2 rather
  than M1, which is a fingerprint of the wobbling mode. In normal signature partner bands, the M1 is dominated over E2 as can be easily
  shown in a one-dimensional cranking approach \cite{WM56}. Further, the phase of the measured electromagnetic ratios of
  B(E2)$_{\textrm {out}}$/B(E2)$_{\textrm {in}}$ and B(M1)$_{\textrm {out}}$/B(E2)$_{\textrm {in}}$ is opposite to that obtained in the cranking model and conforms with the
  phase obtained in the particle-rotor approach with the geometry of a wobbling excitation mode \cite{IK02}. In a subsequent
  work \cite{WM42}, $n_{\omega}=2$ band was also identified and it was shown that the transitions from this band to the $n_{\omega}=1$
  band are almost twice stronger than the transitions from the $n_{\omega}$=1 to the $n_{\omega}$=0 band, and the direct transitions
  from $n_{\omega}=2$ band to $n_{\omega}=0$ band are weak. These results of  the transition probabilities
  are consistent with those expected for a harmonic motion \cite{BM75,IK02,IK03}. 

  The wobbling band structures have also been identified in the TSD band structures of $^{161,165,167}$Lu and $^{167}$Tm nuclides
  \cite{WM44,WM45,WM2,WM46}. For normal deformed nuclei, the wobbling band structures have been reported for
  $^{131}$Cs, $^{135}$Pr, $^{151}$Eu, $^{183}$Au, $^{133}$Ba, $^{105}$Pd, $^{133}$La,
$^{187}$Au and $^{127}$Xe  nuclides \cite{WM54,WM47,WM48,WM6,WM12}, although there is some controversy regarding the observation of this mode in some of these
 isotopes  \cite{LV2022,BCM23}. The original prediction of Bohr and Mottelson was for
  an even-even system, however, so far there is no experimental evidence of the observation of wobbling mode in an
  even-even system. For $^{112}$Ru, it has been proposed that $\gamma$-band may have the wobbling mode characteristics
  as the odd-spin members
  of this band are situated below the average of the neighbouring even-spin members \cite{JH10,SF24ch}. Two-quasiparticle
  band structures in even-even systems of $^{130}$Ba and
  $^{136}$Nd have also been proposed as possible candidates of
  wobbling motion \cite{PETRACHE2019,WANG2020,Chen2019,Chen2021,LV20227}. Further, the wobbling motion has also
  been recently explored for $^{112}$Ru, $^{114}$Pd, $^{150}$Nd and $^{188}$Os even-even systems using the
  five dimensional collective Hamiltonian
  with potential energy surfaces calculated from the relativistic density functional theory \cite{WANG2024859}.
  
  In a recent work \cite{SJ21,SP24}, a systematic investigation was performed using the triaxial projected shell model (TPSM) approach
  for $\gamma$-bands in various
  regions of the Segre chart and it was noted that out of thirty transitional and deformed nuclei
  studied, six nuclei consisting
  of $^{76}$Ge, $^{112}$Ru, $^{188,192}$Os, $^{192}$Pt and $^{232}$Th depicted energy staggering phase with odd-spin members below the even-spin members
  and for all other studied nuclei, the even-spin members are lower than the odd-spin members. The staggering phase with odd-spin
  lower than the even-spin is considered as a fingerprint of the $\gamma-$rigid motion, and the even-spin lower than the
  odd-spin is regarded as a signature of the $\gamma$-soft motion \cite{ZAMFIR1991265,McCutchan2007,ASD1970,Baktash1978,Baktash1980}.

  The purpose of the present work is to explore whether the $\gamma$-bands observed in the aforementioned six nuclei are 
  based on the wobbling mode as expected from the observation that these nuclei have odd-spin members lower than the
  average of the neighbouring even-spin members. These systems have already been studied in our earlier publications \cite{SJ21,SP24}, and
  in the present work we shall present only those results which are relevant to the discussion of the wobbling motion. We shall
  examine the properties of the excited bands, in particular, the inter-band transition probabilities, which were not discussed in our
  previous publications.
  The present manuscript is organised in the following manner. In the next section, the TPSM approach is introduced briefly
  and the details can be found in our previous publications \cite{TPSM1999,TPSM8,TPSM2014,TPSM5,TPSM4,TPSM3,TPSM2,TPSM6,TPSM10}. In section III, the results of the six nuclei are presented
  and discussed, and finally the present work is summarized and concluded in section IV.

\section{Outline of Triaxial Projected Shell Model Approach}

  In recent years, considerable progress has been made in the development of nuclear models to investigate the
  high-spin properties of deformed and transitional nuclei
  \cite{ANGUIANO2001467,Bender2008,dl09,mb09,Yao2009,jm10,Yao2010,Sun2016,Egido2016,Robledo2019,TO20,bb21,Dorian24}. It is
  now feasible to study heavy deformed systems using the
 configuration interaction shell model \cite{TO20} and
energy density functional approaches \cite{Bender2008,DV05}. 
 For instance, a large scale shell model calculations have
 been performed using the Monte-Carlo method and a well deformed $^{166}$Er nucleus has been
 studied in detail \cite{TO19,yt21}. However, these calculations
 are computationally very demanding and it is not possible to perform a systematic analysis within this approach.
 In the nuclear density functional approach, it is now feasible to perform configuration mixing calculations
 with the optimal mean-field determined by solving the Kohn-Sham equations
 \cite{Bender2008,Yao2010,Wang2022}. In this approach, angular-momentum projection
 is required to be performed as the mean-field solution is deformed and the projected states are then used to perform
 the configuration mixing calculations. However,
 it is known that projection formalism in the nuclear density functional approach involves singularities and the
 spurious contributions in the calculated results due to the poles in the projected kernels need to be investigated
 \cite{JD21}.

 In the present work, we have employed the TPSM approach to investigate the wobbling nature of the deformed nuclei. The advantage of
 this approach is that it requires modest computational resources and it is feasible to undertake a systematic analysis of
 a large set of nuclei. As a matter of fact, several systematic studies have been performed and it
 has been demonstrated that TPSM approach provides a unified description of collective and single-particle excitation modes \cite{WM39,SJ24ch}.
 Further, it has been demonstrated recently that TPSM approach provided a comparable description of the observed data
 for $^{166}$Er as that using the large scale shell model Monte Carlo approach \cite{SP24}.
 
 The TPSM approach employs the angular-momentum projected deformed Nilsson states as the basis configurations. These basis states
 are then used to diagonalize the shell model
 Hamiltonian, consisting of monopole pairing, quadrupole pairing, and
 quadrupole-quadrupole interaction terms within the configuration space of three major oscillator shells. The Hamiltonian is
 given by
\begin{equation}
\hat H = \hat H_0 - {1 \over 2} \chi \sum_\mu \hat Q^\dagger_\mu
\hat Q^{}_\mu - G_M \hat P^\dagger \hat P - G_Q \sum_\mu \hat
P^\dagger_\mu\hat P^{}_\mu .
\label{hamham}
\end{equation}
$\hat H_0$ in the above equation is the spherical single-particle potential \cite{Ni69}.
The pairing parameters are adopted from our work  \cite{WM39} with $G_M$ chosen such that the
observed odd-even mass differences are
reproduced for the nuclei of the region, and the quadrupole pairing strength
$G_Q$ is assumed to be 0.18 times $G_M$.
The QQ-force
strength $\chi$ is obtained by the self-consistency 
relation \mbox{$2 \hbar\omega~\epsilon= 3\chi \bra\Phi\ack\hat Q_0\ack\Phi\ket$} \cite{KY95}.
The electromagnetic transition matrix elements are calculated  using the electromagnetic operator ${\cal M}$( E2)$_ {\mu}$
with the effective charges of $1.5e$ for protons and  $0.5e$ for neutrons.

The shell model Hamiltonian of Eq.~\ref{hamham} is then diagonalized
using the multi-quasiparticle 
basis states composed of 0-qp vacuum, two-proton, two-neutron and
four-qp configurations, i.e.,
\begin{eqnarray}\label{basis}
&&\hat P^I_{MK}\ack\Phi\ket;
\hat P^I_{MK}~a^\dagger_{\pi_1} a^\dagger_{\pi_2} \ack\Phi\ket;
\hat P^I_{MK}~a^\dagger_{\nu_1} a^\dagger_{\nu_2} \ack\Phi\ket;\\\nonumber
&&\hat P^I_{MK}~a^\dagger_{\pi_1} a^\dagger_{\pi_2}
a^\dagger_{\nu_1} a^\dagger_{\nu_2} \ack\Phi\ket,
\end{eqnarray}
where $''\pi''$ labels the quasiproton and $''\nu''$ designates the quasineutron states.
$P^I_{MK}$ is the standard three-dimensional angular-momentum projection operator \cite{RS80},
and $\ack\Phi\ket$ represents the triaxially-deformed quasiparticle  vacuum
state. The deformed quasiparticle state, $\ack\Phi\ket$, is obtained by first solving the Nilsson Hamiltonian with the
deformation parameters given in Table \ref{tab1}, and then performing the BCS calculations. The pairing gaps obtained
by solving the BCS equations are presented in Table \ref{tablep}. The pairing constants for the two studied Osmium isotopes have
been slightly adjusted as compared to our previous work \cite{SP24} in order to reproduce the yrast energies more accurately. The results
of the present work for the two isotopes are, therefore, slighly different from those presented in our earlier publications \cite{SP24}.

\section{Results and Discussion}
\begin{table}
\caption{Axial and triaxial quadrupole deformation parameters
$\epsilon$ and $\epsilon'$  employed in the TPSM calculation. Axial deformations are taken from \cite{Raman}
and nonaxial deformations are chosen in such a way that band heads
of the $\gamma $-bands are reproduced.   }
\begin{tabular}{p{0.6cm}p{1.2cm}p{1.2cm}p{1.2cm}p{1.2cm}p{1.2cm}p{1cm}}
\hline    \hline               & $^{76}$Ge       &$^{112}$Ru  & $^{188}$Os& $^{192}$Os  & $^{192}$Pt & $^{232}$Th   \\
   $\epsilon$      &0.200           & 0.289    & 0.183    &  0.164    &  0.150   &   0.210    \\
 $\epsilon'$       &0.160           & 0.130     & 0.088    &  0.085    &  0.087   &  0.085       \\
 $\gamma $          & 39$^{\circ}$            & 24$^{\circ}$         & 26$^{\circ}$        & 27$^{\circ}$       & 30$^{\circ}$       &   22$^{\circ}$        \\
\hline \hline
\end{tabular}\label{tab1}
\end{table}
\begin{table}
\caption{ Neutron ($\Delta_\nu$) and proton ($\Delta_\pi$) pairing gaps for $^{76}$Ge,
  $^{112}$Ru, $^{188,192}$Os, $^{192}$Pt and $^{232}$Th nuclei. These pairing parameters are same as used in our
  earlier work \cite{SP24} except for the two Osmium isotopes for which neutron pairing factors have been slightly
  adjusted to reproduce the near yrast energies more accurately. The earlier pair gaps used are shown in the
brackets.}
\begin{tabular}{p{0.7cm}p{1cm}p{1cm}p{1.5cm}p{1.5cm}p{1cm}p{1cm}}
\hline   \hline                & $^{76}$Ge       &$^{112}$Ru  & $^{188}$Os& $^{192}$Os  & $^{192}$Pt & $^{232}$Th   \\
   $\Delta_{\nu}$      &0.63           & 0.54    & 0.46 (0.40)    &  0.38 (0.33)    &  0.57   &   0.64    \\
 $\Delta_{\pi}$       &0.83           & 0.63     & 0.52     &        0.41     &  0.74   &  0.86       \\
\hline \hline
\end{tabular}\label{tablep}
\end{table}

In the present work, we have performed the TPSM analysis for the six nuclides of $^{76}$Ge,
$^{112}$Ru, $^{188,192}$Os, $^{192}$Pt and $^{232}$Th. As already mentioned in the introduction, 
based on the the staggering pattern of the  $\gamma$-band for the six isotopes 
with odd-spin states lower than the average of the neighbouring even-spin states, these nuclides were
categorized as $\gamma$-rigid \cite{SJ21,SP24}.

\begin{figure}[!ht]
    \begin{center}
        \hspace*{-0.2cm}\includegraphics[width=8.5cm, angle =0]{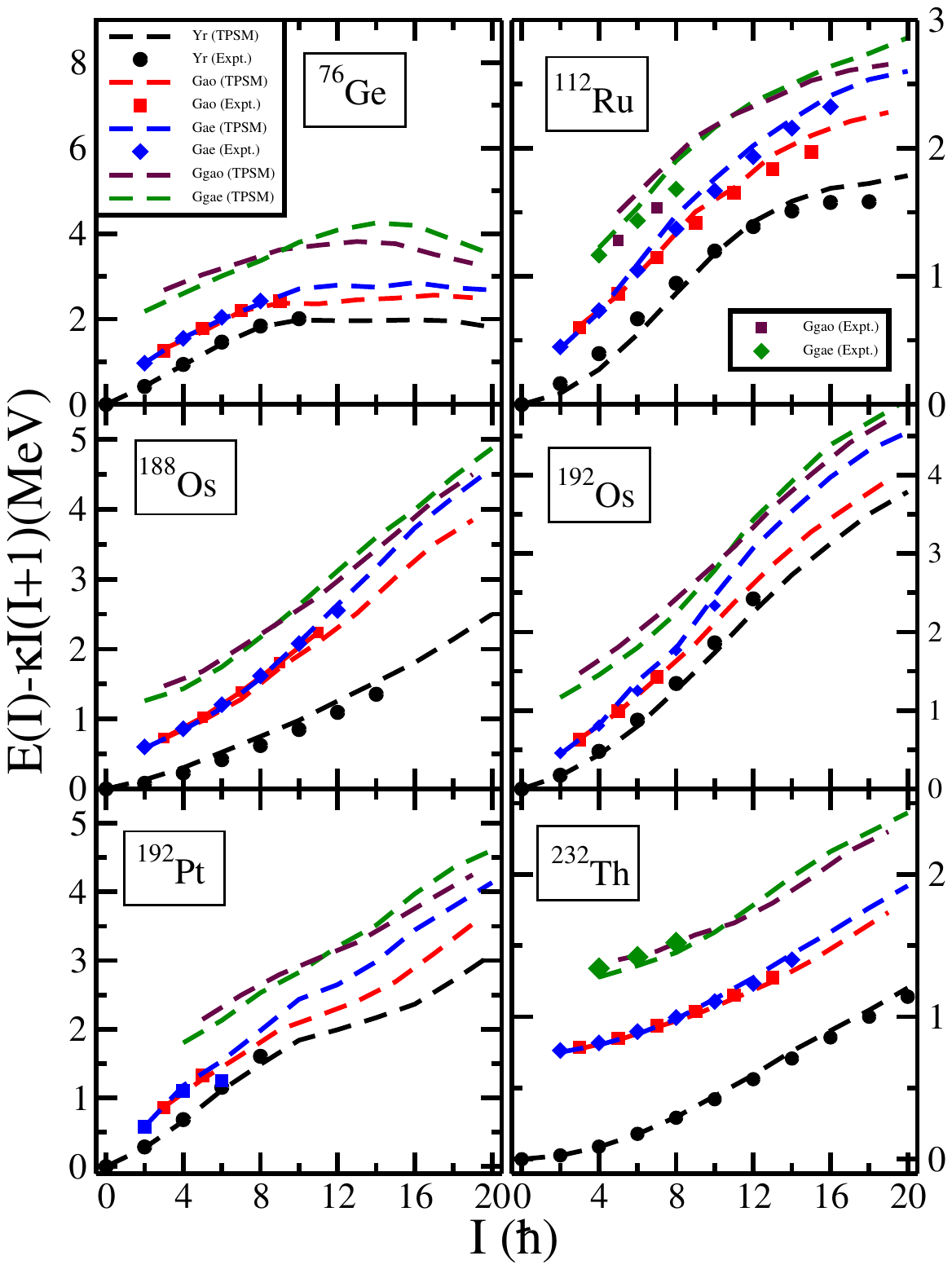}
    \end{center}
    \vspace{-0.5cm}
    \caption{Comparison of the TPSM energies after configuration mixing with the available experimental data  for $^{76}$Ge, $^{112}$Ru, $^{188, 192}$Os, $^{192}$Pt and $^{232}$Th. The curves are labelled as :  Yr - yrast band,  Gae - even-spin $\gamma$-band, Gao - odd-spin $\gamma$-band,
      Ggae - even-spin $\gamma\gamma$-band and Ggao - odd-spin $\gamma\gamma$-band. The value of $\kappa$, shown in y-axis, is
      set as $\kappa$ = 32.32$A^{-5/3}$. Data taken from Refs. \cite{ge76, ZAJAC, WU, KORTEN, martin}.}
  
    \label{Energy_CD_v1}
\end{figure}

\begin{figure}[!ht]
    \begin{center}
        \hspace*{-0.2cm}\includegraphics[width=8.5cm, angle =0]{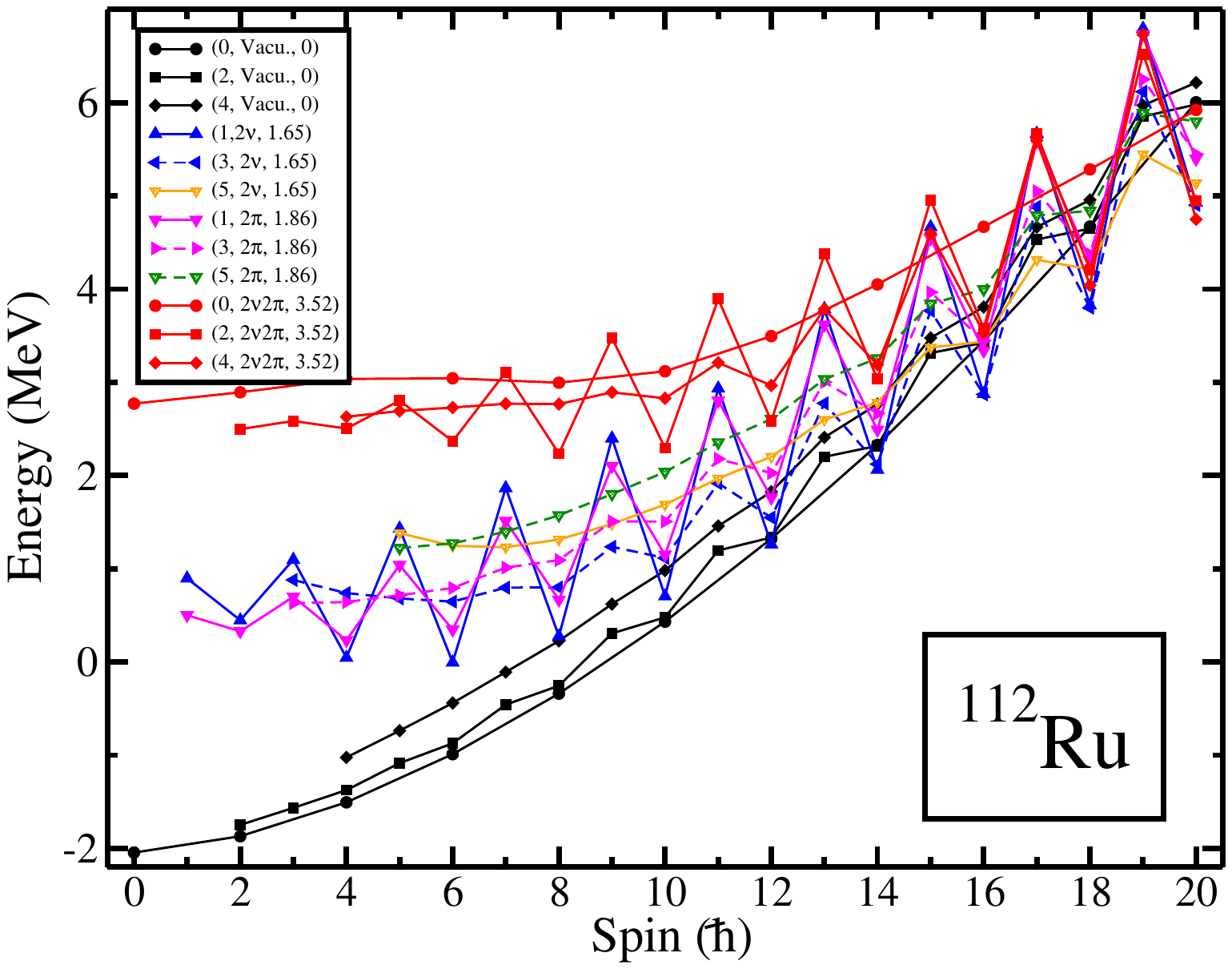}
    \end{center}
    \vspace{-0.5cm}
    \caption{ TPSM projected energies, before band mixing, of $^{112}$Ru nucleus. 
      The curves are labelled by three quantities:  K quantum number, quasiparticle character and energy of the state. For instance,
      (1, 2$\nu$, 1.65) designates two quasineutron state with K= 1 having an intrinsic energy of 1.65 MeV.
  }
    \label{76Ge_BD_ND_3rdjuy_v1}
\end{figure}


\begin{figure}[!ht]
    \begin{center}
        \hspace*{-0.2cm}\includegraphics[width=8.5cm, angle =0]{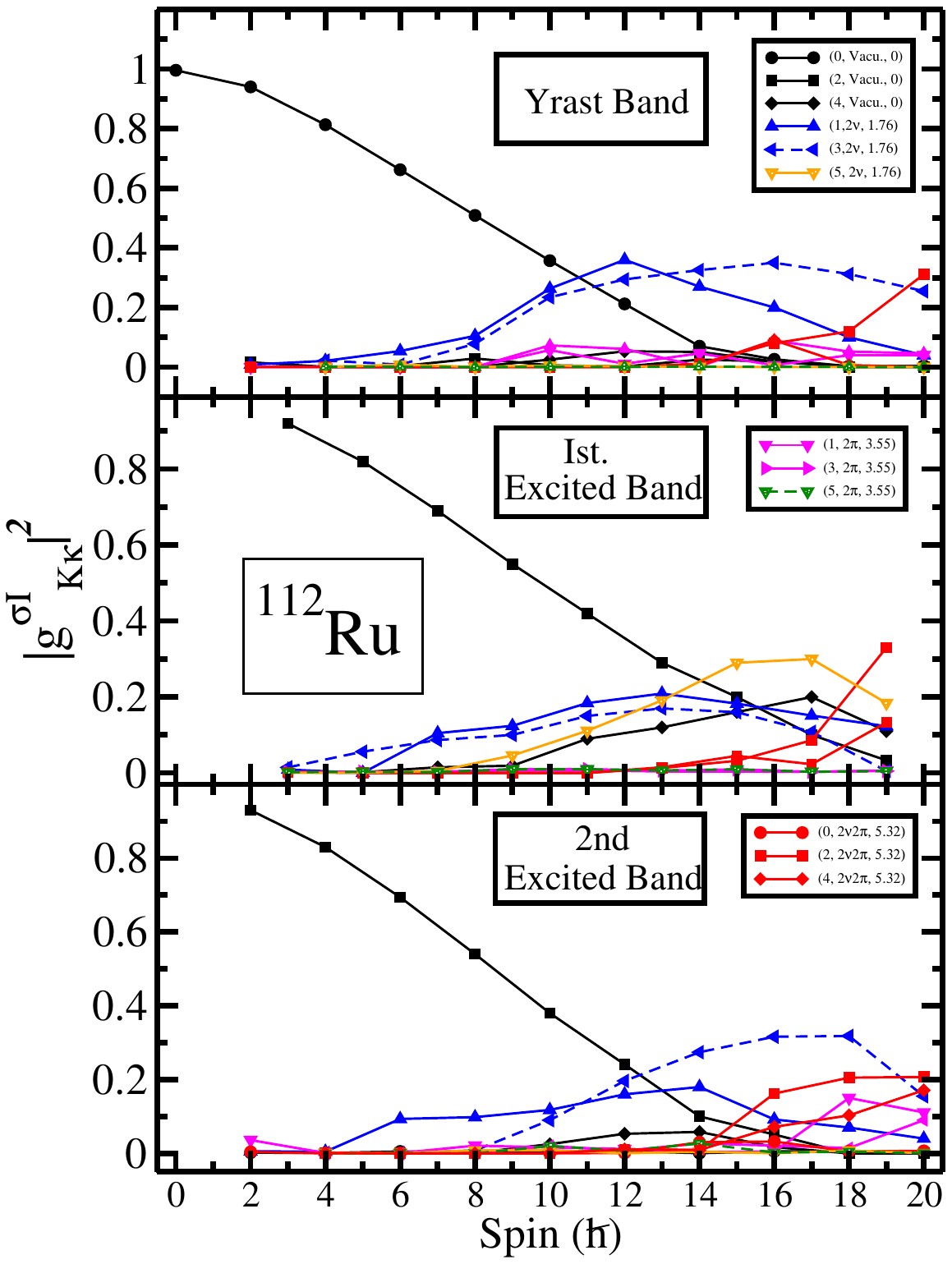}
    \end{center}
    \vspace{-0.5cm}
    \caption{ Probabilities of various projected K-configurations in the wavefunctions of $^{112}$Ru nucleus  after diagonalization. The symbols and colour scheme used for different bands in this figure is consistent with the band diagram figure, Fig. \ref{76Ge_BD_ND_3rdjuy_v1}.    }
    \label{76Ge_wf_SEc1_12july24}
\end{figure}


\begin{figure}[!ht]
    \begin{center}
        \hspace*{-0.2cm}\includegraphics[width=8.5cm, angle =0]{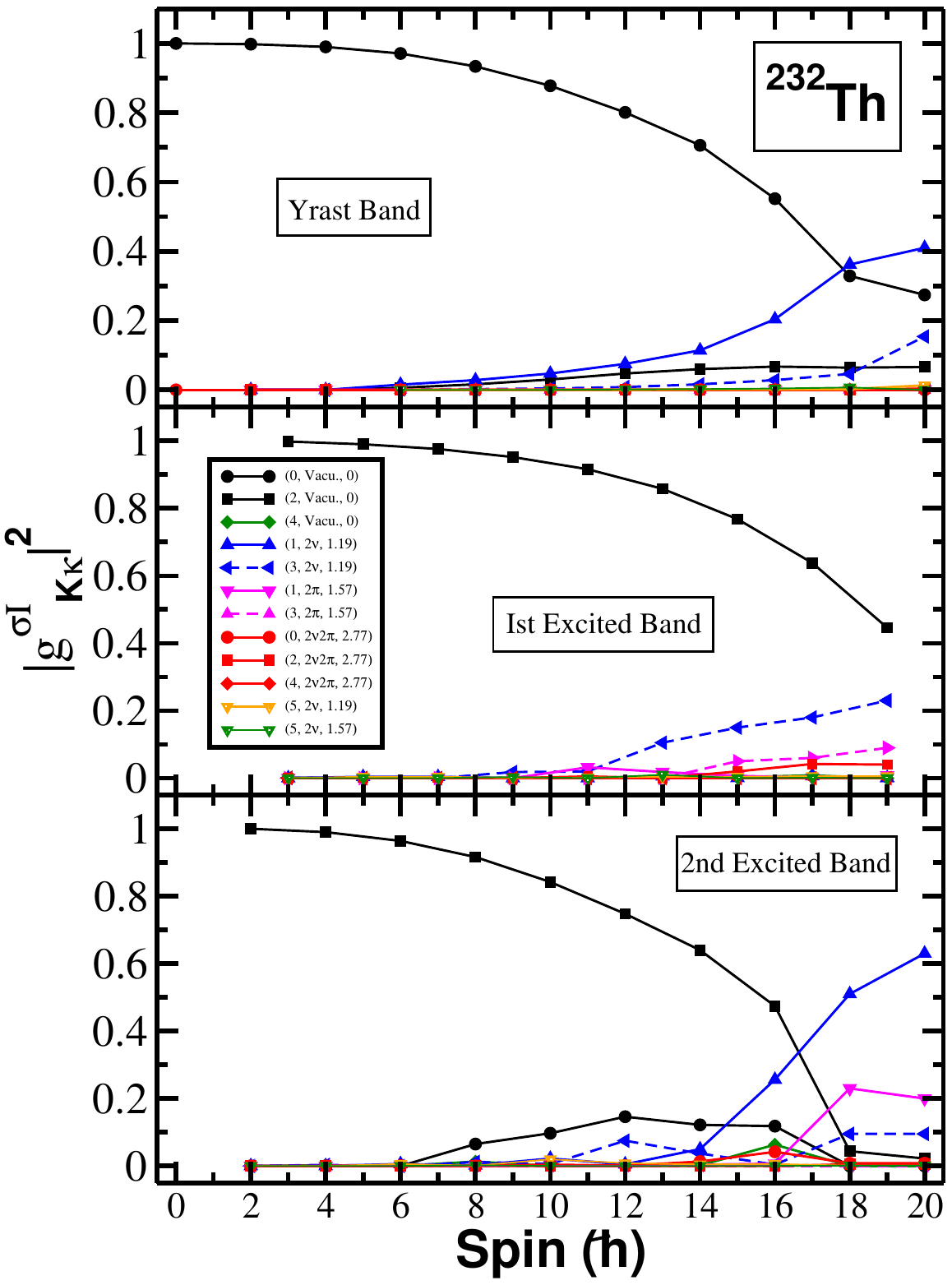}
    \end{center}
    \vspace{-0.5cm}
    \caption{ Probabilities of various projected K-configurations in the wavefunctions of $^{232}$Th nucleus  after diagonalization.    }
    \label{232Th_wave_Sec4}
\end{figure}

\begin{figure}[!ht]
    \begin{center}
        \hspace*{-0.2cm}\includegraphics[width=8.5cm, angle =0]{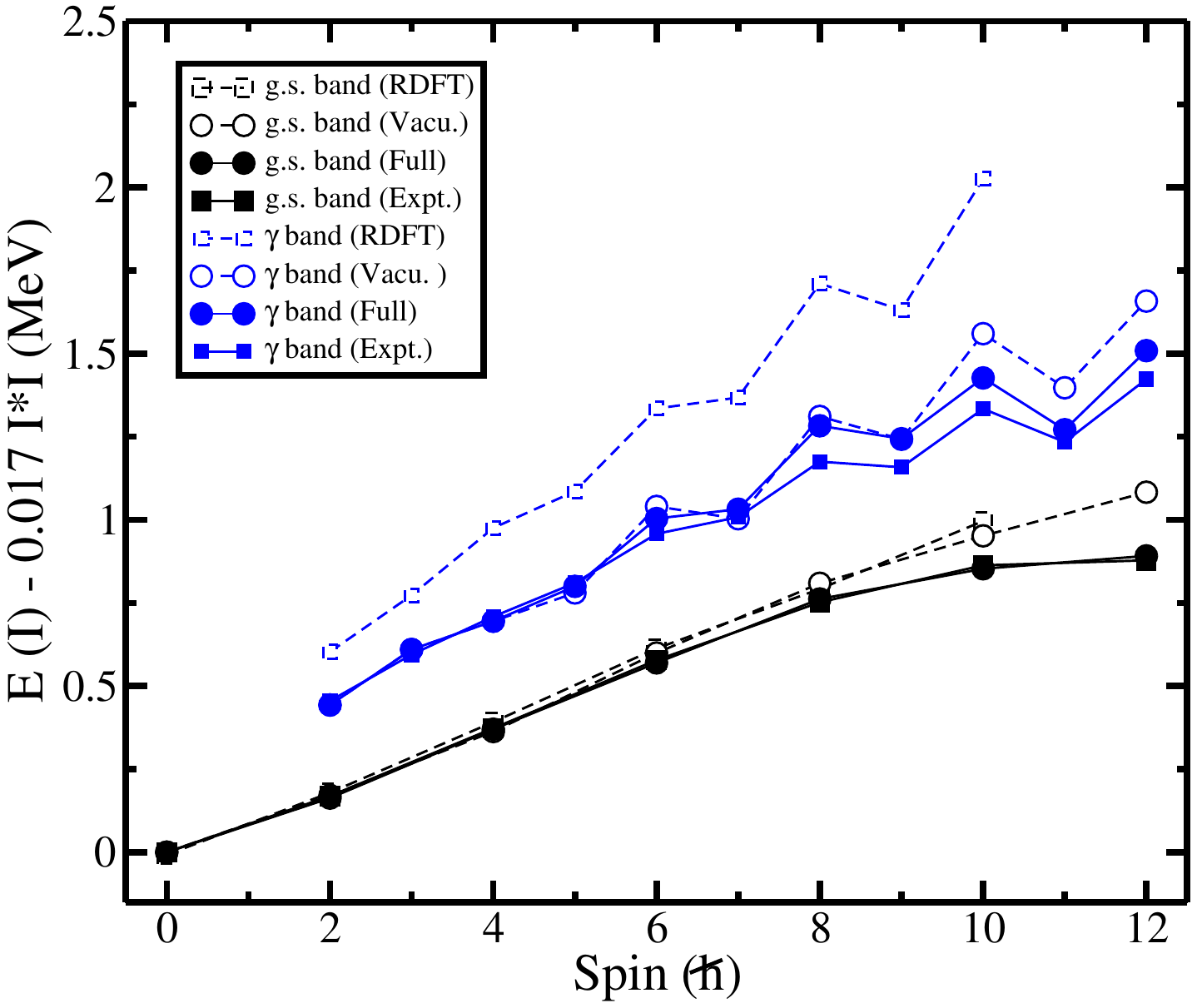}
    \end{center}
    \vspace{-0.5cm}
    \caption{Comparison of the ground-state (g.s.) band and $\gamma$-band calculated by   TPSM  and  relativistic density functional theory (RDFT) \cite{WANG2024859} with the available experimental data  for $^{112}$Ru. The energies are subtracted by rigid rotor energies 0.017${\textrm {I}}^2$, which is same as used in Ref.~ \cite{WANG2024859}.
  }
    \label{112ru_gammaband_RDFT}
\end{figure}

\begin{figure}[!ht]
    \begin{center}
        \hspace*{-0.2cm}\includegraphics[width=8.5cm, angle =0]{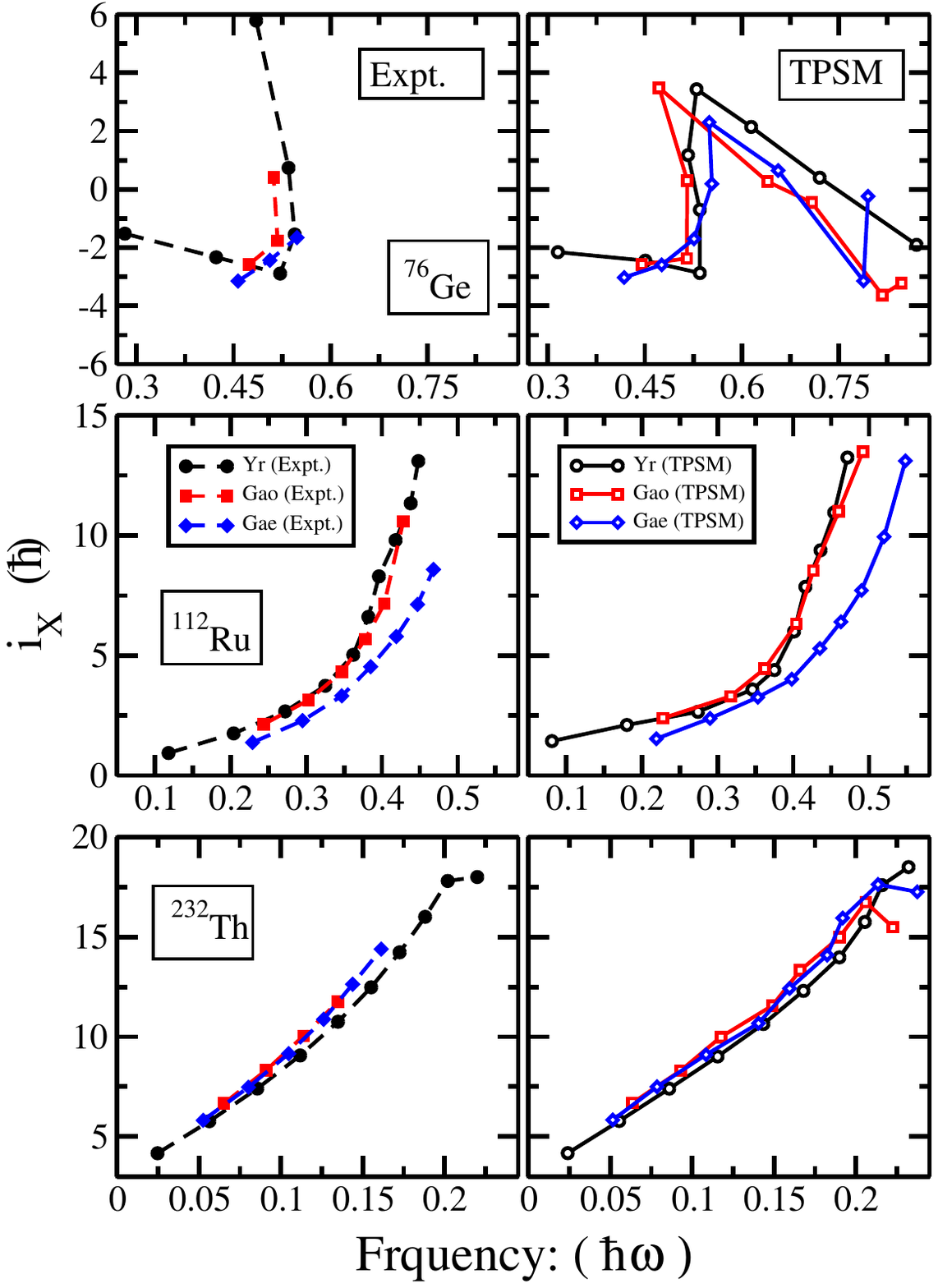}
    \end{center}
    \vspace{-0.5cm}
    \caption{ Comparison of the aligned angular
 momentum obtained from the measured energy levels
 and those calculated
 from the TPSM results for Yr, Gao and Gae bands of $^{76}$Ge,  $^{112}$Ru and $^{232}$Th nuclei.
 The aligned angular-momentum is calculated from the expressions, $i_x=I_x(\omega)-I_{x,ref}(\omega)$,
 where $\hbar\omega=\frac{E_{\gamma}}{I_x^i(\omega)-I_x^f(\omega)}$,  $I_x(\omega)= \sqrt{I(I+1)-K^2}$
 and $I_{x,ref}(\omega)=\omega(J_0+\omega^{2}J_1)$. The Harris reference parameters, $J_0$ and $J_1$, are taken from \cite{j1,Sohler2025,Sohler2012,Sithole2021}.}
    \label{ix_spin_1_July2024}
\end{figure}

\begin{figure}[!ht]
    \begin{center}
      \hspace*{-0.2cm}\includegraphics[width=8.5cm, angle =0]{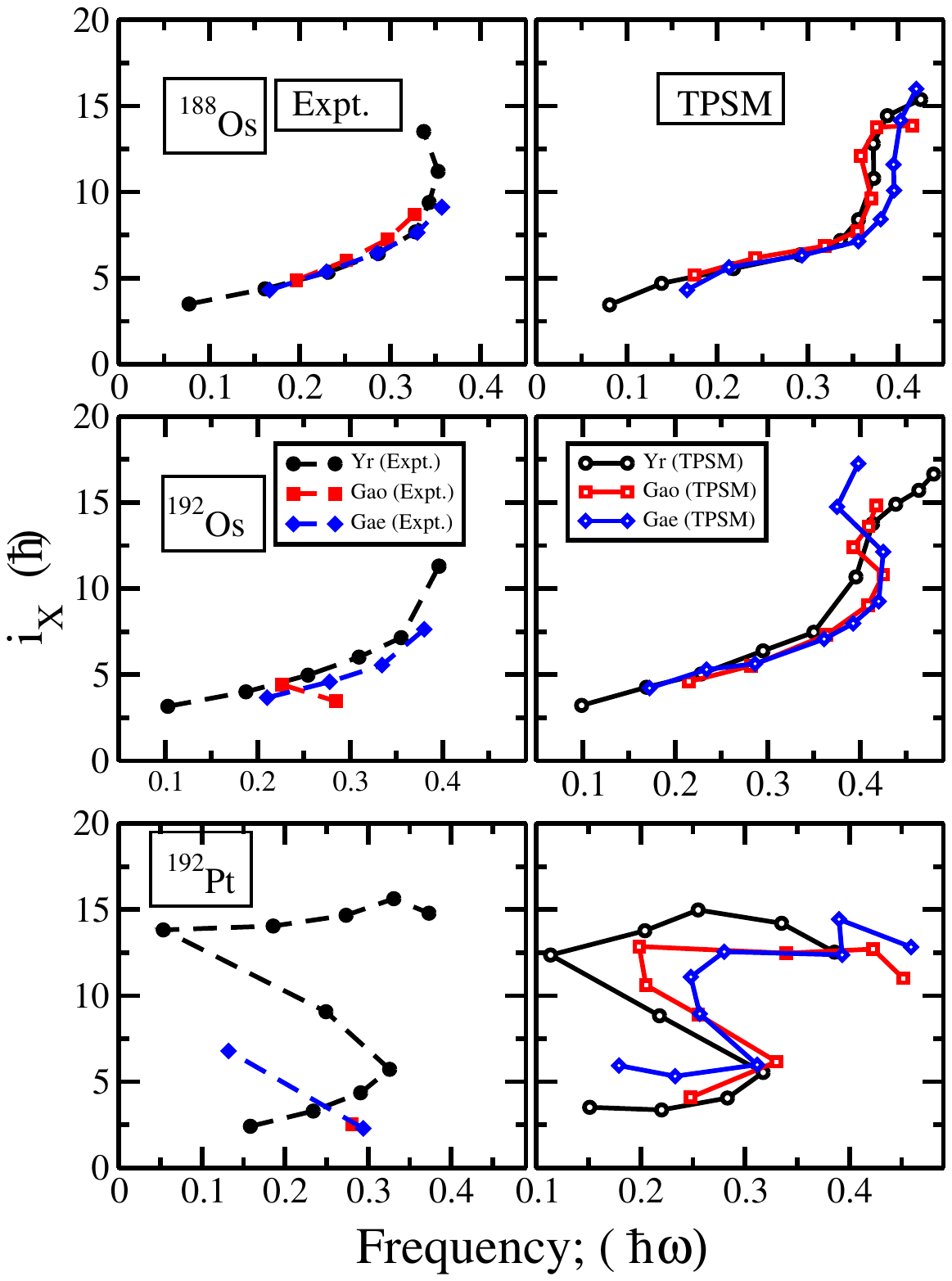}
    \end{center}
    \vspace{-0.5cm}
    \caption{ Comparison of the aligned angular
      momentum obtained from the measured energy levels and those calculated
      from the TPSM results, for  Yr, Gao and Gae bands of  $^{188, 192}$Os and  $^{192}$Pt  nuclei.
    The expressions for the aligned angular momentum is provided in the caption of Fig.~\ref{ix_spin_1_July2024}.}
    \label{ix_spin_2_July2024}
\end{figure}

\begin{figure}[!ht]
    \begin{center}
        \hspace*{-0.2cm}\includegraphics[width=8.5cm, angle =0]{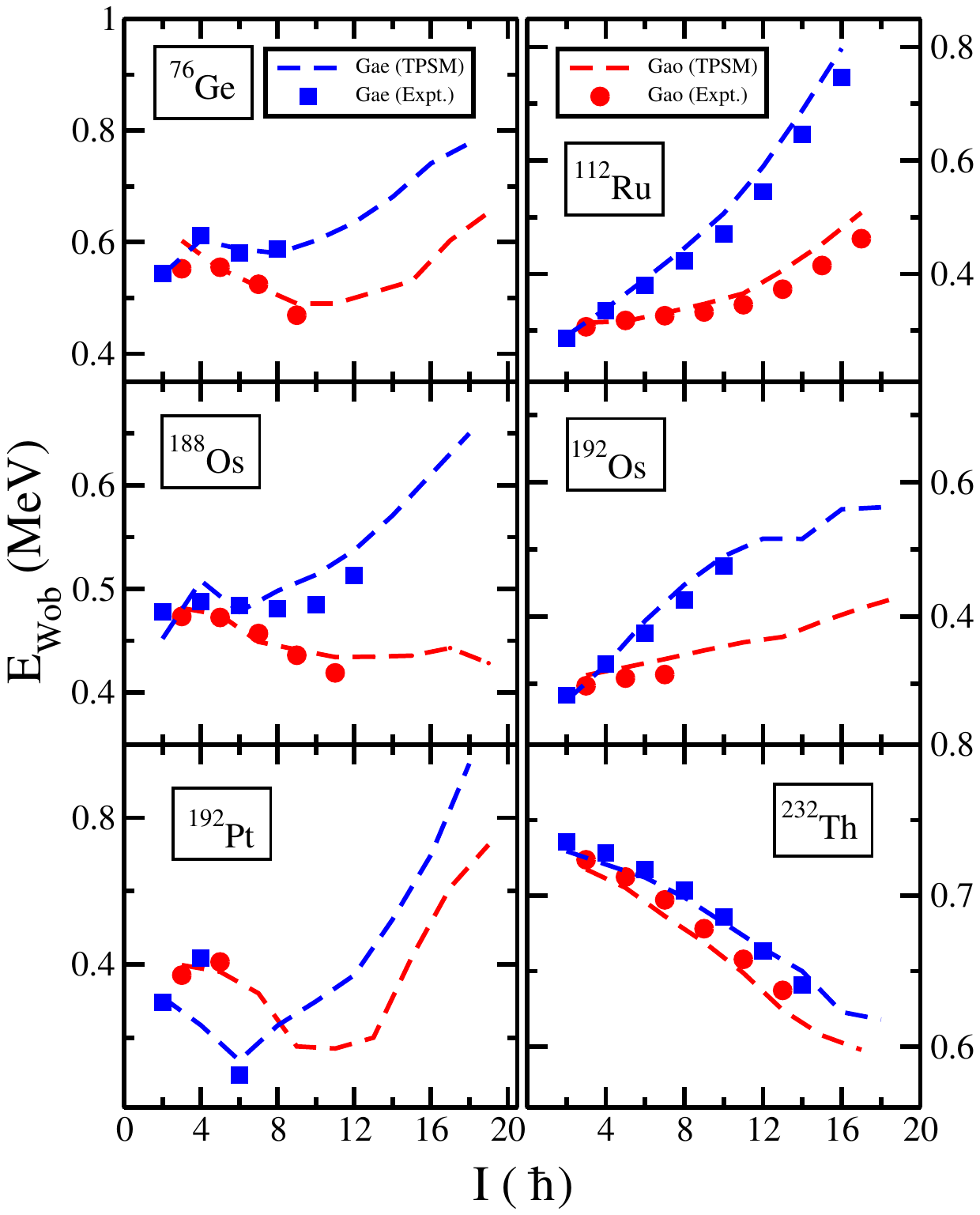}
    \end{center}
    \vspace{-0.5cm}
    \caption{ TPSM wobbling energies are compared with the experimental values obtained for Yr, Gao and Gae bands of
          $^{76}$Ge, $^{112}$Ru, $^{188, 192}$Os, $^{192}$Pt and $^{232}$Th.
  }
    \label{Wobling_v1}
\end{figure}






\begin{figure}[!ht]
    \begin{center}
        \hspace*{-0.2cm}\includegraphics[width=8.5cm, angle =0]{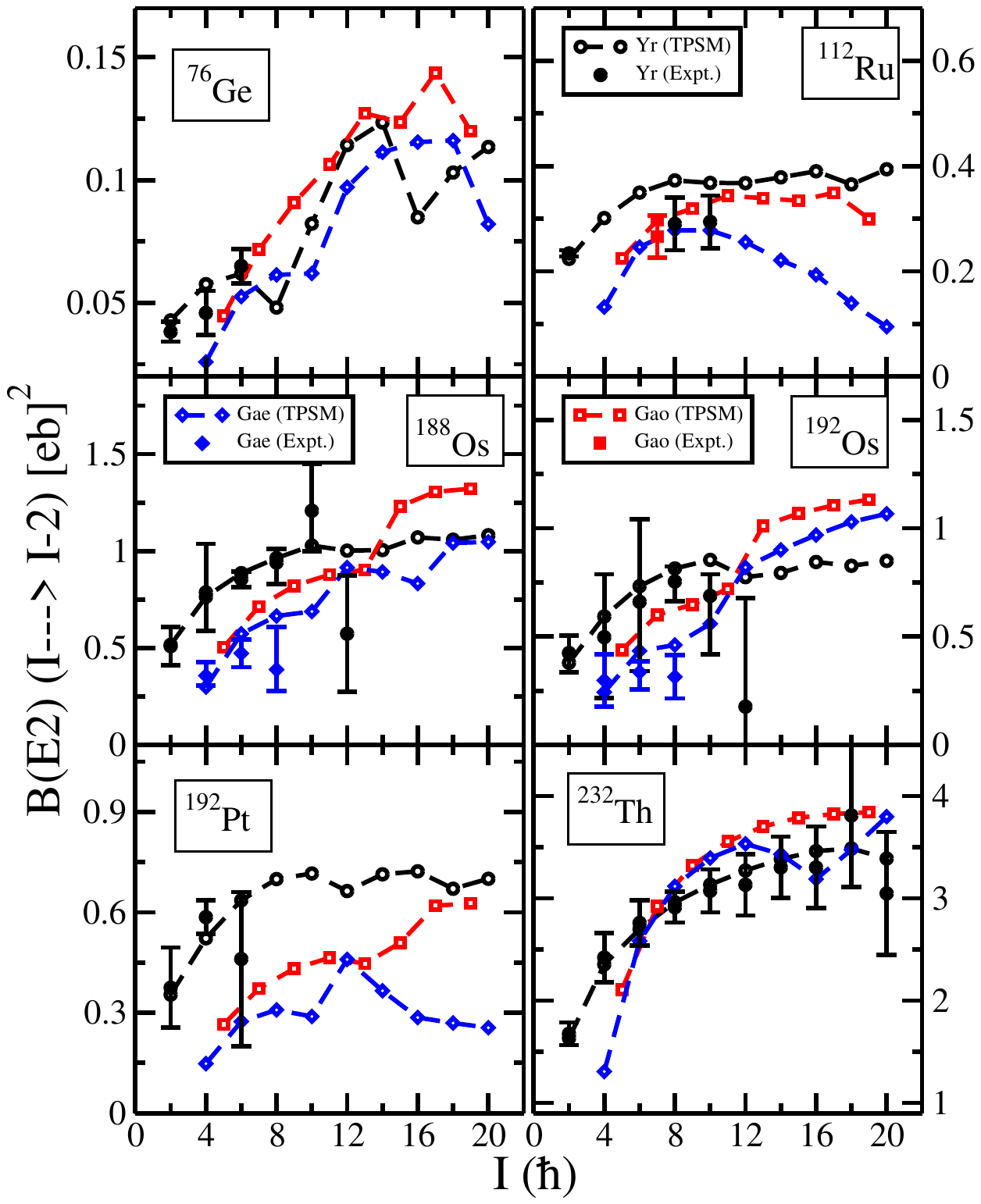}
    \end{center}
    \vspace{-0.5cm}
    \caption{Comparison of the  calculated TPSM values for in-band B(E2) transitions with available experimental data  as  functions of I$(\hbar)$ for  Yr, Gao and Gae bands.}
    \label{Trans_BE2_18Nov_v1}
\end{figure}

\begin{figure}[!ht]
    \begin{center}
        \hspace*{-0.2cm}\includegraphics[width=8.5cm, angle =0]{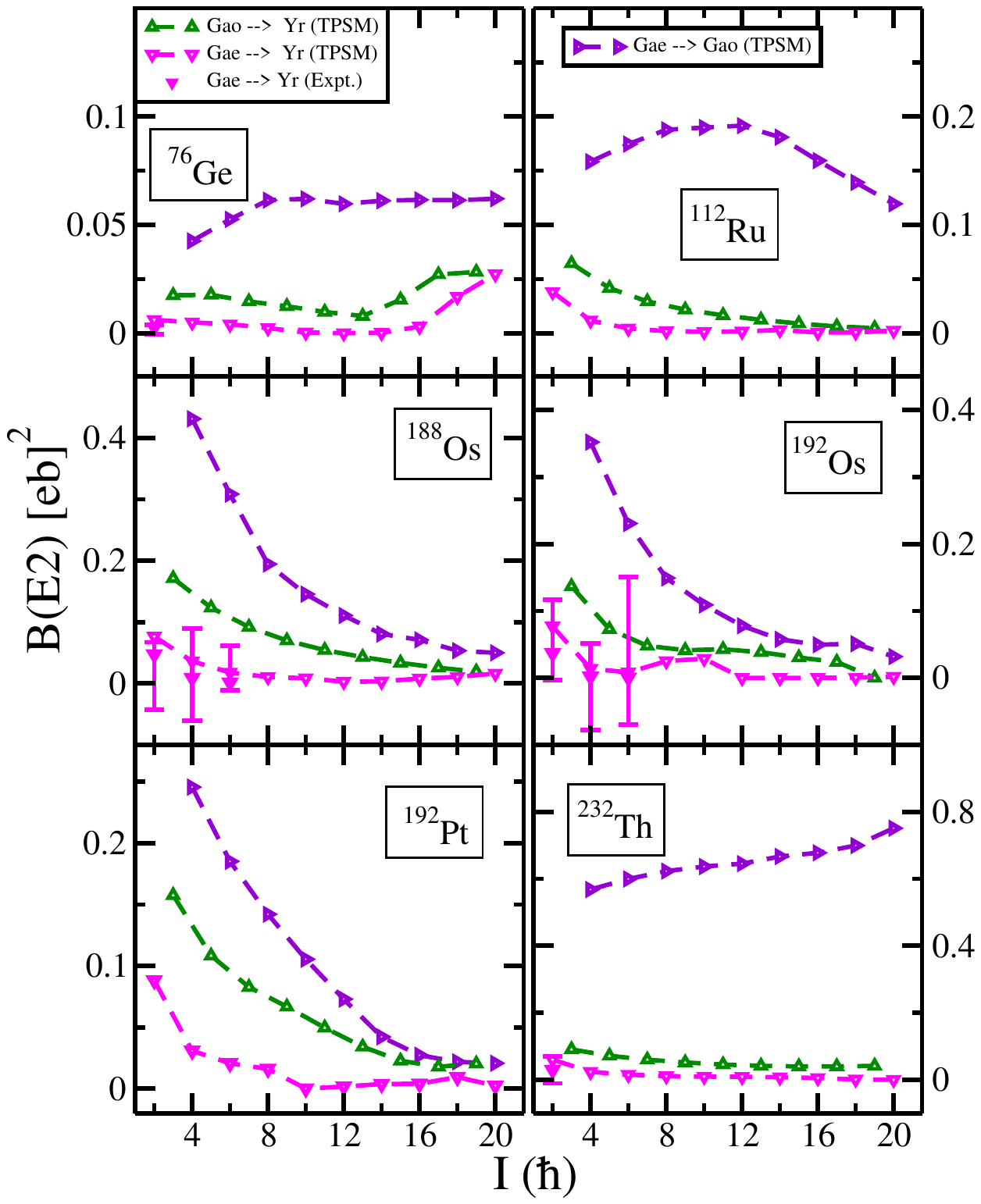}
    \end{center}
    \vspace{-0.5cm}
    \caption{Comparison of the  calculated TPSM values for inter-band B(E2) transitions with available experimental data  as  functions of I$(\hbar)$ for  Yr, Gao and Gae bands.}
    \label{Trans_BE2_18Nov_v2}
\end{figure}







\begin{figure}[!ht]
    \begin{center}
        \hspace*{-0.2cm}\includegraphics[width=8.5cm, angle =0]{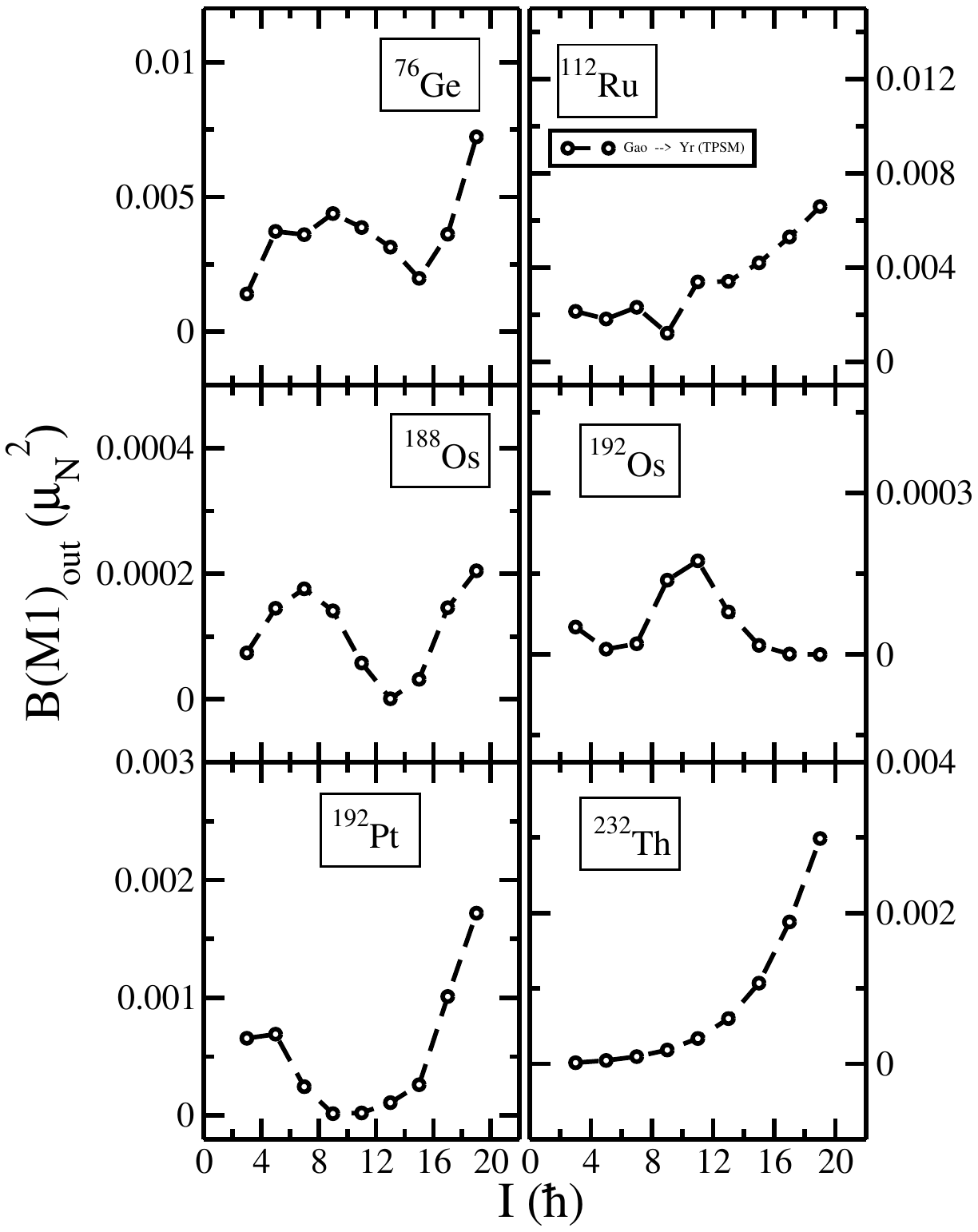}
    \end{center}
    \vspace{-0.5cm}
    \caption{Calculated values from TPSM inter-band B(M1) transitions as  functions of I($\hbar$) for Bands $Gao \rightarrow Yr$. }
    \label{BM1out_v1}
\end{figure}



\begin{figure}[!ht]
    \begin{center}
        \hspace*{-0.2cm}\includegraphics[width=8.5cm, angle =0]{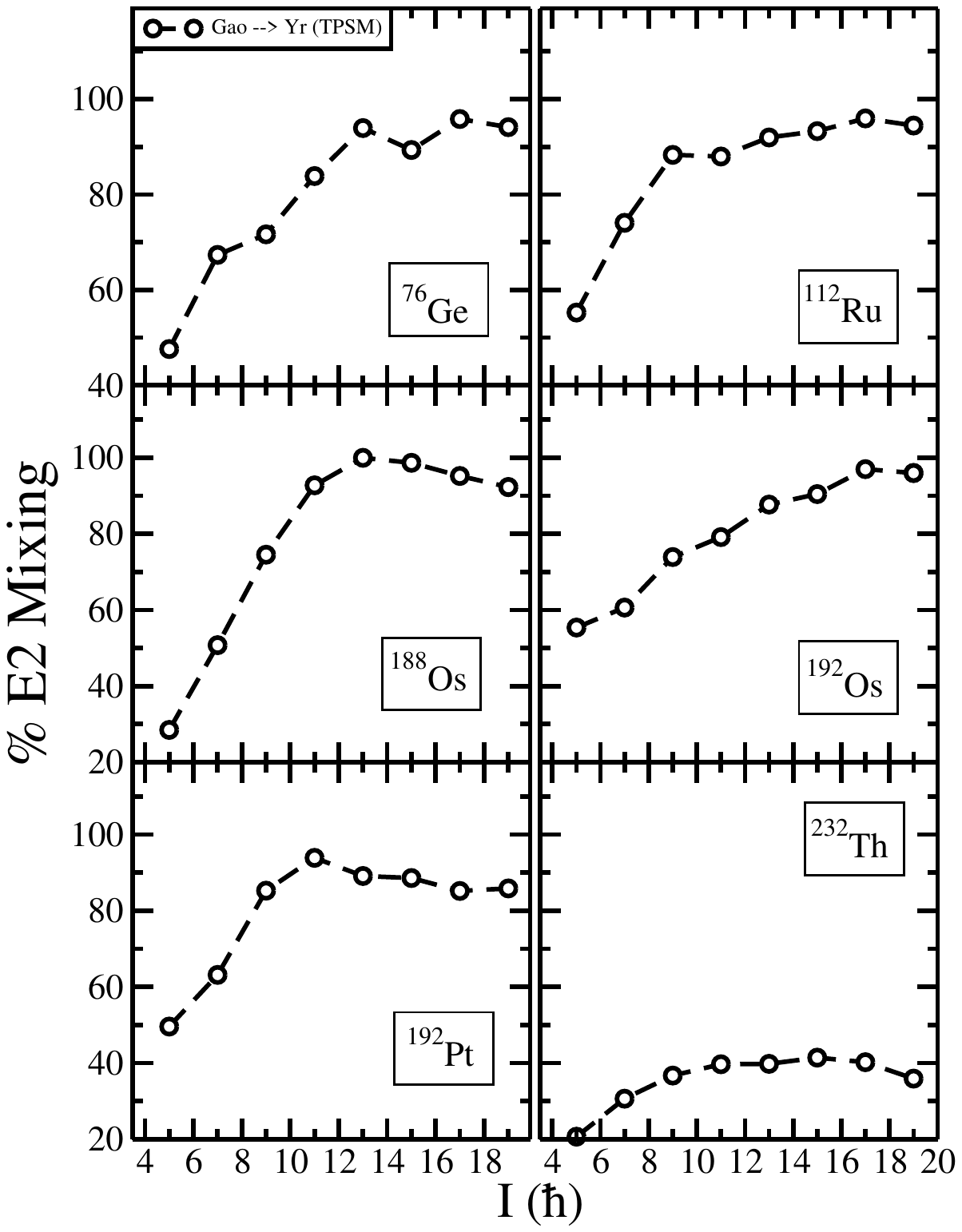}
    \end{center}
    \vspace{-0.5cm}
    \caption{Calculated values from TPSM for E2 mixing $(\%)$ using E2 = $\delta^2/1+\delta^2,\,\,\ \delta = 0.835E_\gamma\frac{<I|E2|I-1>}{<I|M1|I-1>}$ \cite{FD82}  as  functions of I$(\hbar)$ for Bands $Gao \rightarrow Yr$.}
    \label{E2_percentage_fig}
\end{figure}

\begin{figure}[!ht]
    \begin{center}
        \hspace*{-0.2cm}\includegraphics[width=9.5cm, angle =0]{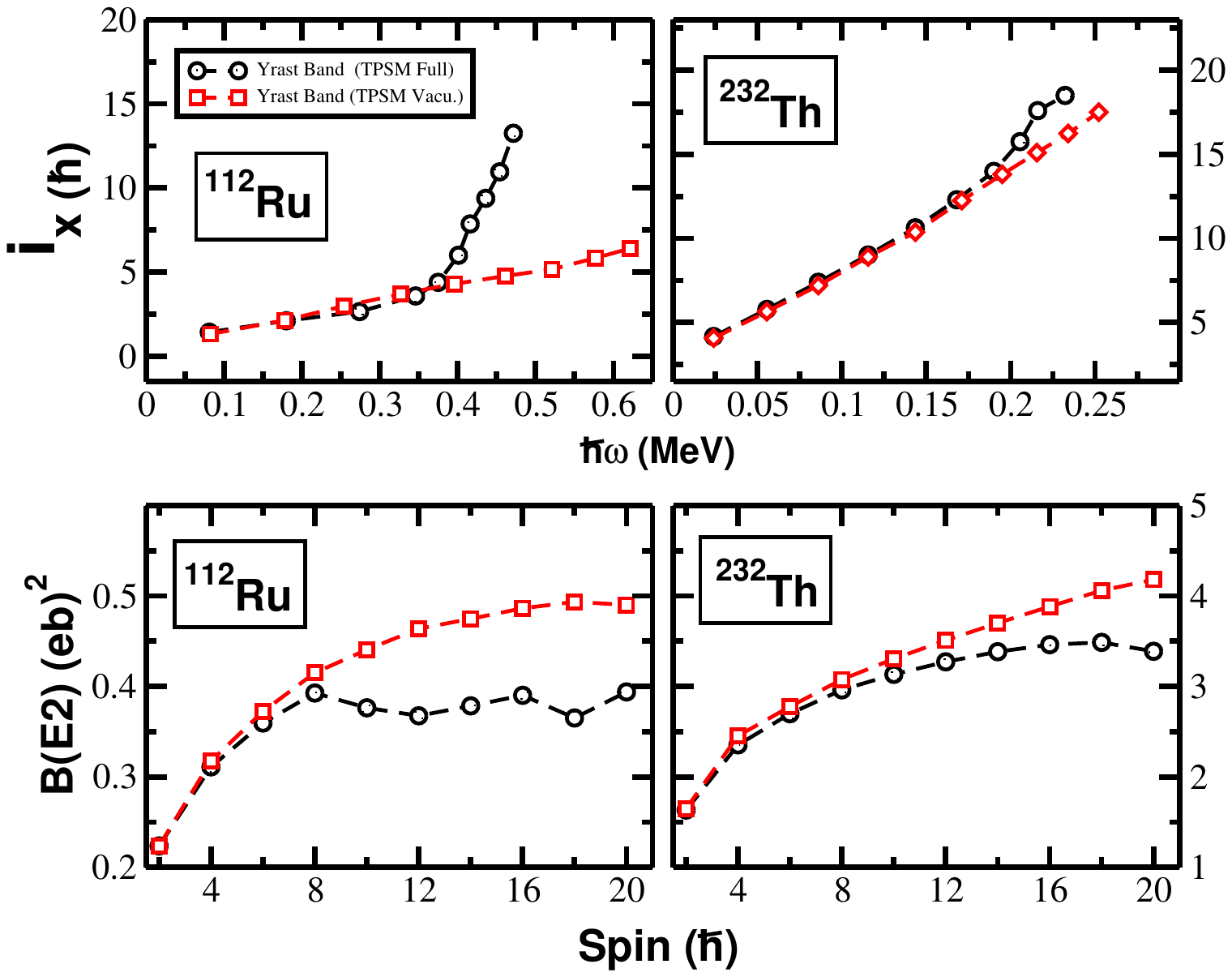}
    \end{center}
    \vspace{-0.5cm}
    \caption{Calculated  aligned angular momentum ($i_x$)  and B(E2) transitions  as  functions of $\hbar\omega$ and  spin, respectively
      for two representative systems of $^{112}$Ru and $^{232}$Th. The expression for $i_x$
      is provided in the caption of Fig.~\ref{ix_spin_1_July2024}.}
    \label{fig13}
\end{figure}


\begin{figure}[htbp]
\centerline{\includegraphics[trim=0cm 0cm 0cm
0cm,width=0.55\textwidth,clip]{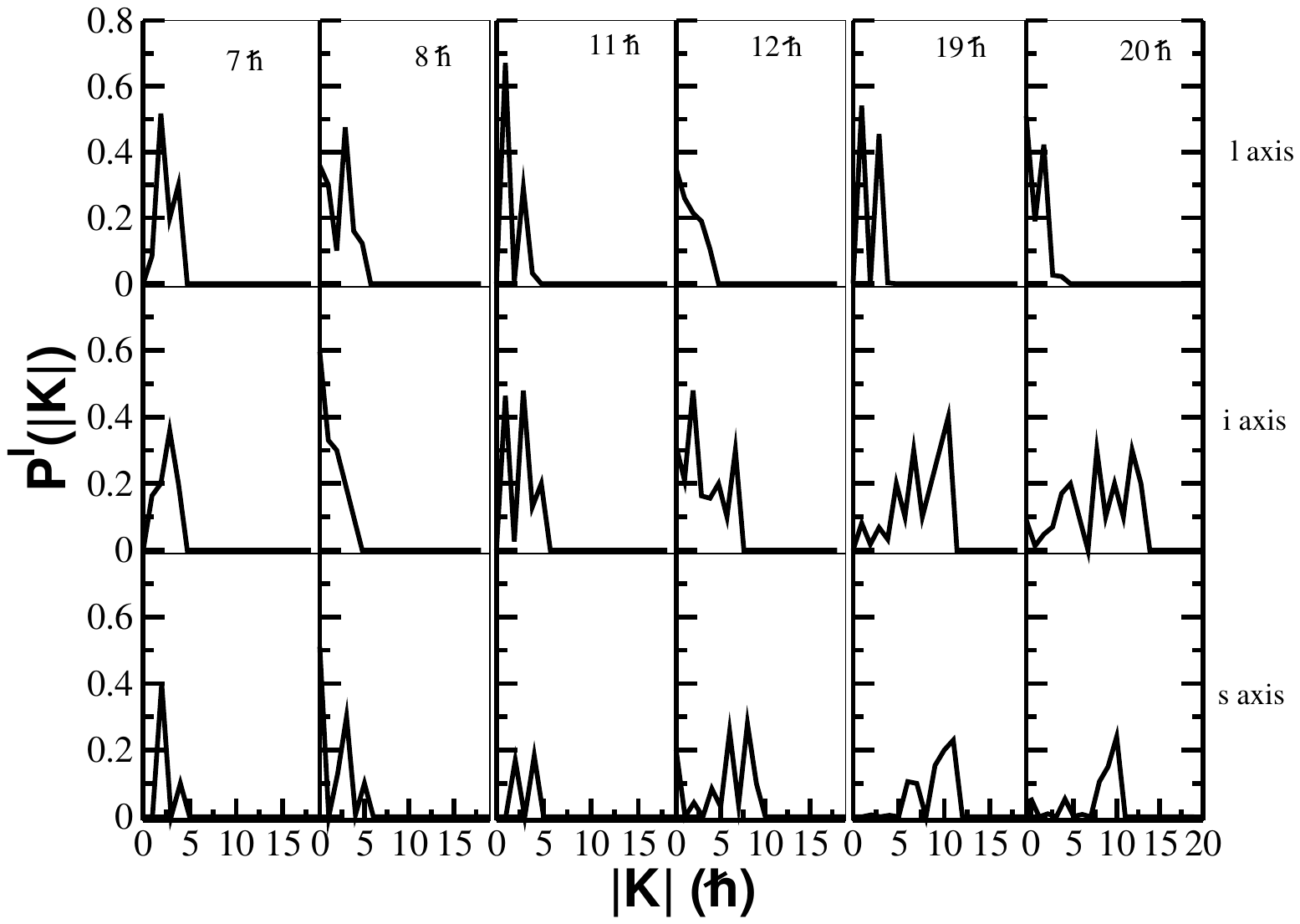}}
\caption{\label{fig:pkk}
  The $K$ distributions of angular momenta on the three principal axes for the zero- (even spin) and one- (odd spin) phonon band at several selected angular momenta. The K distributions on the long, intermediate, and short axes are shown for $^{112}$Ru in the first, second, and third rows, respectively.
  }
 \label{76GE_pk_vs_k}
\end{figure}
\begin{figure}[htbp]
\centerline{\includegraphics[trim=0cm 0cm 0cm
0cm,width=0.55\textwidth,clip]{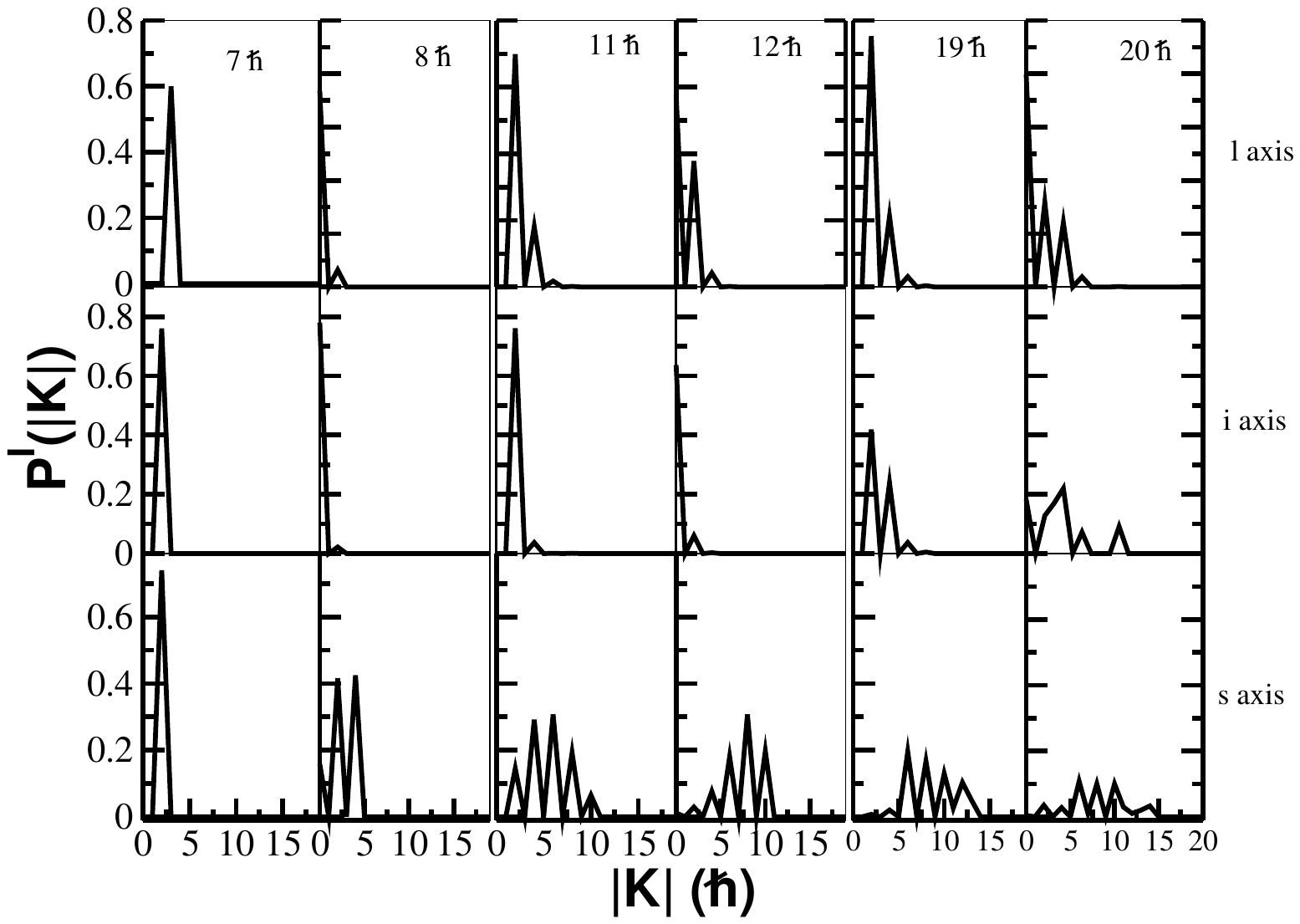}}
\caption{\label{fig:pkk1}
  The $K$ distributions of angular momenta on the three principal axes for the zero- (even spin) and one- (odd spin) phonon band at several selected angular momenta. The K distributions on the long, intermediate, and short axes are shown for $^{232}$Th in the first, second, and third rows, respectively.
  }
 \label{232Th_pk_vs_k}
\end{figure}
 
  We begin the discussion of the results by displaying the energies of the lowest three bands in Fig.~\ref{Energy_CD_v1} for each of
  the six nuclides
  studied. The first excited band is the
  $\gamma$-band and the second excited band is the $\gamma\gamma$-band in the low-spin region of these nuclei \cite{SJ21,SP24}.
  It is noted from the figure that the splitting between even- and odd-spin member branches for both the excited bands
  varies with increasing spin.
  For $^{76}$Ge, the TPSM energies for the odd-spin states of the $\gamma$-band, labelled as $"Gao"$ and what is referred
  to as $n_{\omega}=1$
  in the wobbling terminology, are almost degenerate with the even-spin member till I=7, and then above this spin
  value the odd-spin states are favoured in energy. The reason for this lowering above I=7 is because
  two-quasiparticle band crosses the ground-state band around this spin value, and the odd-spin members above
  I=7 originate from the $\gamma-$band based on the two-quasiparticle configuration. This crossing
  phenomenon has been already studied in our earlier publication \cite{TPSM2014} and will also be discussed in the following.
  It is evident from Fig.~\ref{Energy_CD_v1}
  that TPSM energies are in good agreement with the
  known experimental values, in particular, the lowering of the observed energy of I=9 state of the $\gamma$-band is reproduced
  quite well by the TPSM calculations. For the $\gamma\gamma$-band. the even-spin states are lower than the odd-spin states in the
  low-spin region, however, it is noted that the odd-spin branch crosses the even-spin branch at I=8 and becomes favoured in energy for high-spin states.
  The spectrum of states after the band crossing is similar to what is expected from the wobbling perspective, almost equally spaced
  alternating even- and odd-spin band structures \cite{BM75}. 
 
  For $^{112}$Ru, $^{188}$Os, $^{192}$Os and $^{192}$Pt nuclides in Fig.~\ref{Energy_CD_v1}, the energy of the odd-spin members of the
  $\gamma$-band are again noted to be lower in energy as compared to the even-spin states at high-spin. For
  low-spin states, the odd-spin states are almost degenerate, but for higher spin values the two branches develop a
  splitting. In the case of $^{112}$Ru, the data is available up to I=16 and it is noted that TPSM approach
  reproduces the observed energies quite well. 
  In the case of $^{232}$Th, although the odd-spin branch is slightly lower than
  the even-spin branch, but the two branches remain close in energy and cannot be called as wobbling bands. In the
  limiting case of harmonic spectrum as expected for large angular-momenta, the bands should be equally spaced in the
  wobbling mode. It is noted that
  for $^{76}$Ge,  $^{112}$Ru, $^{188}$Os, $^{192}$Os and $^{192}$Pt, the energy spectra tend to become equally spaced
  at high-spin.

  To investigate the cause for the lowering of the odd-spin members of the $\gamma$-band at high-spin,
  the projected energies for each intrinsic configuration are displayed in Fig.~\ref{76Ge_BD_ND_3rdjuy_v1},
  before configuration mixing,
  for $^{112}$Ru as an illustrative example. We have selected this nucleus as it has different band crossing
  features in comparison to $^{76}$Ge which  was presented in our earlier publication \cite{TPSM2014}. In Fig.~\ref{76Ge_BD_ND_3rdjuy_v1}, the
  crossings have diabatic character and it is easy to visualise the band crossing phenomena. It is noted from 
  Fig.~\ref{76Ge_BD_ND_3rdjuy_v1} that the ground-state band, which is the K=0 projected state from the vacuum configuration,
  is crossed at I=12 by a K=1 configuration belonging to a two-neutron state with quasiparticle energy of
  1.65 MeV. What is interesting is that the $\gamma$-band based on this two-neutron state, having the
  same quasiparticle energy but with K=3, also crosses the ground-state band and it becomes lower than
  the parent configuration for I=16 and 18 states. Above I=18, the four-quasiparticle configuration,
  having two-neutrons and two-protons aligned, becomes lower in energy.
  The odd-spin states at low-spin belong to the $\gamma$-band and at high-spin they originate
  from the K=3 and 5 configurations. The bands based on K=3 and 5 configurations have lower odd-even splitting and
  is the reason that odd-spin members become lower in energy. 
   
  It is observed from Fig.~\ref{Energy_CD_v1} that energies of the odd-spin branch of the $\gamma$-band becomes lower in energy
  as compared to the even-spin branch even before the band crossing spin value of I=12 for $^{112}$Ru. This is in contrast
  to $^{76}$Ge where the band crossing occurs at I=8. To examine this apparent anomaly in $^{112}$Ru, 
  the orthonormalized wavefunction
  amplitudes are displayed in Fig.~\ref{76Ge_wf_SEc1_12july24} for the lowest three bands of $^{112}$Ru. The TPSM  orthonormalized
  amplitude is given by \cite{WANG2020135676,TPSM10}
  \begin{equation}
g^{\sigma}_{K\kappa} = \sum_{K^\prime\kappa^\prime} f^\sigma_{K^\prime\kappa^\prime}\langle K\kappa| \hat{P}^I_{MK^\prime} | \phi_{\kappa^\prime} \rangle ~~~~,
  \end{equation}
  where $|K\kappa \rangle$ is the orthonormal basis set, and $f^\sigma_{K^\prime\kappa}$ is the non-orthonormal amplitude given
  by the TPSM wavefunction
   \begin{equation}
\Psi^{\sigma}_{IM} = \sum_{K\kappa} f^\sigma_{K\kappa}~\hat{P}^I_{MK^\prime} | \phi_{\kappa} \rangle ~~~~.
   \end{equation}
  It is noted from Fig.~\ref{76Ge_wf_SEc1_12july24}
  that the yrast-band at low-spin, shown in the top panel of the
  figure, has dominant K=0 component of the vacuum state. For I=8 and 10, the wavefunctions are mixed, having large
  contributions from K=1 and 3 of the two-neutron aligned configurations apart from the K=0 of the vacuum
  state. For I=12, 14 and 16 states, the wavefunctions have
  dominant K=1 and 3 two-neutron configurations. 
  I=18 has a significant four-quasiparticle configuration and for I=20
  this configuration is the dominant one. 
  The first excited band, shown in the middle panel of Fig.~\ref{76Ge_wf_SEc1_12july24},
  has K=2 configuration based on the vacuum state as the dominant component for the spin states from
  I=3 to 13. However, it is noted that the contributions from K=1, 3 and 5 of the two-neutron aligned
  configuration become increasingly important with spin. I=15 and 17 states have the dominant K=5 component of the two-neutron
  aligned configuration, and I=19 has four-quasiparticle configuration as the dominant component.
  For the second
  excited band, the even-spin states from I=2 to 10 originate from the K=4 state based on the vacuum configuration.
   I=12 state has almost equal contributions from K=4 of the vacuum and K=1 and 3 of the two-quasiparticle state. 
   I=14, 16 and 18 have mixed K=1 and 3 components and also some contribution from the four-quasiparticle state.

  The above wavefunction analysis reveals that yrast band after band crossing, having even-spin states,
  is mainly composed of the $\gamma$-band based on the two-quasiparticle configuration. The first excited band with odd-spin members
  is predominantly made up of $\gamma$-band built on two-quasiparticle configuration, and the second excited band
  is a mixture of K=1 and 3 configurations of the two-quasiparticle state. Thus, after the band crossing, all three bands
  predominantly belong
  to same intrinsic two-quasineutron configuration, but projected different K-states. Fig.~\ref{76Ge_wf_SEc1_12july24} elucidates the reason
  for the lowering of the odd-spin states for $^{112}$Ru even before the band crossing spin of I=12 as the $\gamma$-band
  has large admixtures from the two-quasiparticle configurations over a wide range of spin states. It is noted
  from Fg.~\ref{76Ge_wf_SEc1_12july24} that mixing of the two-quasiparticle configuration becomes significant from I=6 onwards, although the
  band crossing is at I=12.  

  The wavefunction amplitudes for $^{232}$Th are depicted in Fig.~\ref{232Th_wave_Sec4} and it is evident that yrast band
  has the dominant K=0 component of the vacuum configuration till I=14. It is only for I=16 and above that the K=1 state
  from the two-quasiparticle configuration becomes important. For the first excited band,  the dominant component
  is the K=2 state of the vacuum configuration up to the higest spin value studied. The second excited band has the K=4
  dominant state up to I=16. It is clear from the structure of these bands that first- and second- excited bands
  are simply the even- and odd- spin members of the $\gamma$-band based on the ground-state band up to the
  higest spin studied. Therefore, the observed
  bands in $^{232}$Th have normal structures and don't originate from the wobbling motion.

  Recently, $^{112}$Ru, $^{114}$Pd, $^{150}$Nd and $^{188}$Os nuclides have been investigated from the perspective of wobbling motion
  using the five-dimensional collective Hamiltonian with 
  potential, moments of inertia and mass parameters determined from the relativistic density functional
  theory (RDFT) \cite{WANG2024859}. Detailed analysis has been performed for $^{112}$Ru by mapping the quantal motion
  to the topology of the classical orbits using the spin coherent state approach. It has been shown that stable
  wobbling motion exists for $^{112}$Ru with the total angular-momentum vector oscillating about the medium axis.
  The wobbling motion has been investigated \cite{WANG2024859} for the spin states up to I=10, and on comparing with the
  results of the present work, it is noted from Fig.~\ref{Energy_CD_v1} that wobbling motion develops in the
  band crossing region with the odd-spin branch of the $\gamma$-band becoming favoured. The wobbling motion
  discussed in Ref.~\cite{WANG2024859} in the low-spin region is not consistent with the present work.
  Fig.~\ref{112ru_gammaband_RDFT} compares the
  results of the present work with the RDFT calculations. In this figure, TPSM calculations are presented
  with and without quasiparticle excitations. For the set, labelled as ``Vacu.'' in Fig.~\ref{112ru_gammaband_RDFT},
  the projection is performed from the vacuum configuration only, and ``Full'' denotes the results with the
  inclusion of quasiparticle excitations as presented in Fig.~\ref{Energy_CD_v1}. It is evident from the figure, on comparing the TPSM
  results with and without quasiparticle excitation, that quasiparticle contribution becomes important from
  I=6 onwards. It is also observed that RDFT results show large deviations from the experimental energies
  and these differences increase with spin. The reason for these deviations is now clear as the quasiparticle states are not
  considered in Ref.~\cite{WANG2024859} with the energy surfaces calculated and mapped for the ground-state configuration only.
  The results of RDFT for the ground-state band almost coincide with the TPSM ``Vacu.'' calculations. However, for the
  $\gamma$-band the RDFT results deviate even from the ``Vacu.'' values. The reason for this difference
  could be due to the shape fluctuations included in the five-dimensional collective Hamiltonian framework used in the RDFT
  approach.

  We shall now examine the alignment properties of the band structures as for the wobbling mode the
  alignment ($i_x$) and other properties  should be similar. This is in comparison to the normal band structures  for which the
  alignments of the two signature bands are different as their projections along the rotational axis differ. For
  wobbling bands, it is the rotational axis that moves away from the principle axis, giving rise to the
  excited wobbling band structures. The TPSM calculated alignments for the six nuclides are compared with the corresponding
  values evaluated from the experimental data in Figs.~\ref{ix_spin_1_July2024} and \ref{ix_spin_2_July2024}. The
  alignment plot for $^{76}$Ge, shown in Fig.~\ref{ix_spin_1_July2024},
  depicts a large peak at about $\hbar \omega = 0.55$ MeV and is due to the alignment of two-neutrons which is evident
  from the band diagram of Ref.~\cite{TPSM2014}. The alignment rises very sharply which indicates that the interaction between the
  ground- and two-quasiparticle bands is weak. For $^{112}$Ru, $i_x$ increases slowly which implies that the interaction
  between the two bands is strong. It is noted that even for $^{232}$Th, the alignments are similar for three bands and expresses the
  fact that there could be several other reasons for the similarity of the alignments, apart from the wobbling motion.

  The alignment for $^{188}$Os, $^{192}$Os and $^{192}$Pt, shown in Fig.~\ref{ix_spin_2_July2024}, increases rapidly for
  all the three cases. It is evident from the two figures that TPSM reproduces the alignments extracted from the data
  reasonably well.
  Unfortunately data is not available for high-spin states to observe a complete curvature as predicted by the TPSM
  calculations for the two Osmium isotopes. What is important from the wobbling point of view is that predicted TPSM
  alignments for the three bands are quite similar for all the cases. 
  
  We shall further examine the wobbling properties of the band structures and investigate the wobbling energy, defined as $:$
  \begin{equation}
  E_{wob}(I) = E_{n_{\omega}=1}(I) - \frac{[E_{n_{\omega}=0}(I+1) + E_{n_{\omega}=0}(I-1)]}{2}~~~.
  \end{equation}
  In the above equation, $n_{\omega}=0$ corresponds to the yrast-band,   $n_{\omega}=1$ is the odd-spin member of the
  $\gamma$-band and $n_{\omega}=2$ is the even-spin member of the $\gamma$-band.
  The wobbling energy is plotted in Fig.~\ref{Wobling_v1} for the nuclei studied in the present work. It has been shown that this
  quantity provides important information regarding the nature of the wobbling motion
  \cite{IK02,IK03,sf14}.
  It was predicted that the wobbling energy will increase for even-even systems
  with spin, what is referred to as the longitudinal motion \cite{sf14}.
  For odd-mass systems, the orientation of the odd-particle determines the nature of the wobbling motion.
  In case the odd-particle is aligned along the short-axis, the wobbling energy will decrease with spin, and is
  referred to as the transverse wobbling \cite{sf14}.
  In all the odd-mass 
  nuclei for which wobbling mode has been identified, except for $^{133}$La,
$^{187}$Au ($+ive$ parity) and $^{127}$Xe, the wobbling energy decreases with $I$ and the wobbling motion has 
  transverse character \cite{WM54,WM47,WM48,WM6,WM12}. It is evident from Fig.~\ref{Wobling_v1}
  that wobbling motion increases for all the cases studied in the present work, except
  for $Gao$ ($n_\omega=1$) band of $^{188}$Os. The wobbling frequency also
  decreases for $^{232}$Th, but is of no significance as wobbling motion is ruled out for this system.

  The important characteristic feature of the wobbling mode is the dominating inter-band B(E2)
  transitions \cite{BM75}. In the normal signature partner bands of one-dimensional cranking
  model, the B(M1) transitions are dominant \cite{WM56}. It has been shown using the particle-rotor model
  that B(E2) transitions from $n_{\omega}=1$ 
  to $n_{\omega}=0$ will be comparable to the in-band B(E2) transitions, whereas the B(M1) transitions
  should be quite weak for the wobbling motion. Further, due to the harmonic spectrum of the wobbling mode, the
  transitions from $n_{\omega}=2$ to $n_{\omega}=1$ should be twice that of the transitions from
  $n_{\omega}=1$ to $n_{\omega}=0$, and the direct transition from
  $n_{\omega}=2$  to $n_{\omega}=0$ should be negligible \cite{BM75,IK02,IK03}.

  We shall first discuss the in-band transitions before presenting the results for inter-band transitions as they are
  known for many states and the predictions of the TPSM approach can be assessed. The TPSM calculated in-band transitions
  are depicted in Fig.~\ref{Trans_BE2_18Nov_v1} for the three bands and are compared with the available measured values.
  The TPSM transition probabilities for the yrast band have already been presented 
  in our earlier publication \cite{SJ21,SP24}, and we shall only discuss the results for the two excited bands. It is noted from
  Fig.~\ref{Trans_BE2_18Nov_v1} that one transition has been measured for the first excited band of $^{112}$Ru
  and TPSM value is noted to be in good agreement with this known quantity. For the two Osmium isotopes, lowest few
  transitions have been measured for the even-spin branch of the $\gamma$-band ($Gae$) and the TPSM predicted transitions
  are again noted to be in fair agreement. It is expected that since the wobbling motion is built on the same intrinsic configuration, the transition probabilities should be similar for the three band structures.
  Fig.~\ref{Trans_BE2_18Nov_v1}
  exhibits that except for $^{192}$Pt, three band structures have similar B(E2) transition probabilities.

  The inter-band B(E2) transitions are displayed in Fig.~\ref{Trans_BE2_18Nov_v2} for the six isotopes and it is evident that
  these transitions approximately follow the pattern expected for the wobbling motion. First of the all, the transitions
  from  $Gao \rightarrow Yr$ 
  are comparable in magnitude to the in-band transitions of Fig.~\ref{Trans_BE2_18Nov_v1}. Secondly, the transitions
  from  $Gae \rightarrow Gao$  
  are almost twice in magnitude as compared to the transitions from
  $Gao \rightarrow Yr$. Furthermore, the direct transition
  from  $Gae \rightarrow Yr$ 
  are quite small as expected for the harmonic motion. For  $^{76}$Ge and $^{112}$Ru, harmonic motion is
  followed up to high-spin with  transitions $Gae \rightarrow Gao$  
  twice in magnitude as compared to the transitions from
  $Gao \rightarrow Yr$.  
  However, for $^{188,192}$Os and $^{192}$Pt isotopes anharmonicity is noted to develop for high-spin states and only
  for intermediate spin values, harmonic motion is observed. A possible explanation for these anharmonicities found
  for the heavier mass nuclei is that level density increases rapidly with spin for these systems as compared to the medium
  mass nuclei. The larger level density leads to higher mixing of the states and possible damping of the wobbling mode. Clearly,
  this statement is not true in general as harmonic wobbling motion is predicted in $^{163}$Lu \cite{WM42} up to
  a large spin value. The persistence of the harmonic motion will also depend on the intrinsic shell structure
  as for the Lu-isotopes, it is the TSD band structure that exhibits the wobbling mode.

  The inter-band B(M1) transitions from from $Gao \rightarrow Yr$ are depicted in Fig.~\ref{BM1out_v1}  and as expected
  for the wobbling motion, these transitions are quite weak.
  We have also evaluated the $\%$E2 mixing ratios and are displayed in
  Fig.~\ref{E2_percentage_fig} for the six  nuclides. This quantity,  defined in the caption of Fig.~\ref{E2_percentage_fig}, provides information on the E2
  content of the inter-band transitions. For transition with  $\%$E2 ratios more than $60\%$, it can be stated that transitions
  have dominant E2 character. It is evident from the figure that in all the nuclides except for $^{232}$Th,  $\%$E2 ratios
  are more than $80\%$ in the intermediate- and high-spin regions. The TPSM analysis,
  therefore, predicts that the band structures
  observed in five nuclei of $^{76}$Ge, $^{112}$Ru, $^{188}$Os, $^{192}$Os and $^{192}$Pt originate from the wobbling motion.
  It will be interesting to
  perform the lifetime measurements for these band structures to confirm the predictions of the present model study. 

  It is also interesting to investigate the effect of the quasiparticle excitations on the calculated alignments and the B(E2)
  transitions. In Fig.~\ref{fig13}, the results of alignment and the B(E2) transitions are shown for the yrast bands of $^{112}$Ru and $^{232}$Th
  by considering the vacuum configuration only and also with full basis space. It is evident from the figure that the results
  for $^{112}$Ru deviate considerably for the high-spin states due to  the crossing of the two-quasiparticle state around I=12. However,
  for $^{232}$Th, the results with and without quasiparticle states are similar, and deviate only for the highest spin states
  studied in the present work. This is due to the occurrence of the delayed bandcrossing for this system.
  
  To further elucidate whether the wobbling motion is longitudinal or transverse in nature,
  we have evaluated the K-distribution, defined as \cite{BQ09}
  \begin{equation}
P^{I}(|K|) = \sum _{\kappa}|g^{I}(K,\kappa)|^2 + |g^{I}(-K,\kappa)|^2.
  \end{equation}
  The above quantity is displayed in Fig.~ \ref{76GE_pk_vs_k} for $^{112}$Ru as an illustrative example. In the original
  TPSM code, z-axis is chosen as the quantization axis and this corresponds to the panels in Fig.~\ref{76GE_pk_vs_k}  labelled as l-axis.
  Changing $\gamma=-\gamma$  and $\gamma \rightarrow -\gamma - 120^0$  leaves the results invariant, but moves the quantization
  l-axis to i-axis and to s-axis,
  respectively. The TPSM parameters of $\epsilon$ and $\epsilon'$ in terms of the standard $\gamma$ parameter are given by :
  $\epsilon = \epsilon_0 \cos{\gamma}$ and $\epsilon' = \epsilon_0 \sin{\gamma}$.
          
  It is noted from Fig.~\ref{76GE_pk_vs_k} that K-distribution has maximum component along the i-axis and signifies that
  rotational motion is along this axis, what is referred to as the longitudinal wobbling mode.  It is also observed that even-spin
  members have non-zero K=0 amplitudes, and for odd-spin members this amplitude is zero. This implies that wavefunctions are symmetric
  and antisymmetric as expected for $n_{\omega}=0$ and $n_{\omega}=1$ oscillator states.  For high-angular-momentum states,
  the amplitudes for higher K-values is quite large, although the contribution from the aligned two-particles cannot exceed
  ten as neutrons are in the $1h_{11/2}$ subshell.  This suggests that the whole system rotates about the i-axis and motion
  has the wobbling character.

  We have also evaluated the K-distribution for $^{232}$Th as this system shows decreasing trend of wobbling
  energy with spin in contrast to all other nuclides studied. The K-distribution for $^{232}$Th is presented
  in Fig.~\ref{232Th_pk_vs_k} and is noted that it has a maximum component along the short, s- axis for the high-spin states.
  This provides an explanation for the decreasing trend of wobbling energy with spin for this system.

  \section{Summary and conclusions}

  In the present work, we have performed a detailed analysis of six nuclei, $^{76}$Ge, $^{112}$Ru, $^{188,192}$Os, $^{192}$Pt and $^{232}$Th,
  to explore whether the rotational motion for these systems can be categorised as originating from the wobbling motion. The $\gamma$-band
  structures of these nuclei were investigated in our earlier study \cite{SJ21,SP24} and it was shown that energy staggering of the
  six nuclei have odd-spin states lower than the even-spin states, and were grouped as $\gamma$-rigid. This is in comparison to twenty-four
  other nuclei where the staggering was opposite and were classified as $\gamma$-soft systems. As the $\gamma$-rigid nuclei are
  expected to develop wobbling motion, we decided to carry out a detailed study of the six nuclei.

  The energies and electromagnetic transitions were evaluated for the six nuclei, and it was shown
  that band structures become almost equally
  spaced with alternating even- and odd-spin states, which is a characteristic feature of the wobbling motion,
  after the band crossing for all the nuclei except for $^{232}$Th. We have studied the aligned angular-momentum for the
  three bands and it has been shown that they are similar for the three bands, corresponding to $n_{\omega}$=0, 1 and 2
  wobbling modes. The most important feature of the wobbling mode is that inter-band transitions are dominated by B(E2) rather
  than B(M1) as for the normal signature partner bands. It has been shown that B(E2) transitions from $n_{\omega}$=1 to
  $n_{\omega}$=0 are comparable to the in-band transitions, and more importantly the B(E2) transitions from $n_{\omega}$=2 to
  $n_{\omega}$=1 are quite large as compared to the transitions from $n_{\omega}$=1 to $n_{\omega}$=0. The calculated $\%$E2 ratio of
  more than $80\%$ for all the nuclei studied except for $^{232}$Th clearly
  indicates that the inter-band transitions are dominated by E2 rather than M1. 
  Further, we have also calculated the
  K-distributions and it has been shown that the wobbling motion has the longitudinal character.

  The important inference drawn from the
  present work is that  wobbling motion is caused by the quasiparticle alignment. It has been demonstrated that
  band structures with alternating
  even-and odd-spin states, which is a characteristic feature of the wobbling motion, start emerging after the band crossing.
  The transition from normal rotational motion at low-spin to wobbling motion at high-spin is predicted to be smooth. This is evident from the
  steady increase of the $\%$E2 ratios with spin. 

\section{ACKNOWLEDGEMENTS}
The authors would like to acknowledge S.P. Rouoof and Aneeqa Bashir for their help in the preparation of some of the figures presented in the manuscript. The authors are also thankful to Prof. S. Frauendorf for his valuable suggestions.

\bibliographystyle{apsrev4-2}
\bibliography{references}

\begin{thebibliography}{93}%
\makeatletter
\providecommand \@ifxundefined [1]{%
 \@ifx{#1\undefined}
}%
\providecommand \@ifnum [1]{%
 \ifnum #1\expandafter \@firstoftwo
 \else \expandafter \@secondoftwo
 \fi
}%
\providecommand \@ifx [1]{%
 \ifx #1\expandafter \@firstoftwo
 \else \expandafter \@secondoftwo
 \fi
}%
\providecommand \natexlab [1]{#1}%
\providecommand \enquote  [1]{``#1''}%
\providecommand \bibnamefont  [1]{#1}%
\providecommand \bibfnamefont [1]{#1}%
\providecommand \citenamefont [1]{#1}%
\providecommand \href@noop [0]{\@secondoftwo}%
\providecommand \href [0]{\begingroup \@sanitize@url \@href}%
\providecommand \@href[1]{\@@startlink{#1}\@@href}%
\providecommand \@@href[1]{\endgroup#1\@@endlink}%
\providecommand \@sanitize@url [0]{\catcode `\\12\catcode `\$12\catcode
  `\&12\catcode `\#12\catcode `\^12\catcode `\_12\catcode `\%12\relax}%
\providecommand \@@startlink[1]{}%
\providecommand \@@endlink[0]{}%
\providecommand \url  [0]{\begingroup\@sanitize@url \@url }%
\providecommand \@url [1]{\endgroup\@href {#1}{\urlprefix }}%
\providecommand \urlprefix  [0]{URL }%
\providecommand \Eprint [0]{\href }%
\providecommand \doibase [0]{https://doi.org/}%
\providecommand \selectlanguage [0]{\@gobble}%
\providecommand \bibinfo  [0]{\@secondoftwo}%
\providecommand \bibfield  [0]{\@secondoftwo}%
\providecommand \translation [1]{[#1]}%
\providecommand \BibitemOpen [0]{}%
\providecommand \bibitemStop [0]{}%
\providecommand \bibitemNoStop [0]{.\EOS\space}%
\providecommand \EOS [0]{\spacefactor3000\relax}%
\providecommand \BibitemShut  [1]{\csname bibitem#1\endcsname}%
\let\auto@bib@innerbib\@empty
\bibitem [{\citenamefont {Bohr}\ and\ \citenamefont {Mottelson}(1998)}]{BM75}%
  \BibitemOpen
  \bibfield  {author} {\bibinfo {author} {\bibfnamefont {A.}~\bibnamefont
  {Bohr}}\ and\ \bibinfo {author} {\bibfnamefont {B.~R.}\ \bibnamefont
  {Mottelson}},\ }\href {https://doi.org/10.1142/3530} {\emph {\bibinfo {title}
  {Nuclear Structure}}}\ (\bibinfo  {publisher} {World Scientific Publishing
  Company},\ \bibinfo {year} {1998})\BibitemShut {NoStop}%
\bibitem [{\citenamefont {Casten}(2001)}]{RF00}%
  \BibitemOpen
  \bibfield  {author} {\bibinfo {author} {\bibfnamefont {R.~F.}\ \bibnamefont
  {Casten}},\ }\href
  {https://doi.org/10.1093/acprof:oso/9780198507246.001.0001} {\emph {\bibinfo
  {title} {Nuclear Structure from a Simple Perspective}}}\ (\bibinfo
  {publisher} {Oxford University Press},\ \bibinfo {year} {2001})\BibitemShut
  {NoStop}%
\bibitem [{\citenamefont {Hamamoto}\ and\ \citenamefont
  {Mottelson}(1983)}]{IH83}%
  \BibitemOpen
  \bibfield  {author} {\bibinfo {author} {\bibfnamefont {I.}~\bibnamefont
  {Hamamoto}}\ and\ \bibinfo {author} {\bibfnamefont {B.}~\bibnamefont
  {Mottelson}},\ }\href
  {https://doi.org/https://doi.org/10.1016/0370-2693(83)91000-6} {\bibfield
  {journal} {\bibinfo  {journal} {Phys. Lett. B}\ }\textbf {\bibinfo {volume}
  {127}},\ \bibinfo {pages} {281} (\bibinfo {year} {1983})}\BibitemShut
  {NoStop}%
\bibitem [{\citenamefont {Alaga}(1955)}]{Alaga1955}%
  \BibitemOpen
  \bibfield  {author} {\bibinfo {author} {\bibfnamefont {G.}~\bibnamefont
  {Alaga}},\ }\href {https://doi.org/10.1103/PhysRev.100.432} {\bibfield
  {journal} {\bibinfo  {journal} {Phys. Rev.}\ }\textbf {\bibinfo {volume}
  {100}},\ \bibinfo {pages} {432} (\bibinfo {year} {1955})}\BibitemShut
  {NoStop}%
\bibitem [{\citenamefont {Åberg}(1990)}]{Aberg1990}%
  \BibitemOpen
  \bibfield  {author} {\bibinfo {author} {\bibfnamefont {S.}~\bibnamefont
  {Åberg}},\ }\href
  {https://doi.org/https://doi.org/10.1016/0375-9474(90)91132-B} {\bibfield
  {journal} {\bibinfo  {journal} {Nuclear Physics A}\ }\textbf {\bibinfo
  {volume} {520}},\ \bibinfo {pages} {c35} (\bibinfo {year}
  {1990})}\BibitemShut {NoStop}%
\bibitem [{\citenamefont {Ragnarsson}(1989)}]{IR1989}%
  \BibitemOpen
  \bibfield  {author} {\bibinfo {author} {\bibfnamefont {I.}~\bibnamefont
  {Ragnarsson}},\ }\href {https://doi.org/10.1103/PhysRevLett.62.2084}
  {\bibfield  {journal} {\bibinfo  {journal} {Phys. Rev. Lett.}\ }\textbf
  {\bibinfo {volume} {62}},\ \bibinfo {pages} {2084} (\bibinfo {year}
  {1989})}\BibitemShut {NoStop}%
\bibitem [{\citenamefont {Bengtsson}(1989)}]{TB1989}%
  \BibitemOpen
  \bibfield  {author} {\bibinfo {author} {\bibfnamefont {T.}~\bibnamefont
  {Bengtsson}},\ }\href
  {https://doi.org/https://doi.org/10.1016/0375-9474(89)90216-9} {\bibfield
  {journal} {\bibinfo  {journal} {Nuclear Physics A}\ }\textbf {\bibinfo
  {volume} {496}},\ \bibinfo {pages} {56} (\bibinfo {year} {1989})}\BibitemShut
  {NoStop}%
\bibitem [{\citenamefont {Bengtsson}\ and\ \citenamefont {Ryde}(2004)}]{RB04}%
  \BibitemOpen
  \bibfield  {author} {\bibinfo {author} {\bibfnamefont {R.}~\bibnamefont
  {Bengtsson}}\ and\ \bibinfo {author} {\bibfnamefont {H.}~\bibnamefont
  {Ryde}},\ }\href {https://doi.org/10.1140/epja/i2004-10046-4} {\bibfield
  {journal} {\bibinfo  {journal} {Eur. Phys. J. A}\ }\textbf {\bibinfo {volume}
  {22}},\ \bibinfo {pages} {355} (\bibinfo {year} {2004})}\BibitemShut
  {NoStop}%
\bibitem [{\citenamefont {Frauendorf}\ and\ \citenamefont {{Jie
  Meng}}(1997)}]{SF97}%
  \BibitemOpen
  \bibfield  {author} {\bibinfo {author} {\bibfnamefont {S.}~\bibnamefont
  {Frauendorf}}\ and\ \bibinfo {author} {\bibnamefont {{Jie Meng}}},\ }\href
  {https://doi.org/https://doi.org/10.1016/S0375-9474(97)00004-3} {\bibfield
  {journal} {\bibinfo  {journal} {Nuclear Physics A}\ }\textbf {\bibinfo
  {volume} {617}},\ \bibinfo {pages} {131} (\bibinfo {year}
  {1997})}\BibitemShut {NoStop}%
\bibitem [{\citenamefont {Koike}\ \emph {et~al.}(2004)\citenamefont {Koike},
  \citenamefont {Starosta},\ and\ \citenamefont {Hamamoto}}]{TS04}%
  \BibitemOpen
  \bibfield  {author} {\bibinfo {author} {\bibfnamefont {T.}~\bibnamefont
  {Koike}}, \bibinfo {author} {\bibfnamefont {K.}~\bibnamefont {Starosta}},\
  and\ \bibinfo {author} {\bibfnamefont {I.}~\bibnamefont {Hamamoto}},\ }\href
  {https://doi.org/10.1103/PhysRevLett.93.172502} {\bibfield  {journal}
  {\bibinfo  {journal} {Phys. Rev. Lett.}\ }\textbf {\bibinfo {volume} {93}},\
  \bibinfo {pages} {172502} (\bibinfo {year} {2004})}\BibitemShut {NoStop}%
\bibitem [{\citenamefont {Dimitrov}\ \emph {et~al.}(2000)\citenamefont
  {Dimitrov}, \citenamefont {Frauendorf},\ and\ \citenamefont
  {D\"onau}}]{VI00}%
  \BibitemOpen
  \bibfield  {author} {\bibinfo {author} {\bibfnamefont {V.~I.}\ \bibnamefont
  {Dimitrov}}, \bibinfo {author} {\bibfnamefont {S.}~\bibnamefont
  {Frauendorf}},\ and\ \bibinfo {author} {\bibfnamefont {F.}~\bibnamefont
  {D\"onau}},\ }\href {https://doi.org/10.1103/PhysRevLett.84.5732} {\bibfield
  {journal} {\bibinfo  {journal} {Phys. Rev. Lett.}\ }\textbf {\bibinfo
  {volume} {84}},\ \bibinfo {pages} {5732} (\bibinfo {year}
  {2000})}\BibitemShut {NoStop}%
\bibitem [{\citenamefont {Olbratowski}\ \emph {et~al.}(2004)\citenamefont
  {Olbratowski}, \citenamefont {Dobaczewski}, \citenamefont {Dudek},\ and\
  \citenamefont {P\l{}\'ociennik}}]{PO04}%
  \BibitemOpen
  \bibfield  {author} {\bibinfo {author} {\bibfnamefont {P.}~\bibnamefont
  {Olbratowski}}, \bibinfo {author} {\bibfnamefont {J.}~\bibnamefont
  {Dobaczewski}}, \bibinfo {author} {\bibfnamefont {J.}~\bibnamefont {Dudek}},\
  and\ \bibinfo {author} {\bibfnamefont {W.}~\bibnamefont {P\l{}\'ociennik}},\
  }\href {https://doi.org/10.1103/PhysRevLett.93.052501} {\bibfield  {journal}
  {\bibinfo  {journal} {Phys. Rev. Lett.}\ }\textbf {\bibinfo {volume} {93}},\
  \bibinfo {pages} {052501} (\bibinfo {year} {2004})}\BibitemShut {NoStop}%
\bibitem [{\citenamefont {Starosta}\ \emph {et~al.}(2001)\citenamefont
  {Starosta}, \citenamefont {Koike}, \citenamefont {Chiara}, \citenamefont
  {Fossan}, \citenamefont {LaFosse}, \citenamefont {Hecht}, \citenamefont
  {Beausang}, \citenamefont {Caprio}, \citenamefont {Cooper}, \citenamefont
  {Kr\"ucken}, \citenamefont {Novak}, \citenamefont {Zamfir}, \citenamefont
  {Zyromski}, \citenamefont {Hartley}, \citenamefont {Balabanski},
  \citenamefont {Zhang}, \citenamefont {Frauendorf},\ and\ \citenamefont
  {Dimitrov}}]{KS01}%
  \BibitemOpen
  \bibfield  {author} {\bibinfo {author} {\bibfnamefont {K.}~\bibnamefont
  {Starosta}}, \bibinfo {author} {\bibfnamefont {T.}~\bibnamefont {Koike}},
  \bibinfo {author} {\bibfnamefont {C.~J.}\ \bibnamefont {Chiara}}, \bibinfo
  {author} {\bibfnamefont {D.~B.}\ \bibnamefont {Fossan}}, \bibinfo {author}
  {\bibfnamefont {D.~R.}\ \bibnamefont {LaFosse}}, \bibinfo {author}
  {\bibfnamefont {A.~A.}\ \bibnamefont {Hecht}}, \bibinfo {author}
  {\bibfnamefont {C.~W.}\ \bibnamefont {Beausang}}, \bibinfo {author}
  {\bibfnamefont {M.~A.}\ \bibnamefont {Caprio}}, \bibinfo {author}
  {\bibfnamefont {J.~R.}\ \bibnamefont {Cooper}}, \bibinfo {author}
  {\bibfnamefont {R.}~\bibnamefont {Kr\"ucken}}, \bibinfo {author}
  {\bibfnamefont {J.~R.}\ \bibnamefont {Novak}}, \bibinfo {author}
  {\bibfnamefont {N.~V.}\ \bibnamefont {Zamfir}}, \bibinfo {author}
  {\bibfnamefont {K.~E.}\ \bibnamefont {Zyromski}}, \bibinfo {author}
  {\bibfnamefont {D.~J.}\ \bibnamefont {Hartley}}, \bibinfo {author}
  {\bibfnamefont {D.~L.}\ \bibnamefont {Balabanski}}, \bibinfo {author}
  {\bibfnamefont {J.-y.}\ \bibnamefont {Zhang}}, \bibinfo {author}
  {\bibfnamefont {S.}~\bibnamefont {Frauendorf}},\ and\ \bibinfo {author}
  {\bibfnamefont {V.~I.}\ \bibnamefont {Dimitrov}},\ }\href
  {https://doi.org/10.1103/PhysRevLett.86.971} {\bibfield  {journal} {\bibinfo
  {journal} {Phys. Rev. Lett.}\ }\textbf {\bibinfo {volume} {86}},\ \bibinfo
  {pages} {971} (\bibinfo {year} {2001})}\BibitemShut {NoStop}%
\bibitem [{\citenamefont {Zhu}\ \emph {et~al.}(2003)\citenamefont {Zhu},
  \citenamefont {Garg}, \citenamefont {Nayak}, \citenamefont {Ghugre},
  \citenamefont {Pattabiraman}, \citenamefont {Fossan}, \citenamefont {Koike},
  \citenamefont {Starosta}, \citenamefont {Vaman}, \citenamefont {Janssens},
  \citenamefont {Chakrawarthy}, \citenamefont {Whitehead}, \citenamefont
  {Macchiavelli},\ and\ \citenamefont {Frauendorf}}]{SU03}%
  \BibitemOpen
  \bibfield  {author} {\bibinfo {author} {\bibfnamefont {S.}~\bibnamefont
  {Zhu}}, \bibinfo {author} {\bibfnamefont {U.}~\bibnamefont {Garg}}, \bibinfo
  {author} {\bibfnamefont {B.~K.}\ \bibnamefont {Nayak}}, \bibinfo {author}
  {\bibfnamefont {S.~S.}\ \bibnamefont {Ghugre}}, \bibinfo {author}
  {\bibfnamefont {N.~S.}\ \bibnamefont {Pattabiraman}}, \bibinfo {author}
  {\bibfnamefont {D.~B.}\ \bibnamefont {Fossan}}, \bibinfo {author}
  {\bibfnamefont {T.}~\bibnamefont {Koike}}, \bibinfo {author} {\bibfnamefont
  {K.}~\bibnamefont {Starosta}}, \bibinfo {author} {\bibfnamefont
  {C.}~\bibnamefont {Vaman}}, \bibinfo {author} {\bibfnamefont {R.~V.~F.}\
  \bibnamefont {Janssens}}, \bibinfo {author} {\bibfnamefont {R.~S.}\
  \bibnamefont {Chakrawarthy}}, \bibinfo {author} {\bibfnamefont
  {M.}~\bibnamefont {Whitehead}}, \bibinfo {author} {\bibfnamefont {A.~O.}\
  \bibnamefont {Macchiavelli}},\ and\ \bibinfo {author} {\bibfnamefont
  {S.}~\bibnamefont {Frauendorf}},\ }\href
  {https://doi.org/10.1103/PhysRevLett.91.132501} {\bibfield  {journal}
  {\bibinfo  {journal} {Phys. Rev. Lett.}\ }\textbf {\bibinfo {volume} {91}},\
  \bibinfo {pages} {132501} (\bibinfo {year} {2003})}\BibitemShut {NoStop}%
\bibitem [{\citenamefont {Grodner}\ \emph {et~al.}(2006)\citenamefont
  {Grodner}, \citenamefont {Srebrny}, \citenamefont {Pasternak}, \citenamefont
  {Zalewska}, \citenamefont {Morek}, \citenamefont {Droste}, \citenamefont
  {Mierzejewski}, \citenamefont {Kowalczyk}, \citenamefont {Kownacki},
  \citenamefont {Kisieli\ifmmode~\acute{n}\else \'{n}\fi{}ski}, \citenamefont
  {Rohozi\ifmmode~\acute{n}\else \'{n}\fi{}ski}, \citenamefont {Koike},
  \citenamefont {Starosta}, \citenamefont {Kordyasz}, \citenamefont
  {Napiorkowski}, \citenamefont
  {Woli\ifmmode\acute{n}\else\'{n}\fi{}ska-Cichocka}, \citenamefont
  {Ruchowska}, \citenamefont {P\l{}\'ociennik},\ and\ \citenamefont
  {Perkowski}}]{EG06}%
  \BibitemOpen
  \bibfield  {author} {\bibinfo {author} {\bibfnamefont {E.}~\bibnamefont
  {Grodner}}, \bibinfo {author} {\bibfnamefont {J.}~\bibnamefont {Srebrny}},
  \bibinfo {author} {\bibfnamefont {A.~A.}\ \bibnamefont {Pasternak}}, \bibinfo
  {author} {\bibfnamefont {I.}~\bibnamefont {Zalewska}}, \bibinfo {author}
  {\bibfnamefont {T.}~\bibnamefont {Morek}}, \bibinfo {author} {\bibfnamefont
  {C.}~\bibnamefont {Droste}}, \bibinfo {author} {\bibfnamefont
  {J.}~\bibnamefont {Mierzejewski}}, \bibinfo {author} {\bibfnamefont
  {M.}~\bibnamefont {Kowalczyk}}, \bibinfo {author} {\bibfnamefont
  {J.}~\bibnamefont {Kownacki}}, \bibinfo {author} {\bibfnamefont
  {M.}~\bibnamefont {Kisieli\ifmmode~\acute{n}\else \'{n}\fi{}ski}}, \bibinfo
  {author} {\bibfnamefont {S.~G.}\ \bibnamefont {Rohozi\ifmmode~\acute{n}\else
  \'{n}\fi{}ski}}, \bibinfo {author} {\bibfnamefont {T.}~\bibnamefont {Koike}},
  \bibinfo {author} {\bibfnamefont {K.}~\bibnamefont {Starosta}}, \bibinfo
  {author} {\bibfnamefont {A.}~\bibnamefont {Kordyasz}}, \bibinfo {author}
  {\bibfnamefont {P.~J.}\ \bibnamefont {Napiorkowski}}, \bibinfo {author}
  {\bibfnamefont {M.}~\bibnamefont
  {Woli\ifmmode\acute{n}\else\'{n}\fi{}ska-Cichocka}}, \bibinfo {author}
  {\bibfnamefont {E.}~\bibnamefont {Ruchowska}}, \bibinfo {author}
  {\bibfnamefont {W.}~\bibnamefont {P\l{}\'ociennik}},\ and\ \bibinfo {author}
  {\bibfnamefont {J.}~\bibnamefont {Perkowski}},\ }\href
  {https://doi.org/10.1103/PhysRevLett.97.172501} {\bibfield  {journal}
  {\bibinfo  {journal} {Phys. Rev. Lett.}\ }\textbf {\bibinfo {volume} {97}},\
  \bibinfo {pages} {172501} (\bibinfo {year} {2006})}\BibitemShut {NoStop}%
\bibitem [{\citenamefont {\O{}deg\aa{}rd}\ \emph {et~al.}(2001)\citenamefont
  {\O{}deg\aa{}rd}, \citenamefont {Hagemann}, \citenamefont {Jensen},
  \citenamefont {Bergstr\"om}, \citenamefont {Herskind}, \citenamefont
  {Sletten}, \citenamefont {T\"orm\"anen}, \citenamefont {Wilson},
  \citenamefont {Tj\o{}m}, \citenamefont {Hamamoto}, \citenamefont {Spohr},
  \citenamefont {H\"ubel}, \citenamefont {G\"orgen}, \citenamefont
  {Sch\"onwasser}, \citenamefont {Bracco}, \citenamefont {Leoni}, \citenamefont
  {Maj}, \citenamefont {Petrache}, \citenamefont {Bednarczyk},\ and\
  \citenamefont {Curien}}]{WM1}%
  \BibitemOpen
  \bibfield  {author} {\bibinfo {author} {\bibfnamefont {S.~W.}\ \bibnamefont
  {\O{}deg\aa{}rd}}, \bibinfo {author} {\bibfnamefont {G.~B.}\ \bibnamefont
  {Hagemann}}, \bibinfo {author} {\bibfnamefont {D.~R.}\ \bibnamefont
  {Jensen}}, \bibinfo {author} {\bibfnamefont {M.}~\bibnamefont {Bergstr\"om}},
  \bibinfo {author} {\bibfnamefont {B.}~\bibnamefont {Herskind}}, \bibinfo
  {author} {\bibfnamefont {G.}~\bibnamefont {Sletten}}, \bibinfo {author}
  {\bibfnamefont {S.}~\bibnamefont {T\"orm\"anen}}, \bibinfo {author}
  {\bibfnamefont {J.~N.}\ \bibnamefont {Wilson}}, \bibinfo {author}
  {\bibfnamefont {P.~O.}\ \bibnamefont {Tj\o{}m}}, \bibinfo {author}
  {\bibfnamefont {I.}~\bibnamefont {Hamamoto}}, \bibinfo {author}
  {\bibfnamefont {K.}~\bibnamefont {Spohr}}, \bibinfo {author} {\bibfnamefont
  {H.}~\bibnamefont {H\"ubel}}, \bibinfo {author} {\bibfnamefont
  {A.}~\bibnamefont {G\"orgen}}, \bibinfo {author} {\bibfnamefont
  {G.}~\bibnamefont {Sch\"onwasser}}, \bibinfo {author} {\bibfnamefont
  {A.}~\bibnamefont {Bracco}}, \bibinfo {author} {\bibfnamefont
  {S.}~\bibnamefont {Leoni}}, \bibinfo {author} {\bibfnamefont
  {A.}~\bibnamefont {Maj}}, \bibinfo {author} {\bibfnamefont {C.~M.}\
  \bibnamefont {Petrache}}, \bibinfo {author} {\bibfnamefont {P.}~\bibnamefont
  {Bednarczyk}},\ and\ \bibinfo {author} {\bibfnamefont {D.}~\bibnamefont
  {Curien}},\ }\href {https://doi.org/10.1103/PhysRevLett.86.5866} {\bibfield
  {journal} {\bibinfo  {journal} {Phys. Rev. Lett.}\ }\textbf {\bibinfo
  {volume} {86}},\ \bibinfo {pages} {5866} (\bibinfo {year}
  {2001})}\BibitemShut {NoStop}%
\bibitem [{\citenamefont {Schnack-Petersen}\ \emph {et~al.}(1995)\citenamefont
  {Schnack-Petersen}, \citenamefont {Bengtsson}, \citenamefont {Bark},
  \citenamefont {Bosetti}, \citenamefont {Brockstedt}, \citenamefont
  {Carlsson}, \citenamefont {Ekström}, \citenamefont {Hagemann}, \citenamefont
  {Herskind}, \citenamefont {Ingebretsen}, \citenamefont {Jensen},
  \citenamefont {Leoni}, \citenamefont {Nordlund}, \citenamefont {Ryde},
  \citenamefont {Tjøm},\ and\ \citenamefont {Yang}}]{WM41}%
  \BibitemOpen
  \bibfield  {author} {\bibinfo {author} {\bibfnamefont {H.}~\bibnamefont
  {Schnack-Petersen}}, \bibinfo {author} {\bibfnamefont {R.}~\bibnamefont
  {Bengtsson}}, \bibinfo {author} {\bibfnamefont {R.}~\bibnamefont {Bark}},
  \bibinfo {author} {\bibfnamefont {P.}~\bibnamefont {Bosetti}}, \bibinfo
  {author} {\bibfnamefont {A.}~\bibnamefont {Brockstedt}}, \bibinfo {author}
  {\bibfnamefont {H.}~\bibnamefont {Carlsson}}, \bibinfo {author}
  {\bibfnamefont {L.}~\bibnamefont {Ekström}}, \bibinfo {author}
  {\bibfnamefont {G.}~\bibnamefont {Hagemann}}, \bibinfo {author}
  {\bibfnamefont {B.}~\bibnamefont {Herskind}}, \bibinfo {author}
  {\bibfnamefont {F.}~\bibnamefont {Ingebretsen}}, \bibinfo {author}
  {\bibfnamefont {H.}~\bibnamefont {Jensen}}, \bibinfo {author} {\bibfnamefont
  {S.}~\bibnamefont {Leoni}}, \bibinfo {author} {\bibfnamefont
  {A.}~\bibnamefont {Nordlund}}, \bibinfo {author} {\bibfnamefont
  {H.}~\bibnamefont {Ryde}}, \bibinfo {author} {\bibfnamefont {P.}~\bibnamefont
  {Tjøm}},\ and\ \bibinfo {author} {\bibfnamefont {C.}~\bibnamefont {Yang}},\
  }\href {https://doi.org/https://doi.org/10.1016/0375-9474(95)00363-6}
  {\bibfield  {journal} {\bibinfo  {journal} {Nuclear Physics A}\ }\textbf
  {\bibinfo {volume} {594}},\ \bibinfo {pages} {175} (\bibinfo {year}
  {1995})}\BibitemShut {NoStop}%
\bibitem [{\citenamefont {Frauendorf}(2001)}]{WM56}%
  \BibitemOpen
  \bibfield  {author} {\bibinfo {author} {\bibfnamefont {S.}~\bibnamefont
  {Frauendorf}},\ }\href {https://doi.org/10.1103/RevModPhys.73.463} {\bibfield
   {journal} {\bibinfo  {journal} {Rev. Mod. Phys.}\ }\textbf {\bibinfo
  {volume} {73}},\ \bibinfo {pages} {463} (\bibinfo {year} {2001})}\BibitemShut
  {NoStop}%
\bibitem [{\citenamefont {Hamamoto}(2002)}]{IK02}%
  \BibitemOpen
  \bibfield  {author} {\bibinfo {author} {\bibfnamefont {I.}~\bibnamefont
  {Hamamoto}},\ }\href
  {https://doi.org/https://doi.org/10.1103/PhysRevC.65.044305} {\bibfield
  {journal} {\bibinfo  {journal} {Phys. Rev. C}\ }\textbf {\bibinfo {volume}
  {65}},\ \bibinfo {pages} {044305} (\bibinfo {year} {2002})}\BibitemShut
  {NoStop}%
\bibitem [{\citenamefont {Jensen}\ \emph {et~al.}(2002)\citenamefont {Jensen},
  \citenamefont {Hagemann}, \citenamefont {Hamamoto}, \citenamefont
  {\O{}deg\aa{}rd}, \citenamefont {Herskind}, \citenamefont {Sletten},
  \citenamefont {Wilson}, \citenamefont {Spohr}, \citenamefont {H\"ubel},
  \citenamefont {Bringel}, \citenamefont {Neu\ss{}er}, \citenamefont
  {Sch\"onwa\ss{}er}, \citenamefont {Singh}, \citenamefont {Ma}, \citenamefont
  {Amro}, \citenamefont {Bracco}, \citenamefont {Leoni}, \citenamefont
  {Benzoni}, \citenamefont {Maj}, \citenamefont {Petrache}, \citenamefont
  {Bianco}, \citenamefont {Bednarczyk},\ and\ \citenamefont {Curien}}]{WM42}%
  \BibitemOpen
  \bibfield  {author} {\bibinfo {author} {\bibfnamefont {D.~R.}\ \bibnamefont
  {Jensen}}, \bibinfo {author} {\bibfnamefont {G.~B.}\ \bibnamefont
  {Hagemann}}, \bibinfo {author} {\bibfnamefont {I.}~\bibnamefont {Hamamoto}},
  \bibinfo {author} {\bibfnamefont {S.~W.}\ \bibnamefont {\O{}deg\aa{}rd}},
  \bibinfo {author} {\bibfnamefont {B.}~\bibnamefont {Herskind}}, \bibinfo
  {author} {\bibfnamefont {G.}~\bibnamefont {Sletten}}, \bibinfo {author}
  {\bibfnamefont {J.~N.}\ \bibnamefont {Wilson}}, \bibinfo {author}
  {\bibfnamefont {K.}~\bibnamefont {Spohr}}, \bibinfo {author} {\bibfnamefont
  {H.}~\bibnamefont {H\"ubel}}, \bibinfo {author} {\bibfnamefont
  {P.}~\bibnamefont {Bringel}}, \bibinfo {author} {\bibfnamefont
  {A.}~\bibnamefont {Neu\ss{}er}}, \bibinfo {author} {\bibfnamefont
  {G.}~\bibnamefont {Sch\"onwa\ss{}er}}, \bibinfo {author} {\bibfnamefont
  {A.~K.}\ \bibnamefont {Singh}}, \bibinfo {author} {\bibfnamefont {W.~C.}\
  \bibnamefont {Ma}}, \bibinfo {author} {\bibfnamefont {H.}~\bibnamefont
  {Amro}}, \bibinfo {author} {\bibfnamefont {A.}~\bibnamefont {Bracco}},
  \bibinfo {author} {\bibfnamefont {S.}~\bibnamefont {Leoni}}, \bibinfo
  {author} {\bibfnamefont {G.}~\bibnamefont {Benzoni}}, \bibinfo {author}
  {\bibfnamefont {A.}~\bibnamefont {Maj}}, \bibinfo {author} {\bibfnamefont
  {C.~M.}\ \bibnamefont {Petrache}}, \bibinfo {author} {\bibfnamefont {G.~L.}\
  \bibnamefont {Bianco}}, \bibinfo {author} {\bibfnamefont {P.}~\bibnamefont
  {Bednarczyk}},\ and\ \bibinfo {author} {\bibfnamefont {D.}~\bibnamefont
  {Curien}},\ }\href {https://doi.org/10.1103/PhysRevLett.89.142503} {\bibfield
   {journal} {\bibinfo  {journal} {Phys. Rev. Lett.}\ }\textbf {\bibinfo
  {volume} {89}},\ \bibinfo {pages} {142503} (\bibinfo {year}
  {2002})}\BibitemShut {NoStop}%
\bibitem [{\citenamefont {Hamamoto}\ and\ \citenamefont
  {Hagemann}(2003)}]{IK03}%
  \BibitemOpen
  \bibfield  {author} {\bibinfo {author} {\bibfnamefont {I.}~\bibnamefont
  {Hamamoto}}\ and\ \bibinfo {author} {\bibfnamefont {G.~B.}\ \bibnamefont
  {Hagemann}},\ }\href
  {https://doi.org/https://doi.org/10.1103/PhysRevC.67.014319} {\bibfield
  {journal} {\bibinfo  {journal} {Phys. Rev. C}\ }\textbf {\bibinfo {volume}
  {67}},\ \bibinfo {pages} {014319} (\bibinfo {year} {2003})}\BibitemShut
  {NoStop}%
\bibitem [{\citenamefont {Schönwaßer}\ \emph {et~al.}(2003)\citenamefont
  {Schönwaßer}, \citenamefont {Hübel}, \citenamefont {Hagemann},
  \citenamefont {Bednarczyk}, \citenamefont {Benzoni}, \citenamefont {Bracco},
  \citenamefont {Bringel}, \citenamefont {Chapman}, \citenamefont {Curien},
  \citenamefont {Domscheit}, \citenamefont {Herskind}, \citenamefont {Jensen},
  \citenamefont {Leoni}, \citenamefont {{Lo Bianco}}, \citenamefont {Ma},
  \citenamefont {Maj}, \citenamefont {Neußer}, \citenamefont {Ødegård},
  \citenamefont {Petrache}, \citenamefont {Roßbach}, \citenamefont {Ryde},
  \citenamefont {Spohr},\ and\ \citenamefont {Singh}}]{WM44}%
  \BibitemOpen
  \bibfield  {author} {\bibinfo {author} {\bibfnamefont {G.}~\bibnamefont
  {Schönwaßer}}, \bibinfo {author} {\bibfnamefont {H.}~\bibnamefont
  {Hübel}}, \bibinfo {author} {\bibfnamefont {G.}~\bibnamefont {Hagemann}},
  \bibinfo {author} {\bibfnamefont {P.}~\bibnamefont {Bednarczyk}}, \bibinfo
  {author} {\bibfnamefont {G.}~\bibnamefont {Benzoni}}, \bibinfo {author}
  {\bibfnamefont {A.}~\bibnamefont {Bracco}}, \bibinfo {author} {\bibfnamefont
  {P.}~\bibnamefont {Bringel}}, \bibinfo {author} {\bibfnamefont
  {R.}~\bibnamefont {Chapman}}, \bibinfo {author} {\bibfnamefont
  {D.}~\bibnamefont {Curien}}, \bibinfo {author} {\bibfnamefont
  {J.}~\bibnamefont {Domscheit}}, \bibinfo {author} {\bibfnamefont
  {B.}~\bibnamefont {Herskind}}, \bibinfo {author} {\bibfnamefont
  {D.}~\bibnamefont {Jensen}}, \bibinfo {author} {\bibfnamefont
  {S.}~\bibnamefont {Leoni}}, \bibinfo {author} {\bibfnamefont
  {G.}~\bibnamefont {{Lo Bianco}}}, \bibinfo {author} {\bibfnamefont
  {W.}~\bibnamefont {Ma}}, \bibinfo {author} {\bibfnamefont {A.}~\bibnamefont
  {Maj}}, \bibinfo {author} {\bibfnamefont {A.}~\bibnamefont {Neußer}},
  \bibinfo {author} {\bibfnamefont {S.}~\bibnamefont {Ødegård}}, \bibinfo
  {author} {\bibfnamefont {C.}~\bibnamefont {Petrache}}, \bibinfo {author}
  {\bibfnamefont {D.}~\bibnamefont {Roßbach}}, \bibinfo {author}
  {\bibfnamefont {H.}~\bibnamefont {Ryde}}, \bibinfo {author} {\bibfnamefont
  {K.}~\bibnamefont {Spohr}},\ and\ \bibinfo {author} {\bibfnamefont
  {A.}~\bibnamefont {Singh}},\ }\href
  {https://doi.org/https://doi.org/10.1016/S0370-2693(02)03095-2} {\bibfield
  {journal} {\bibinfo  {journal} {Phys. Lett. B}\ }\textbf {\bibinfo {volume}
  {552}},\ \bibinfo {pages} {9} (\bibinfo {year} {2003})}\BibitemShut {NoStop}%
\bibitem [{\citenamefont {{H Amro and W.C Ma and G.B Hagemann and R.M Diamond
  and J Domscheit and P Fallon and A Görgen and B Herskind and H Hübel and
  D.R Jensen and Y Li and A.O Macchiavelli and D Roux and G Sletten and J
  Thompson and D Ward and I Wiedenhöver and J.N Wilson and J.A
  Winger}}(2003)}]{WM45}%
  \BibitemOpen
  \bibfield  {author} {\bibinfo {author} {\bibnamefont {{H Amro and W.C Ma and
  G.B Hagemann and R.M Diamond and J Domscheit and P Fallon and A Görgen and B
  Herskind and H Hübel and D.R Jensen and Y Li and A.O Macchiavelli and D Roux
  and G Sletten and J Thompson and D Ward and I Wiedenhöver and J.N Wilson and
  J.A Winger}}},\ }\href
  {https://doi.org/https://doi.org/10.1016/S0370-2693(02)03199-4} {\bibfield
  {journal} {\bibinfo  {journal} {Phys. Lett. B}\ }\textbf {\bibinfo {volume}
  {553}},\ \bibinfo {pages} {197} (\bibinfo {year} {2003})}\BibitemShut
  {NoStop}%
\bibitem [{\citenamefont {Bringel}\ \emph {et~al.}(2005)\citenamefont
  {Bringel}, \citenamefont {Hagemann}, \citenamefont {H{\"u}bel}, \citenamefont
  {Al-khatib}, \citenamefont {Bednarczyk}, \citenamefont {B{\"u}rger},
  \citenamefont {Curien}, \citenamefont {Gangopadhyay}, \citenamefont
  {Herskind}, \citenamefont {Jensen}, \citenamefont {Joss}, \citenamefont
  {Kr{\"o}ll}, \citenamefont {Lo~Bianco}, \citenamefont {Lunardi},
  \citenamefont {Ma}, \citenamefont {Nenoff}, \citenamefont
  {Neu{\ss}er-Neffgen}, \citenamefont {Petrache}, \citenamefont
  {Sch{\"o}nwasser}, \citenamefont {Simpson}, \citenamefont {Singh},
  \citenamefont {Singh},\ and\ \citenamefont {Sletten}}]{WM2}%
  \BibitemOpen
  \bibfield  {author} {\bibinfo {author} {\bibfnamefont {P.}~\bibnamefont
  {Bringel}}, \bibinfo {author} {\bibfnamefont {G.~B.}\ \bibnamefont
  {Hagemann}}, \bibinfo {author} {\bibfnamefont {H.}~\bibnamefont {H{\"u}bel}},
  \bibinfo {author} {\bibfnamefont {A.}~\bibnamefont {Al-khatib}}, \bibinfo
  {author} {\bibfnamefont {P.}~\bibnamefont {Bednarczyk}}, \bibinfo {author}
  {\bibfnamefont {A.}~\bibnamefont {B{\"u}rger}}, \bibinfo {author}
  {\bibfnamefont {D.}~\bibnamefont {Curien}}, \bibinfo {author} {\bibfnamefont
  {G.}~\bibnamefont {Gangopadhyay}}, \bibinfo {author} {\bibfnamefont
  {B.}~\bibnamefont {Herskind}}, \bibinfo {author} {\bibfnamefont {D.~R.}\
  \bibnamefont {Jensen}}, \bibinfo {author} {\bibfnamefont {D.~T.}\
  \bibnamefont {Joss}}, \bibinfo {author} {\bibfnamefont {T.}~\bibnamefont
  {Kr{\"o}ll}}, \bibinfo {author} {\bibfnamefont {G.}~\bibnamefont
  {Lo~Bianco}}, \bibinfo {author} {\bibfnamefont {S.}~\bibnamefont {Lunardi}},
  \bibinfo {author} {\bibfnamefont {W.~C.}\ \bibnamefont {Ma}}, \bibinfo
  {author} {\bibfnamefont {N.}~\bibnamefont {Nenoff}}, \bibinfo {author}
  {\bibfnamefont {A.}~\bibnamefont {Neu{\ss}er-Neffgen}}, \bibinfo {author}
  {\bibfnamefont {C.~M.}\ \bibnamefont {Petrache}}, \bibinfo {author}
  {\bibfnamefont {G.}~\bibnamefont {Sch{\"o}nwasser}}, \bibinfo {author}
  {\bibfnamefont {J.}~\bibnamefont {Simpson}}, \bibinfo {author} {\bibfnamefont
  {A.~K.}\ \bibnamefont {Singh}}, \bibinfo {author} {\bibfnamefont
  {N.}~\bibnamefont {Singh}},\ and\ \bibinfo {author} {\bibfnamefont
  {G.}~\bibnamefont {Sletten}},\ }\href
  {https://doi.org/10.1140/epja/i2005-10005-7} {\bibfield  {journal} {\bibinfo
  {journal} {Eur. Phys. J. A}\ }\textbf {\bibinfo {volume} {24}},\ \bibinfo
  {pages} {167} (\bibinfo {year} {2005})}\BibitemShut {NoStop}%
\bibitem [{\citenamefont {Hartley}\ \emph {et~al.}(2009)\citenamefont
  {Hartley}, \citenamefont {Janssens}, \citenamefont {Riedinger}, \citenamefont
  {Riley}, \citenamefont {Aguilar}, \citenamefont {Carpenter}, \citenamefont
  {Chiara}, \citenamefont {Chowdhury}, \citenamefont {Darby}, \citenamefont
  {Garg}, \citenamefont {Ijaz}, \citenamefont {Kondev}, \citenamefont
  {Lakshmi}, \citenamefont {Lauritsen}, \citenamefont {Ludington},
  \citenamefont {Ma}, \citenamefont {McCutchan}, \citenamefont {Mukhopadhyay},
  \citenamefont {Pifer}, \citenamefont {Seyfried}, \citenamefont {Stefanescu},
  \citenamefont {Tandel}, \citenamefont {Tandel}, \citenamefont {Vanhoy},
  \citenamefont {Wang}, \citenamefont {Zhu}, \citenamefont {Hamamoto},\ and\
  \citenamefont {Frauendorf}}]{WM46}%
  \BibitemOpen
  \bibfield  {author} {\bibinfo {author} {\bibfnamefont {D.~J.}\ \bibnamefont
  {Hartley}}, \bibinfo {author} {\bibfnamefont {R.~V.~F.}\ \bibnamefont
  {Janssens}}, \bibinfo {author} {\bibfnamefont {L.~L.}\ \bibnamefont
  {Riedinger}}, \bibinfo {author} {\bibfnamefont {M.~A.}\ \bibnamefont
  {Riley}}, \bibinfo {author} {\bibfnamefont {A.}~\bibnamefont {Aguilar}},
  \bibinfo {author} {\bibfnamefont {M.~P.}\ \bibnamefont {Carpenter}}, \bibinfo
  {author} {\bibfnamefont {C.~J.}\ \bibnamefont {Chiara}}, \bibinfo {author}
  {\bibfnamefont {P.}~\bibnamefont {Chowdhury}}, \bibinfo {author}
  {\bibfnamefont {I.~G.}\ \bibnamefont {Darby}}, \bibinfo {author}
  {\bibfnamefont {U.}~\bibnamefont {Garg}}, \bibinfo {author} {\bibfnamefont
  {Q.~A.}\ \bibnamefont {Ijaz}}, \bibinfo {author} {\bibfnamefont {F.~G.}\
  \bibnamefont {Kondev}}, \bibinfo {author} {\bibfnamefont {S.}~\bibnamefont
  {Lakshmi}}, \bibinfo {author} {\bibfnamefont {T.}~\bibnamefont {Lauritsen}},
  \bibinfo {author} {\bibfnamefont {A.}~\bibnamefont {Ludington}}, \bibinfo
  {author} {\bibfnamefont {W.~C.}\ \bibnamefont {Ma}}, \bibinfo {author}
  {\bibfnamefont {E.~A.}\ \bibnamefont {McCutchan}}, \bibinfo {author}
  {\bibfnamefont {S.}~\bibnamefont {Mukhopadhyay}}, \bibinfo {author}
  {\bibfnamefont {R.}~\bibnamefont {Pifer}}, \bibinfo {author} {\bibfnamefont
  {E.~P.}\ \bibnamefont {Seyfried}}, \bibinfo {author} {\bibfnamefont
  {I.}~\bibnamefont {Stefanescu}}, \bibinfo {author} {\bibfnamefont {S.~K.}\
  \bibnamefont {Tandel}}, \bibinfo {author} {\bibfnamefont {U.}~\bibnamefont
  {Tandel}}, \bibinfo {author} {\bibfnamefont {J.~R.}\ \bibnamefont {Vanhoy}},
  \bibinfo {author} {\bibfnamefont {X.}~\bibnamefont {Wang}}, \bibinfo {author}
  {\bibfnamefont {S.}~\bibnamefont {Zhu}}, \bibinfo {author} {\bibfnamefont
  {I.}~\bibnamefont {Hamamoto}},\ and\ \bibinfo {author} {\bibfnamefont
  {S.}~\bibnamefont {Frauendorf}},\ }\href
  {https://doi.org/10.1103/PhysRevC.80.041304} {\bibfield  {journal} {\bibinfo
  {journal} {Phys. Rev. C}\ }\textbf {\bibinfo {volume} {80}},\ \bibinfo
  {pages} {041304} (\bibinfo {year} {2009})}\BibitemShut {NoStop}%
\bibitem [{\citenamefont {Biswas}\ \emph {et~al.}(2019)\citenamefont {Biswas},
  \citenamefont {Palit}, \citenamefont {Frauendorf}, \citenamefont {Garg},
  \citenamefont {Li}, \citenamefont {Bhat}, \citenamefont {Sheikh},
  \citenamefont {Sethi}, \citenamefont {Saha}, \citenamefont {Singh},
  \citenamefont {Choudhury}, \citenamefont {Matta}, \citenamefont {Ayangeakaa},
  \citenamefont {Dar}, \citenamefont {Singh},\ and\ \citenamefont
  {Sihotra}}]{WM54}%
  \BibitemOpen
  \bibfield  {author} {\bibinfo {author} {\bibfnamefont {S.}~\bibnamefont
  {Biswas}}, \bibinfo {author} {\bibfnamefont {R.}~\bibnamefont {Palit}},
  \bibinfo {author} {\bibfnamefont {S.}~\bibnamefont {Frauendorf}}, \bibinfo
  {author} {\bibfnamefont {U.}~\bibnamefont {Garg}}, \bibinfo {author}
  {\bibfnamefont {W.}~\bibnamefont {Li}}, \bibinfo {author} {\bibfnamefont
  {G.~H.}\ \bibnamefont {Bhat}}, \bibinfo {author} {\bibfnamefont {J.~A.}\
  \bibnamefont {Sheikh}}, \bibinfo {author} {\bibfnamefont {J.}~\bibnamefont
  {Sethi}}, \bibinfo {author} {\bibfnamefont {S.}~\bibnamefont {Saha}},
  \bibinfo {author} {\bibfnamefont {P.}~\bibnamefont {Singh}}, \bibinfo
  {author} {\bibfnamefont {D.}~\bibnamefont {Choudhury}}, \bibinfo {author}
  {\bibfnamefont {J.~T.}\ \bibnamefont {Matta}}, \bibinfo {author}
  {\bibfnamefont {A.~D.}\ \bibnamefont {Ayangeakaa}}, \bibinfo {author}
  {\bibfnamefont {W.~A.}\ \bibnamefont {Dar}}, \bibinfo {author} {\bibfnamefont
  {V.}~\bibnamefont {Singh}},\ and\ \bibinfo {author} {\bibfnamefont
  {S.}~\bibnamefont {Sihotra}},\ }\href
  {https://doi.org/10.1140/epja/i2019-12856-5} {\bibfield  {journal} {\bibinfo
  {journal} {Eur. Phys. J. A}\ }\textbf {\bibinfo {volume} {55}},\ \bibinfo
  {pages} {159} (\bibinfo {year} {2019})}\BibitemShut {NoStop}%
\bibitem [{\citenamefont {Matta}\ \emph {et~al.}(2015)\citenamefont {Matta},
  \citenamefont {Garg}, \citenamefont {Li}, \citenamefont {Frauendorf},
  \citenamefont {Ayangeakaa}, \citenamefont {Patel}, \citenamefont {Schlax},
  \citenamefont {Palit}, \citenamefont {Saha}, \citenamefont {Sethi},
  \citenamefont {Trivedi}, \citenamefont {Ghugre}, \citenamefont {Raut},
  \citenamefont {Sinha}, \citenamefont {Janssens}, \citenamefont {Zhu},
  \citenamefont {Carpenter}, \citenamefont {Lauritsen}, \citenamefont
  {Seweryniak}, \citenamefont {Chiara}, \citenamefont {Kondev}, \citenamefont
  {Hartley}, \citenamefont {Petrache}, \citenamefont {Mukhopadhyay},
  \citenamefont {Lakshmi}, \citenamefont {Raju}, \citenamefont
  {Madhusudhana~Rao}, \citenamefont {Tandel}, \citenamefont {Ray},\ and\
  \citenamefont {D\"onau}}]{WM47}%
  \BibitemOpen
  \bibfield  {author} {\bibinfo {author} {\bibfnamefont {J.~T.}\ \bibnamefont
  {Matta}}, \bibinfo {author} {\bibfnamefont {U.}~\bibnamefont {Garg}},
  \bibinfo {author} {\bibfnamefont {W.}~\bibnamefont {Li}}, \bibinfo {author}
  {\bibfnamefont {S.}~\bibnamefont {Frauendorf}}, \bibinfo {author}
  {\bibfnamefont {A.~D.}\ \bibnamefont {Ayangeakaa}}, \bibinfo {author}
  {\bibfnamefont {D.}~\bibnamefont {Patel}}, \bibinfo {author} {\bibfnamefont
  {K.~W.}\ \bibnamefont {Schlax}}, \bibinfo {author} {\bibfnamefont
  {R.}~\bibnamefont {Palit}}, \bibinfo {author} {\bibfnamefont
  {S.}~\bibnamefont {Saha}}, \bibinfo {author} {\bibfnamefont {J.}~\bibnamefont
  {Sethi}}, \bibinfo {author} {\bibfnamefont {T.}~\bibnamefont {Trivedi}},
  \bibinfo {author} {\bibfnamefont {S.~S.}\ \bibnamefont {Ghugre}}, \bibinfo
  {author} {\bibfnamefont {R.}~\bibnamefont {Raut}}, \bibinfo {author}
  {\bibfnamefont {A.~K.}\ \bibnamefont {Sinha}}, \bibinfo {author}
  {\bibfnamefont {R.~V.~F.}\ \bibnamefont {Janssens}}, \bibinfo {author}
  {\bibfnamefont {S.}~\bibnamefont {Zhu}}, \bibinfo {author} {\bibfnamefont
  {M.~P.}\ \bibnamefont {Carpenter}}, \bibinfo {author} {\bibfnamefont
  {T.}~\bibnamefont {Lauritsen}}, \bibinfo {author} {\bibfnamefont
  {D.}~\bibnamefont {Seweryniak}}, \bibinfo {author} {\bibfnamefont {C.~J.}\
  \bibnamefont {Chiara}}, \bibinfo {author} {\bibfnamefont {F.~G.}\
  \bibnamefont {Kondev}}, \bibinfo {author} {\bibfnamefont {D.~J.}\
  \bibnamefont {Hartley}}, \bibinfo {author} {\bibfnamefont {C.~M.}\
  \bibnamefont {Petrache}}, \bibinfo {author} {\bibfnamefont {S.}~\bibnamefont
  {Mukhopadhyay}}, \bibinfo {author} {\bibfnamefont {D.~V.}\ \bibnamefont
  {Lakshmi}}, \bibinfo {author} {\bibfnamefont {M.~K.}\ \bibnamefont {Raju}},
  \bibinfo {author} {\bibfnamefont {P.~V.}\ \bibnamefont {Madhusudhana~Rao}},
  \bibinfo {author} {\bibfnamefont {S.~K.}\ \bibnamefont {Tandel}}, \bibinfo
  {author} {\bibfnamefont {S.}~\bibnamefont {Ray}},\ and\ \bibinfo {author}
  {\bibfnamefont {F.}~\bibnamefont {D\"onau}},\ }\href
  {https://doi.org/10.1103/PhysRevLett.114.082501} {\bibfield  {journal}
  {\bibinfo  {journal} {Phys. Rev. Lett.}\ }\textbf {\bibinfo {volume} {114}},\
  \bibinfo {pages} {082501} (\bibinfo {year} {2015})}\BibitemShut {NoStop}%
\bibitem [{\citenamefont {Sensharma}\ \emph {et~al.}(2019)\citenamefont
  {Sensharma}, \citenamefont {Garg}, \citenamefont {Zhu}, \citenamefont
  {Ayangeakaa}, \citenamefont {Frauendorf}, \citenamefont {Li}, \citenamefont
  {Bhat}, \citenamefont {Sheikh}, \citenamefont {Carpenter}, \citenamefont
  {Chen}, \citenamefont {Cozzi}, \citenamefont {Ghugre}, \citenamefont {Gupta},
  \citenamefont {Hartley}, \citenamefont {Howard}, \citenamefont {Janssens},
  \citenamefont {Kondev}, \citenamefont {McMaken}, \citenamefont {Palit},
  \citenamefont {Sethi}, \citenamefont {Seweryniak},\ and\ \citenamefont
  {Singh}}]{WM48}%
  \BibitemOpen
  \bibfield  {author} {\bibinfo {author} {\bibfnamefont {N.}~\bibnamefont
  {Sensharma}}, \bibinfo {author} {\bibfnamefont {U.}~\bibnamefont {Garg}},
  \bibinfo {author} {\bibfnamefont {S.}~\bibnamefont {Zhu}}, \bibinfo {author}
  {\bibfnamefont {A.}~\bibnamefont {Ayangeakaa}}, \bibinfo {author}
  {\bibfnamefont {S.}~\bibnamefont {Frauendorf}}, \bibinfo {author}
  {\bibfnamefont {W.}~\bibnamefont {Li}}, \bibinfo {author} {\bibfnamefont
  {G.}~\bibnamefont {Bhat}}, \bibinfo {author} {\bibfnamefont {J.}~\bibnamefont
  {Sheikh}}, \bibinfo {author} {\bibfnamefont {M.}~\bibnamefont {Carpenter}},
  \bibinfo {author} {\bibfnamefont {Q.}~\bibnamefont {Chen}}, \bibinfo {author}
  {\bibfnamefont {J.}~\bibnamefont {Cozzi}}, \bibinfo {author} {\bibfnamefont
  {S.}~\bibnamefont {Ghugre}}, \bibinfo {author} {\bibfnamefont
  {Y.}~\bibnamefont {Gupta}}, \bibinfo {author} {\bibfnamefont
  {D.}~\bibnamefont {Hartley}}, \bibinfo {author} {\bibfnamefont
  {K.}~\bibnamefont {Howard}}, \bibinfo {author} {\bibfnamefont
  {R.}~\bibnamefont {Janssens}}, \bibinfo {author} {\bibfnamefont
  {F.}~\bibnamefont {Kondev}}, \bibinfo {author} {\bibfnamefont
  {T.}~\bibnamefont {McMaken}}, \bibinfo {author} {\bibfnamefont
  {R.}~\bibnamefont {Palit}}, \bibinfo {author} {\bibfnamefont
  {J.}~\bibnamefont {Sethi}}, \bibinfo {author} {\bibfnamefont
  {D.}~\bibnamefont {Seweryniak}},\ and\ \bibinfo {author} {\bibfnamefont
  {R.}~\bibnamefont {Singh}},\ }\href
  {https://doi.org/https://doi.org/10.1016/j.physletb.2019.03.038} {\bibfield
  {journal} {\bibinfo  {journal} {Phys. Lett. B}\ }\textbf {\bibinfo {volume}
  {792}},\ \bibinfo {pages} {170} (\bibinfo {year} {2019})}\BibitemShut
  {NoStop}%
\bibitem [{\citenamefont {Mukherjee}\ \emph {et~al.}(2023)\citenamefont
  {Mukherjee}, \citenamefont {Bhattacharya}, \citenamefont {Trivedi},
  \citenamefont {Tiwari}, \citenamefont {Singh}, \citenamefont {Muralithar},
  \citenamefont {Yashraj}, \citenamefont {Katre}, \citenamefont {Kumar},
  \citenamefont {Palit}, \citenamefont {Chakraborty}, \citenamefont {Jehangir},
  \citenamefont {Nazir}, \citenamefont {Rouoof}, \citenamefont {Bhat},
  \citenamefont {Sheikh}, \citenamefont {Rather}, \citenamefont {Raut},
  \citenamefont {Ghugre}, \citenamefont {Ali}, \citenamefont {Rajbanshi},
  \citenamefont {Nag}, \citenamefont {Tiwary}, \citenamefont {Sharma},
  \citenamefont {Kumar}, \citenamefont {Yadav},\ and\ \citenamefont
  {Jain}}]{WM6}%
  \BibitemOpen
  \bibfield  {author} {\bibinfo {author} {\bibfnamefont {A.}~\bibnamefont
  {Mukherjee}}, \bibinfo {author} {\bibfnamefont {S.}~\bibnamefont
  {Bhattacharya}}, \bibinfo {author} {\bibfnamefont {T.}~\bibnamefont
  {Trivedi}}, \bibinfo {author} {\bibfnamefont {S.}~\bibnamefont {Tiwari}},
  \bibinfo {author} {\bibfnamefont {R.~P.}\ \bibnamefont {Singh}}, \bibinfo
  {author} {\bibfnamefont {S.}~\bibnamefont {Muralithar}}, \bibinfo {author}
  {\bibnamefont {Yashraj}}, \bibinfo {author} {\bibfnamefont {K.}~\bibnamefont
  {Katre}}, \bibinfo {author} {\bibfnamefont {R.}~\bibnamefont {Kumar}},
  \bibinfo {author} {\bibfnamefont {R.}~\bibnamefont {Palit}}, \bibinfo
  {author} {\bibfnamefont {S.}~\bibnamefont {Chakraborty}}, \bibinfo {author}
  {\bibfnamefont {S.}~\bibnamefont {Jehangir}}, \bibinfo {author}
  {\bibfnamefont {N.}~\bibnamefont {Nazir}}, \bibinfo {author} {\bibfnamefont
  {S.~P.}\ \bibnamefont {Rouoof}}, \bibinfo {author} {\bibfnamefont {G.~H.}\
  \bibnamefont {Bhat}}, \bibinfo {author} {\bibfnamefont {J.~A.}\ \bibnamefont
  {Sheikh}}, \bibinfo {author} {\bibfnamefont {N.}~\bibnamefont {Rather}},
  \bibinfo {author} {\bibfnamefont {R.}~\bibnamefont {Raut}}, \bibinfo {author}
  {\bibfnamefont {S.~S.}\ \bibnamefont {Ghugre}}, \bibinfo {author}
  {\bibfnamefont {S.}~\bibnamefont {Ali}}, \bibinfo {author} {\bibfnamefont
  {S.}~\bibnamefont {Rajbanshi}}, \bibinfo {author} {\bibfnamefont
  {S.}~\bibnamefont {Nag}}, \bibinfo {author} {\bibfnamefont {S.~S.}\
  \bibnamefont {Tiwary}}, \bibinfo {author} {\bibfnamefont {A.}~\bibnamefont
  {Sharma}}, \bibinfo {author} {\bibfnamefont {S.}~\bibnamefont {Kumar}},
  \bibinfo {author} {\bibfnamefont {S.}~\bibnamefont {Yadav}},\ and\ \bibinfo
  {author} {\bibfnamefont {A.~K.}\ \bibnamefont {Jain}},\ }\href
  {https://doi.org/10.1103/PhysRevC.107.054310} {\bibfield  {journal} {\bibinfo
   {journal} {Phys. Rev. C}\ }\textbf {\bibinfo {volume} {107}},\ \bibinfo
  {pages} {054310} (\bibinfo {year} {2023})}\BibitemShut {NoStop}%
\bibitem [{\citenamefont {Tim\'ar}\ \emph {et~al.}(2019)\citenamefont
  {Tim\'ar}, \citenamefont {Chen}, \citenamefont {Kruzsicz}, \citenamefont
  {Sohler}, \citenamefont {Kuti}, \citenamefont {Zhang}, \citenamefont {Meng},
  \citenamefont {Joshi}, \citenamefont {Wadsworth}, \citenamefont {Starosta},
  \citenamefont {Algora}, \citenamefont {Bednarczyk}, \citenamefont {Curien},
  \citenamefont {Dombr\'adi}, \citenamefont {Duch\^ene}, \citenamefont {Gizon},
  \citenamefont {Gizon}, \citenamefont {Jenkins}, \citenamefont {Koike},
  \citenamefont {Krasznahorkay}, \citenamefont {Moln\'ar}, \citenamefont
  {Nyak\'o}, \citenamefont {Paul}, \citenamefont {Rainovski}, \citenamefont
  {Scheurer}, \citenamefont {Simons}, \citenamefont {Vaman},\ and\
  \citenamefont {Zolnai}}]{WM12}%
  \BibitemOpen
  \bibfield  {author} {\bibinfo {author} {\bibfnamefont {J.}~\bibnamefont
  {Tim\'ar}}, \bibinfo {author} {\bibfnamefont {Q.~B.}\ \bibnamefont {Chen}},
  \bibinfo {author} {\bibfnamefont {B.}~\bibnamefont {Kruzsicz}}, \bibinfo
  {author} {\bibfnamefont {D.}~\bibnamefont {Sohler}}, \bibinfo {author}
  {\bibfnamefont {I.}~\bibnamefont {Kuti}}, \bibinfo {author} {\bibfnamefont
  {S.~Q.}\ \bibnamefont {Zhang}}, \bibinfo {author} {\bibfnamefont
  {J.}~\bibnamefont {Meng}}, \bibinfo {author} {\bibfnamefont {P.}~\bibnamefont
  {Joshi}}, \bibinfo {author} {\bibfnamefont {R.}~\bibnamefont {Wadsworth}},
  \bibinfo {author} {\bibfnamefont {K.}~\bibnamefont {Starosta}}, \bibinfo
  {author} {\bibfnamefont {A.}~\bibnamefont {Algora}}, \bibinfo {author}
  {\bibfnamefont {P.}~\bibnamefont {Bednarczyk}}, \bibinfo {author}
  {\bibfnamefont {D.}~\bibnamefont {Curien}}, \bibinfo {author} {\bibfnamefont
  {Z.}~\bibnamefont {Dombr\'adi}}, \bibinfo {author} {\bibfnamefont
  {G.}~\bibnamefont {Duch\^ene}}, \bibinfo {author} {\bibfnamefont
  {A.}~\bibnamefont {Gizon}}, \bibinfo {author} {\bibfnamefont
  {J.}~\bibnamefont {Gizon}}, \bibinfo {author} {\bibfnamefont {D.~G.}\
  \bibnamefont {Jenkins}}, \bibinfo {author} {\bibfnamefont {T.}~\bibnamefont
  {Koike}}, \bibinfo {author} {\bibfnamefont {A.}~\bibnamefont
  {Krasznahorkay}}, \bibinfo {author} {\bibfnamefont {J.}~\bibnamefont
  {Moln\'ar}}, \bibinfo {author} {\bibfnamefont {B.~M.}\ \bibnamefont
  {Nyak\'o}}, \bibinfo {author} {\bibfnamefont {E.~S.}\ \bibnamefont {Paul}},
  \bibinfo {author} {\bibfnamefont {G.}~\bibnamefont {Rainovski}}, \bibinfo
  {author} {\bibfnamefont {J.~N.}\ \bibnamefont {Scheurer}}, \bibinfo {author}
  {\bibfnamefont {A.~J.}\ \bibnamefont {Simons}}, \bibinfo {author}
  {\bibfnamefont {C.}~\bibnamefont {Vaman}},\ and\ \bibinfo {author}
  {\bibfnamefont {L.}~\bibnamefont {Zolnai}},\ }\href
  {https://doi.org/10.1103/PhysRevLett.122.062501} {\bibfield  {journal}
  {\bibinfo  {journal} {Phys. Rev. Lett.}\ }\textbf {\bibinfo {volume} {122}},\
  \bibinfo {pages} {062501} (\bibinfo {year} {2019})}\BibitemShut {NoStop}%
\bibitem [{\citenamefont {Lv}\ \emph {et~al.}(2022{\natexlab{a}})\citenamefont
  {Lv}, \citenamefont {Petrache}, \citenamefont {Lawrie}, \citenamefont {Guo},
  \citenamefont {Astier}, \citenamefont {Zheng}, \citenamefont {Ong},
  \citenamefont {Wang}, \citenamefont {Zhou}, \citenamefont {Sun},
  \citenamefont {Greenlees}, \citenamefont {Badran}, \citenamefont {Calverley},
  \citenamefont {Cox}, \citenamefont {Grahn}, \citenamefont {Hilton},
  \citenamefont {Julin}, \citenamefont {Juutinen}, \citenamefont {Konki},
  \citenamefont {Pakarinen}, \citenamefont {Papadakis}, \citenamefont
  {Partanen}, \citenamefont {Rahkila}, \citenamefont {Ruotsalainen},
  \citenamefont {Sandzelius}, \citenamefont {Sarén}, \citenamefont {Scholey},
  \citenamefont {Sorri}, \citenamefont {Stolze}, \citenamefont {Uusitalo},
  \citenamefont {Cederwall}, \citenamefont {Ertoprak}, \citenamefont {Liu},
  \citenamefont {Kuti}, \citenamefont {Timár}, \citenamefont {Tucholski},
  \citenamefont {Srebrny},\ and\ \citenamefont {Andreoiu}}]{LV2022}%
  \BibitemOpen
  \bibfield  {author} {\bibinfo {author} {\bibfnamefont {B.}~\bibnamefont
  {Lv}}, \bibinfo {author} {\bibfnamefont {C.}~\bibnamefont {Petrache}},
  \bibinfo {author} {\bibfnamefont {E.}~\bibnamefont {Lawrie}}, \bibinfo
  {author} {\bibfnamefont {S.}~\bibnamefont {Guo}}, \bibinfo {author}
  {\bibfnamefont {A.}~\bibnamefont {Astier}}, \bibinfo {author} {\bibfnamefont
  {K.}~\bibnamefont {Zheng}}, \bibinfo {author} {\bibfnamefont
  {H.}~\bibnamefont {Ong}}, \bibinfo {author} {\bibfnamefont {J.}~\bibnamefont
  {Wang}}, \bibinfo {author} {\bibfnamefont {X.}~\bibnamefont {Zhou}}, \bibinfo
  {author} {\bibfnamefont {Z.}~\bibnamefont {Sun}}, \bibinfo {author}
  {\bibfnamefont {P.}~\bibnamefont {Greenlees}}, \bibinfo {author}
  {\bibfnamefont {H.}~\bibnamefont {Badran}}, \bibinfo {author} {\bibfnamefont
  {T.}~\bibnamefont {Calverley}}, \bibinfo {author} {\bibfnamefont
  {D.}~\bibnamefont {Cox}}, \bibinfo {author} {\bibfnamefont {T.}~\bibnamefont
  {Grahn}}, \bibinfo {author} {\bibfnamefont {J.}~\bibnamefont {Hilton}},
  \bibinfo {author} {\bibfnamefont {R.}~\bibnamefont {Julin}}, \bibinfo
  {author} {\bibfnamefont {S.}~\bibnamefont {Juutinen}}, \bibinfo {author}
  {\bibfnamefont {J.}~\bibnamefont {Konki}}, \bibinfo {author} {\bibfnamefont
  {J.}~\bibnamefont {Pakarinen}}, \bibinfo {author} {\bibfnamefont
  {P.}~\bibnamefont {Papadakis}}, \bibinfo {author} {\bibfnamefont
  {J.}~\bibnamefont {Partanen}}, \bibinfo {author} {\bibfnamefont
  {P.}~\bibnamefont {Rahkila}}, \bibinfo {author} {\bibfnamefont
  {P.}~\bibnamefont {Ruotsalainen}}, \bibinfo {author} {\bibfnamefont
  {M.}~\bibnamefont {Sandzelius}}, \bibinfo {author} {\bibfnamefont
  {J.}~\bibnamefont {Sarén}}, \bibinfo {author} {\bibfnamefont
  {C.}~\bibnamefont {Scholey}}, \bibinfo {author} {\bibfnamefont
  {J.}~\bibnamefont {Sorri}}, \bibinfo {author} {\bibfnamefont
  {S.}~\bibnamefont {Stolze}}, \bibinfo {author} {\bibfnamefont
  {J.}~\bibnamefont {Uusitalo}}, \bibinfo {author} {\bibfnamefont
  {B.}~\bibnamefont {Cederwall}}, \bibinfo {author} {\bibfnamefont
  {A.}~\bibnamefont {Ertoprak}}, \bibinfo {author} {\bibfnamefont
  {H.}~\bibnamefont {Liu}}, \bibinfo {author} {\bibfnamefont {I.}~\bibnamefont
  {Kuti}}, \bibinfo {author} {\bibfnamefont {J.}~\bibnamefont {Timár}},
  \bibinfo {author} {\bibfnamefont {A.}~\bibnamefont {Tucholski}}, \bibinfo
  {author} {\bibfnamefont {J.}~\bibnamefont {Srebrny}},\ and\ \bibinfo {author}
  {\bibfnamefont {C.}~\bibnamefont {Andreoiu}},\ }\href
  {https://doi.org/https://doi.org/10.1016/j.physletb.2021.136840} {\bibfield
  {journal} {\bibinfo  {journal} {Phys. Lett. B}\ }\textbf {\bibinfo {volume}
  {824}},\ \bibinfo {pages} {136840} (\bibinfo {year}
  {2022}{\natexlab{a}})}\BibitemShut {NoStop}%
\bibitem [{\citenamefont {Lv}\ and\ \citenamefont {Petrache}(2023)}]{BCM23}%
  \BibitemOpen
  \bibfield  {author} {\bibinfo {author} {\bibfnamefont {B.}~\bibnamefont
  {Lv}}\ and\ \bibinfo {author} {\bibfnamefont {C.~M.}\ \bibnamefont
  {Petrache}},\ }\bibfield  {journal} {\bibinfo  {journal} {Symmetry}\ }\textbf
  {\bibinfo {volume} {15}},\ \href {https://doi.org/10.3390/sym15051075}
  {10.3390/sym15051075} (\bibinfo {year} {2023})\BibitemShut {NoStop}%
\bibitem [{\citenamefont {Hamilton}\ \emph {et~al.}(2010)\citenamefont
  {Hamilton}, \citenamefont {Zhu}, \citenamefont {Luo}, \citenamefont
  {Ramayya}, \citenamefont {Frauendorf}, \citenamefont {Rasmussen},
  \citenamefont {Hwang}, \citenamefont {Liu}, \citenamefont {Ter-Akopian},
  \citenamefont {Daniel},\ and\ \citenamefont {Oganessian}}]{JH10}%
  \BibitemOpen
  \bibfield  {author} {\bibinfo {author} {\bibfnamefont {J.}~\bibnamefont
  {Hamilton}}, \bibinfo {author} {\bibfnamefont {S.}~\bibnamefont {Zhu}},
  \bibinfo {author} {\bibfnamefont {Y.}~\bibnamefont {Luo}}, \bibinfo {author}
  {\bibfnamefont {A.}~\bibnamefont {Ramayya}}, \bibinfo {author} {\bibfnamefont
  {S.}~\bibnamefont {Frauendorf}}, \bibinfo {author} {\bibfnamefont
  {J.}~\bibnamefont {Rasmussen}}, \bibinfo {author} {\bibfnamefont
  {J.}~\bibnamefont {Hwang}}, \bibinfo {author} {\bibfnamefont
  {S.}~\bibnamefont {Liu}}, \bibinfo {author} {\bibfnamefont {G.}~\bibnamefont
  {Ter-Akopian}}, \bibinfo {author} {\bibfnamefont {A.}~\bibnamefont
  {Daniel}},\ and\ \bibinfo {author} {\bibfnamefont {Y.}~\bibnamefont
  {Oganessian}},\ }\href
  {https://doi.org/https://doi.org/10.1016/j.nuclphysa.2010.01.010} {\bibfield
  {journal} {\bibinfo  {journal} {Nuclear Physics A}\ }\textbf {\bibinfo
  {volume} {834}},\ \bibinfo {pages} {28c} (\bibinfo {year} {2010})},\ \bibinfo
  {note} {the 10th International Conference on Nucleus-Nucleus Collisions
  (NN2009)}\BibitemShut {NoStop}%
\bibitem [{\citenamefont {Frauendorf}(2024)}]{SF24ch}%
  \BibitemOpen
  \bibfield  {author} {\bibinfo {author} {\bibfnamefont {S.}~\bibnamefont
  {Frauendorf}},\ }\href {https://arxiv.org/abs/2405.02747} {\bibinfo {title}
  {Wobbling motion in triaxial nuclei}} (\bibinfo {year} {2024}),\ \Eprint
  {https://arxiv.org/abs/2405.02747} {arXiv:2405.02747 [nucl-th]} \BibitemShut
  {NoStop}%
\bibitem [{\citenamefont {Petrache}\ \emph {et~al.}(2019)\citenamefont
  {Petrache}, \citenamefont {Walker}, \citenamefont {Guo}, \citenamefont
  {Chen}, \citenamefont {Frauendorf}, \citenamefont {Liu}, \citenamefont
  {Wyss}, \citenamefont {Mengoni}, \citenamefont {Qiang}, \citenamefont
  {Astier}, \citenamefont {Dupont}, \citenamefont {Li}, \citenamefont {Lv},
  \citenamefont {Zheng}, \citenamefont {Bazzacco}, \citenamefont {Boso},
  \citenamefont {Goasduff}, \citenamefont {Recchia}, \citenamefont {Testov},
  \citenamefont {Galtarossa}, \citenamefont {Jaworski}, \citenamefont {Napoli},
  \citenamefont {Riccetto}, \citenamefont {Siciliano}, \citenamefont
  {Valiente-Dobon}, \citenamefont {Liu}, \citenamefont {Zhou}, \citenamefont
  {Wang}, \citenamefont {Andreoiu}, \citenamefont {Garcia}, \citenamefont
  {Ortner}, \citenamefont {Whitmore}, \citenamefont {Bäck}, \citenamefont
  {Cederwall}, \citenamefont {Lawrie}, \citenamefont {Kuti}, \citenamefont
  {Sohler}, \citenamefont {Timár}, \citenamefont {Marchlewski}, \citenamefont
  {Srebrny},\ and\ \citenamefont {Tucholski}}]{PETRACHE2019}%
  \BibitemOpen
  \bibfield  {author} {\bibinfo {author} {\bibfnamefont {C.~M.}\ \bibnamefont
  {Petrache}}, \bibinfo {author} {\bibfnamefont {P.~M.}\ \bibnamefont
  {Walker}}, \bibinfo {author} {\bibfnamefont {S.}~\bibnamefont {Guo}},
  \bibinfo {author} {\bibfnamefont {Q.~B.}\ \bibnamefont {Chen}}, \bibinfo
  {author} {\bibfnamefont {S.}~\bibnamefont {Frauendorf}}, \bibinfo {author}
  {\bibfnamefont {Y.~X.}\ \bibnamefont {Liu}}, \bibinfo {author} {\bibfnamefont
  {R.~A.}\ \bibnamefont {Wyss}}, \bibinfo {author} {\bibfnamefont
  {D.}~\bibnamefont {Mengoni}}, \bibinfo {author} {\bibfnamefont {Y.~H.}\
  \bibnamefont {Qiang}}, \bibinfo {author} {\bibfnamefont {A.}~\bibnamefont
  {Astier}}, \bibinfo {author} {\bibfnamefont {E.}~\bibnamefont {Dupont}},
  \bibinfo {author} {\bibfnamefont {R.}~\bibnamefont {Li}}, \bibinfo {author}
  {\bibfnamefont {B.~F.}\ \bibnamefont {Lv}}, \bibinfo {author} {\bibfnamefont
  {K.~K.}\ \bibnamefont {Zheng}}, \bibinfo {author} {\bibfnamefont
  {D.}~\bibnamefont {Bazzacco}}, \bibinfo {author} {\bibfnamefont
  {A.}~\bibnamefont {Boso}}, \bibinfo {author} {\bibfnamefont {A.}~\bibnamefont
  {Goasduff}}, \bibinfo {author} {\bibfnamefont {F.}~\bibnamefont {Recchia}},
  \bibinfo {author} {\bibfnamefont {D.}~\bibnamefont {Testov}}, \bibinfo
  {author} {\bibfnamefont {F.}~\bibnamefont {Galtarossa}}, \bibinfo {author}
  {\bibfnamefont {G.}~\bibnamefont {Jaworski}}, \bibinfo {author}
  {\bibfnamefont {D.~R.}\ \bibnamefont {Napoli}}, \bibinfo {author}
  {\bibfnamefont {S.}~\bibnamefont {Riccetto}}, \bibinfo {author}
  {\bibfnamefont {M.}~\bibnamefont {Siciliano}}, \bibinfo {author}
  {\bibfnamefont {J.~J.}\ \bibnamefont {Valiente-Dobon}}, \bibinfo {author}
  {\bibfnamefont {M.}~\bibnamefont {Liu}}, \bibinfo {author} {\bibfnamefont
  {X.}~\bibnamefont {Zhou}}, \bibinfo {author} {\bibfnamefont {J.~G.}\
  \bibnamefont {Wang}}, \bibinfo {author} {\bibfnamefont {C.}~\bibnamefont
  {Andreoiu}}, \bibinfo {author} {\bibfnamefont {F.~H.}\ \bibnamefont
  {Garcia}}, \bibinfo {author} {\bibfnamefont {K.}~\bibnamefont {Ortner}},
  \bibinfo {author} {\bibfnamefont {K.}~\bibnamefont {Whitmore}}, \bibinfo
  {author} {\bibfnamefont {T.}~\bibnamefont {Bäck}}, \bibinfo {author}
  {\bibfnamefont {B.}~\bibnamefont {Cederwall}}, \bibinfo {author}
  {\bibfnamefont {E.}~\bibnamefont {Lawrie}}, \bibinfo {author} {\bibfnamefont
  {I.}~\bibnamefont {Kuti}}, \bibinfo {author} {\bibfnamefont {D.}~\bibnamefont
  {Sohler}}, \bibinfo {author} {\bibfnamefont {J.}~\bibnamefont {Timár}},
  \bibinfo {author} {\bibfnamefont {T.}~\bibnamefont {Marchlewski}}, \bibinfo
  {author} {\bibfnamefont {J.}~\bibnamefont {Srebrny}},\ and\ \bibinfo {author}
  {\bibfnamefont {A.}~\bibnamefont {Tucholski}},\ }\href
  {https://doi.org/https://doi.org/10.1016/j.physletb.2019.06.040} {\bibfield
  {journal} {\bibinfo  {journal} {Phys. Lett. B}\ }\textbf {\bibinfo {volume}
  {795}},\ \bibinfo {pages} {241} (\bibinfo {year} {2019})}\BibitemShut
  {NoStop}%
\bibitem [{\citenamefont {Wang}\ \emph
  {et~al.}(2020{\natexlab{a}})\citenamefont {Wang}, \citenamefont {Chen},\ and\
  \citenamefont {Zhao}}]{WANG2020}%
  \BibitemOpen
  \bibfield  {author} {\bibinfo {author} {\bibfnamefont {Y.~K.}\ \bibnamefont
  {Wang}}, \bibinfo {author} {\bibfnamefont {F.~Q.}\ \bibnamefont {Chen}},\
  and\ \bibinfo {author} {\bibfnamefont {P.~W.}\ \bibnamefont {Zhao}},\ }\href
  {https://doi.org/https://doi.org/10.1016/j.physletb.2020.135246} {\bibfield
  {journal} {\bibinfo  {journal} {Phys. Lett. B}\ }\textbf {\bibinfo {volume}
  {802}},\ \bibinfo {pages} {135246} (\bibinfo {year}
  {2020}{\natexlab{a}})}\BibitemShut {NoStop}%
\bibitem [{\citenamefont {Chen}\ \emph {et~al.}(2019)\citenamefont {Chen},
  \citenamefont {Frauendorf},\ and\ \citenamefont {Petrache}}]{Chen2019}%
  \BibitemOpen
  \bibfield  {author} {\bibinfo {author} {\bibfnamefont {Q.~B.}\ \bibnamefont
  {Chen}}, \bibinfo {author} {\bibfnamefont {S.}~\bibnamefont {Frauendorf}},\
  and\ \bibinfo {author} {\bibfnamefont {C.~M.}\ \bibnamefont {Petrache}},\
  }\href {https://doi.org/10.1103/PhysRevC.100.061301} {\bibfield  {journal}
  {\bibinfo  {journal} {Phys. Rev. C}\ }\textbf {\bibinfo {volume} {100}},\
  \bibinfo {pages} {061301(R)} (\bibinfo {year} {2019})}\BibitemShut {NoStop}%
\bibitem [{\citenamefont {Chen}\ and\ \citenamefont
  {Petrache}(2021)}]{Chen2021}%
  \BibitemOpen
  \bibfield  {author} {\bibinfo {author} {\bibfnamefont {F.-Q.}\ \bibnamefont
  {Chen}}\ and\ \bibinfo {author} {\bibfnamefont {C.~M.}\ \bibnamefont
  {Petrache}},\ }\href {https://doi.org/10.1103/PhysRevC.103.064319} {\bibfield
   {journal} {\bibinfo  {journal} {Phys. Rev. C}\ }\textbf {\bibinfo {volume}
  {103}},\ \bibinfo {pages} {064319} (\bibinfo {year} {2021})}\BibitemShut
  {NoStop}%
\bibitem [{\citenamefont {Lv}\ \emph {et~al.}(2022{\natexlab{b}})\citenamefont
  {Lv}, \citenamefont {Petrache}, \citenamefont {Budaca}, \citenamefont
  {Astier}, \citenamefont {Zheng}, \citenamefont {Greenlees}, \citenamefont
  {Badran}, \citenamefont {Calverley}, \citenamefont {Cox}, \citenamefont
  {Grahn}, \citenamefont {Hilton}, \citenamefont {Julin}, \citenamefont
  {Juutinen}, \citenamefont {Konki}, \citenamefont {Pakarinen}, \citenamefont
  {Papadakis}, \citenamefont {Partanen}, \citenamefont {Rahkila}, \citenamefont
  {Ruotsalainen}, \citenamefont {Sandzelius}, \citenamefont {Saren},
  \citenamefont {Scholey}, \citenamefont {Sorri}, \citenamefont {Stolze},
  \citenamefont {Uusitalo}, \citenamefont {Cederwall}, \citenamefont
  {Ertoprak}, \citenamefont {Liu}, \citenamefont {Guo}, \citenamefont {Wang},
  \citenamefont {Ong}, \citenamefont {Zhou}, \citenamefont {Sun}, \citenamefont
  {Kuti}, \citenamefont {Tim\'ar}, \citenamefont {Tucholski}, \citenamefont
  {Srebrny},\ and\ \citenamefont {Andreoiu}}]{LV20227}%
  \BibitemOpen
  \bibfield  {author} {\bibinfo {author} {\bibfnamefont {B.~F.}\ \bibnamefont
  {Lv}}, \bibinfo {author} {\bibfnamefont {C.~M.}\ \bibnamefont {Petrache}},
  \bibinfo {author} {\bibfnamefont {R.}~\bibnamefont {Budaca}}, \bibinfo
  {author} {\bibfnamefont {A.}~\bibnamefont {Astier}}, \bibinfo {author}
  {\bibfnamefont {K.~K.}\ \bibnamefont {Zheng}}, \bibinfo {author}
  {\bibfnamefont {P.}~\bibnamefont {Greenlees}}, \bibinfo {author}
  {\bibfnamefont {H.}~\bibnamefont {Badran}}, \bibinfo {author} {\bibfnamefont
  {T.}~\bibnamefont {Calverley}}, \bibinfo {author} {\bibfnamefont {D.~M.}\
  \bibnamefont {Cox}}, \bibinfo {author} {\bibfnamefont {T.}~\bibnamefont
  {Grahn}}, \bibinfo {author} {\bibfnamefont {J.}~\bibnamefont {Hilton}},
  \bibinfo {author} {\bibfnamefont {R.}~\bibnamefont {Julin}}, \bibinfo
  {author} {\bibfnamefont {S.}~\bibnamefont {Juutinen}}, \bibinfo {author}
  {\bibfnamefont {J.}~\bibnamefont {Konki}}, \bibinfo {author} {\bibfnamefont
  {J.}~\bibnamefont {Pakarinen}}, \bibinfo {author} {\bibfnamefont
  {P.}~\bibnamefont {Papadakis}}, \bibinfo {author} {\bibfnamefont
  {J.}~\bibnamefont {Partanen}}, \bibinfo {author} {\bibfnamefont
  {P.}~\bibnamefont {Rahkila}}, \bibinfo {author} {\bibfnamefont
  {P.}~\bibnamefont {Ruotsalainen}}, \bibinfo {author} {\bibfnamefont
  {M.}~\bibnamefont {Sandzelius}}, \bibinfo {author} {\bibfnamefont
  {J.}~\bibnamefont {Saren}}, \bibinfo {author} {\bibfnamefont
  {C.}~\bibnamefont {Scholey}}, \bibinfo {author} {\bibfnamefont
  {J.}~\bibnamefont {Sorri}}, \bibinfo {author} {\bibfnamefont
  {S.}~\bibnamefont {Stolze}}, \bibinfo {author} {\bibfnamefont
  {J.}~\bibnamefont {Uusitalo}}, \bibinfo {author} {\bibfnamefont
  {B.}~\bibnamefont {Cederwall}}, \bibinfo {author} {\bibfnamefont
  {A.}~\bibnamefont {Ertoprak}}, \bibinfo {author} {\bibfnamefont
  {H.}~\bibnamefont {Liu}}, \bibinfo {author} {\bibfnamefont {S.}~\bibnamefont
  {Guo}}, \bibinfo {author} {\bibfnamefont {J.~G.}\ \bibnamefont {Wang}},
  \bibinfo {author} {\bibfnamefont {H.~J.}\ \bibnamefont {Ong}}, \bibinfo
  {author} {\bibfnamefont {X.~H.}\ \bibnamefont {Zhou}}, \bibinfo {author}
  {\bibfnamefont {Z.~Y.}\ \bibnamefont {Sun}}, \bibinfo {author} {\bibfnamefont
  {I.}~\bibnamefont {Kuti}}, \bibinfo {author} {\bibfnamefont {J.}~\bibnamefont
  {Tim\'ar}}, \bibinfo {author} {\bibfnamefont {A.}~\bibnamefont {Tucholski}},
  \bibinfo {author} {\bibfnamefont {J.}~\bibnamefont {Srebrny}},\ and\ \bibinfo
  {author} {\bibfnamefont {C.}~\bibnamefont {Andreoiu}},\ }\href
  {https://doi.org/10.1103/PhysRevC.105.034302} {\bibfield  {journal} {\bibinfo
   {journal} {Phys. Rev. C}\ }\textbf {\bibinfo {volume} {105}},\ \bibinfo
  {pages} {034302} (\bibinfo {year} {2022}{\natexlab{b}})}\BibitemShut
  {NoStop}%
\bibitem [{\citenamefont {Wang}\ and\ \citenamefont
  {Chen}(2024)}]{WANG2024859}%
  \BibitemOpen
  \bibfield  {author} {\bibinfo {author} {\bibfnamefont {Y.~M.}\ \bibnamefont
  {Wang}}\ and\ \bibinfo {author} {\bibfnamefont {Q.~B.}\ \bibnamefont
  {Chen}},\ }\href
  {https://doi.org/https://doi.org/10.1016/j.physletb.2024.139131} {\bibfield
  {journal} {\bibinfo  {journal} {Phys. Lett. B}\ }\textbf {\bibinfo {volume}
  {859}},\ \bibinfo {pages} {139131} (\bibinfo {year} {2024})}\BibitemShut
  {NoStop}%
\bibitem [{\citenamefont {Jehangir}\ \emph
  {et~al.}(2021{\natexlab{a}})\citenamefont {Jehangir}, \citenamefont {Bhat},
  \citenamefont {Sheikh}, \citenamefont {Frauendorf}, \citenamefont {Li},
  \citenamefont {Palit},\ and\ \citenamefont {Rather}}]{SJ21}%
  \BibitemOpen
  \bibfield  {author} {\bibinfo {author} {\bibfnamefont {S.}~\bibnamefont
  {Jehangir}}, \bibinfo {author} {\bibfnamefont {G.~H.}\ \bibnamefont {Bhat}},
  \bibinfo {author} {\bibfnamefont {J.~A.}\ \bibnamefont {Sheikh}}, \bibinfo
  {author} {\bibfnamefont {S.}~\bibnamefont {Frauendorf}}, \bibinfo {author}
  {\bibfnamefont {W.}~\bibnamefont {Li}}, \bibinfo {author} {\bibfnamefont
  {R.}~\bibnamefont {Palit}},\ and\ \bibinfo {author} {\bibfnamefont
  {N.}~\bibnamefont {Rather}},\ }\href
  {https://doi.org/10.1140/epja/s10050-021-00620-7} {\bibfield  {journal}
  {\bibinfo  {journal} {Eur. Phys. J. A}\ }\textbf {\bibinfo {volume} {57}},\
  \bibinfo {pages} {308} (\bibinfo {year} {2021}{\natexlab{a}})}\BibitemShut
  {NoStop}%
\bibitem [{\citenamefont {Rouoof}\ \emph {et~al.}(2024)\citenamefont {Rouoof},
  \citenamefont {Nazir}, \citenamefont {Jehangir}, \citenamefont {Bhat},
  \citenamefont {Sheikh}, \citenamefont {Rather},\ and\ \citenamefont
  {Frauendorf}}]{SP24}%
  \BibitemOpen
  \bibfield  {author} {\bibinfo {author} {\bibfnamefont {S.~P.}\ \bibnamefont
  {Rouoof}}, \bibinfo {author} {\bibfnamefont {N.}~\bibnamefont {Nazir}},
  \bibinfo {author} {\bibfnamefont {S.}~\bibnamefont {Jehangir}}, \bibinfo
  {author} {\bibfnamefont {G.~H.}\ \bibnamefont {Bhat}}, \bibinfo {author}
  {\bibfnamefont {J.~A.}\ \bibnamefont {Sheikh}}, \bibinfo {author}
  {\bibfnamefont {N.}~\bibnamefont {Rather}},\ and\ \bibinfo {author}
  {\bibfnamefont {S.}~\bibnamefont {Frauendorf}},\ }\href
  {https://doi.org/10.1140/epja/s10050-024-01257-y} {\bibfield  {journal}
  {\bibinfo  {journal} {Eur. Phys. J. A}\ }\textbf {\bibinfo {volume} {60}},\
  \bibinfo {pages} {40} (\bibinfo {year} {2024})}\BibitemShut {NoStop}%
\bibitem [{\citenamefont {Zamfir}\ and\ \citenamefont
  {Casten}(1991)}]{ZAMFIR1991265}%
  \BibitemOpen
  \bibfield  {author} {\bibinfo {author} {\bibfnamefont {N.~V.}\ \bibnamefont
  {Zamfir}}\ and\ \bibinfo {author} {\bibfnamefont {R.~F.}\ \bibnamefont
  {Casten}},\ }\href
  {https://doi.org/https://doi.org/10.1016/0370-2693(91)91610-8} {\bibfield
  {journal} {\bibinfo  {journal} {Phys. Lett. B}\ }\textbf {\bibinfo {volume}
  {260}},\ \bibinfo {pages} {265} (\bibinfo {year} {1991})}\BibitemShut
  {NoStop}%
\bibitem [{\citenamefont {McCutchan}\ \emph {et~al.}(2007)\citenamefont
  {McCutchan}, \citenamefont {Bonatsos}, \citenamefont {Zamfir},\ and\
  \citenamefont {Casten}}]{McCutchan2007}%
  \BibitemOpen
  \bibfield  {author} {\bibinfo {author} {\bibfnamefont {E.~A.}\ \bibnamefont
  {McCutchan}}, \bibinfo {author} {\bibfnamefont {D.}~\bibnamefont {Bonatsos}},
  \bibinfo {author} {\bibfnamefont {N.~V.}\ \bibnamefont {Zamfir}},\ and\
  \bibinfo {author} {\bibfnamefont {R.~F.}\ \bibnamefont {Casten}},\ }\href
  {https://doi.org/10.1103/PhysRevC.76.024306} {\bibfield  {journal} {\bibinfo
  {journal} {Phys. Rev. C}\ }\textbf {\bibinfo {volume} {76}},\ \bibinfo
  {pages} {024306} (\bibinfo {year} {2007})}\BibitemShut {NoStop}%
\bibitem [{\citenamefont {Davydov}(1970)}]{ASD1970}%
  \BibitemOpen
  \bibfield  {author} {\bibinfo {author} {\bibfnamefont {A.~S.}\ \bibnamefont
  {Davydov}},\ }in\ \href@noop {} {\emph {\bibinfo {booktitle}
  {Proc.Intern.Conf. on Nuclear Structure (Kingston, Canada), eds. D. A.
  Bromley and E. W. Vogt (University of Toronto Press, Toronto, 1960) p.
  801}}}\ (\bibinfo {year} {1970})\BibitemShut {NoStop}%
\bibitem [{\citenamefont {Baktash}\ \emph {et~al.}(1978)\citenamefont
  {Baktash}, \citenamefont {Saladin}, \citenamefont {O'Brien},\ and\
  \citenamefont {Alessi}}]{Baktash1978}%
  \BibitemOpen
  \bibfield  {author} {\bibinfo {author} {\bibfnamefont {C.}~\bibnamefont
  {Baktash}}, \bibinfo {author} {\bibfnamefont {J.~X.}\ \bibnamefont
  {Saladin}}, \bibinfo {author} {\bibfnamefont {J.~J.}\ \bibnamefont
  {O'Brien}},\ and\ \bibinfo {author} {\bibfnamefont {J.~G.}\ \bibnamefont
  {Alessi}},\ }\href {https://doi.org/10.1103/PhysRevC.18.131} {\bibfield
  {journal} {\bibinfo  {journal} {Phys. Rev. C}\ }\textbf {\bibinfo {volume}
  {18}},\ \bibinfo {pages} {131} (\bibinfo {year} {1978})}\BibitemShut
  {NoStop}%
\bibitem [{\citenamefont {Baktash}\ \emph {et~al.}(1980)\citenamefont
  {Baktash}, \citenamefont {Saladin}, \citenamefont {O'Brien},\ and\
  \citenamefont {Alessi}}]{Baktash1980}%
  \BibitemOpen
  \bibfield  {author} {\bibinfo {author} {\bibfnamefont {C.}~\bibnamefont
  {Baktash}}, \bibinfo {author} {\bibfnamefont {J.~X.}\ \bibnamefont
  {Saladin}}, \bibinfo {author} {\bibfnamefont {J.~J.}\ \bibnamefont
  {O'Brien}},\ and\ \bibinfo {author} {\bibfnamefont {J.~G.}\ \bibnamefont
  {Alessi}},\ }\href {https://doi.org/10.1103/PhysRevC.22.2383} {\bibfield
  {journal} {\bibinfo  {journal} {Phys. Rev. C}\ }\textbf {\bibinfo {volume}
  {22}},\ \bibinfo {pages} {2383} (\bibinfo {year} {1980})}\BibitemShut
  {NoStop}%
\bibitem [{\citenamefont {Sheikh}\ and\ \citenamefont {Hara}(1999)}]{TPSM1999}%
  \BibitemOpen
  \bibfield  {author} {\bibinfo {author} {\bibfnamefont {J.~A.}\ \bibnamefont
  {Sheikh}}\ and\ \bibinfo {author} {\bibfnamefont {K.}~\bibnamefont {Hara}},\
  }\href {https://doi.org/10.1103/PhysRevLett.82.3968} {\bibfield  {journal}
  {\bibinfo  {journal} {Phys. Rev. Lett.}\ }\textbf {\bibinfo {volume} {82}},\
  \bibinfo {pages} {3968} (\bibinfo {year} {1999})}\BibitemShut {NoStop}%
\bibitem [{\citenamefont {Bhat}\ \emph {et~al.}(2012)\citenamefont {Bhat},
  \citenamefont {Sheikh},\ and\ \citenamefont {Palit}}]{TPSM8}%
  \BibitemOpen
  \bibfield  {author} {\bibinfo {author} {\bibfnamefont {G.~H.}\ \bibnamefont
  {Bhat}}, \bibinfo {author} {\bibfnamefont {J.~A.}\ \bibnamefont {Sheikh}},\
  and\ \bibinfo {author} {\bibfnamefont {R.}~\bibnamefont {Palit}},\ }\href
  {https://doi.org/https://doi.org/10.1016/j.physletb.2011.12.035} {\bibfield
  {journal} {\bibinfo  {journal} {Phys. Lett. B}\ }\textbf {\bibinfo {volume}
  {707}},\ \bibinfo {pages} {250} (\bibinfo {year} {2012})}\BibitemShut
  {NoStop}%
\bibitem [{\citenamefont {Bhat}\ \emph {et~al.}(2014)\citenamefont {Bhat},
  \citenamefont {Dar}, \citenamefont {Sheikh},\ and\ \citenamefont
  {Sun}}]{TPSM2014}%
  \BibitemOpen
  \bibfield  {author} {\bibinfo {author} {\bibfnamefont {G.~H.}\ \bibnamefont
  {Bhat}}, \bibinfo {author} {\bibfnamefont {W.~A.}\ \bibnamefont {Dar}},
  \bibinfo {author} {\bibfnamefont {J.~A.}\ \bibnamefont {Sheikh}},\ and\
  \bibinfo {author} {\bibfnamefont {Y.}~\bibnamefont {Sun}},\ }\href
  {https://doi.org/10.1103/PhysRevC.89.014328} {\bibfield  {journal} {\bibinfo
  {journal} {Phys. Rev. C}\ }\textbf {\bibinfo {volume} {89}},\ \bibinfo
  {pages} {014328} (\bibinfo {year} {2014})}\BibitemShut {NoStop}%
\bibitem [{\citenamefont {Jehangir}\ \emph {et~al.}(2018)\citenamefont
  {Jehangir}, \citenamefont {Bhat}, \citenamefont {Sheikh}, \citenamefont
  {Frauendorf}, \citenamefont {Majola}, \citenamefont {Ganai},\ and\
  \citenamefont {Sharpey-Schafer}}]{TPSM5}%
  \BibitemOpen
  \bibfield  {author} {\bibinfo {author} {\bibfnamefont {S.}~\bibnamefont
  {Jehangir}}, \bibinfo {author} {\bibfnamefont {G.~H.}\ \bibnamefont {Bhat}},
  \bibinfo {author} {\bibfnamefont {J.~A.}\ \bibnamefont {Sheikh}}, \bibinfo
  {author} {\bibfnamefont {S.}~\bibnamefont {Frauendorf}}, \bibinfo {author}
  {\bibfnamefont {S.~N.~T.}\ \bibnamefont {Majola}}, \bibinfo {author}
  {\bibfnamefont {P.~A.}\ \bibnamefont {Ganai}},\ and\ \bibinfo {author}
  {\bibfnamefont {J.~F.}\ \bibnamefont {Sharpey-Schafer}},\ }\href
  {https://doi.org/10.1103/PhysRevC.97.014310} {\bibfield  {journal} {\bibinfo
  {journal} {Phys. Rev. C}\ }\textbf {\bibinfo {volume} {97}},\ \bibinfo
  {pages} {014310} (\bibinfo {year} {2018})}\BibitemShut {NoStop}%
\bibitem [{\citenamefont {Jehangir}\ \emph {et~al.}(2020)\citenamefont
  {Jehangir}, \citenamefont {Maqbool}, \citenamefont {Bhat}, \citenamefont
  {Sheikh}, \citenamefont {Palit},\ and\ \citenamefont {Rather}}]{TPSM4}%
  \BibitemOpen
  \bibfield  {author} {\bibinfo {author} {\bibfnamefont {S.}~\bibnamefont
  {Jehangir}}, \bibinfo {author} {\bibfnamefont {I.}~\bibnamefont {Maqbool}},
  \bibinfo {author} {\bibfnamefont {G.~H.}\ \bibnamefont {Bhat}}, \bibinfo
  {author} {\bibfnamefont {J.~A.}\ \bibnamefont {Sheikh}}, \bibinfo {author}
  {\bibfnamefont {R.}~\bibnamefont {Palit}},\ and\ \bibinfo {author}
  {\bibfnamefont {N.}~\bibnamefont {Rather}},\ }\href
  {https://doi.org/10.1140/epja/s10050-020-00206-9} {\bibfield  {journal}
  {\bibinfo  {journal} {Eur. Phys. J. A}\ }\textbf {\bibinfo {volume} {56}},\
  \bibinfo {pages} {197} (\bibinfo {year} {2020})}\BibitemShut {NoStop}%
\bibitem [{\citenamefont {Jehangir}\ \emph
  {et~al.}(2021{\natexlab{b}})\citenamefont {Jehangir}, \citenamefont {Bhat},
  \citenamefont {Rather}, \citenamefont {Sheikh},\ and\ \citenamefont
  {Palit}}]{TPSM3}%
  \BibitemOpen
  \bibfield  {author} {\bibinfo {author} {\bibfnamefont {S.}~\bibnamefont
  {Jehangir}}, \bibinfo {author} {\bibfnamefont {G.~H.}\ \bibnamefont {Bhat}},
  \bibinfo {author} {\bibfnamefont {N.}~\bibnamefont {Rather}}, \bibinfo
  {author} {\bibfnamefont {J.~A.}\ \bibnamefont {Sheikh}},\ and\ \bibinfo
  {author} {\bibfnamefont {R.}~\bibnamefont {Palit}},\ }\href
  {https://doi.org/10.1103/PhysRevC.104.044322} {\bibfield  {journal} {\bibinfo
   {journal} {Phys. Rev. C}\ }\textbf {\bibinfo {volume} {104}},\ \bibinfo
  {pages} {044322} (\bibinfo {year} {2021}{\natexlab{b}})}\BibitemShut
  {NoStop}%
\bibitem [{\citenamefont {Jehangir}\ \emph {et~al.}(2022)\citenamefont
  {Jehangir}, \citenamefont {Nazir}, \citenamefont {Bhat}, \citenamefont
  {Sheikh}, \citenamefont {Rather}, \citenamefont {Chakraborty},\ and\
  \citenamefont {Palit}}]{TPSM2}%
  \BibitemOpen
  \bibfield  {author} {\bibinfo {author} {\bibfnamefont {S.}~\bibnamefont
  {Jehangir}}, \bibinfo {author} {\bibfnamefont {N.}~\bibnamefont {Nazir}},
  \bibinfo {author} {\bibfnamefont {G.~H.}\ \bibnamefont {Bhat}}, \bibinfo
  {author} {\bibfnamefont {J.~A.}\ \bibnamefont {Sheikh}}, \bibinfo {author}
  {\bibfnamefont {N.}~\bibnamefont {Rather}}, \bibinfo {author} {\bibfnamefont
  {S.}~\bibnamefont {Chakraborty}},\ and\ \bibinfo {author} {\bibfnamefont
  {R.}~\bibnamefont {Palit}},\ }\href
  {https://doi.org/10.1103/PhysRevC.105.054310} {\bibfield  {journal} {\bibinfo
   {journal} {Phys. Rev. C}\ }\textbf {\bibinfo {volume} {105}},\ \bibinfo
  {pages} {054310} (\bibinfo {year} {2022})}\BibitemShut {NoStop}%
\bibitem [{\citenamefont {Nazir}\ \emph
  {et~al.}(2023{\natexlab{a}})\citenamefont {Nazir}, \citenamefont {Jehangir},
  \citenamefont {Rouoof}, \citenamefont {Bhat}, \citenamefont {Sheikh},
  \citenamefont {Rather},\ and\ \citenamefont {Frauendorf}}]{TPSM6}%
  \BibitemOpen
  \bibfield  {author} {\bibinfo {author} {\bibfnamefont {N.}~\bibnamefont
  {Nazir}}, \bibinfo {author} {\bibfnamefont {S.}~\bibnamefont {Jehangir}},
  \bibinfo {author} {\bibfnamefont {S.~P.}\ \bibnamefont {Rouoof}}, \bibinfo
  {author} {\bibfnamefont {G.~H.}\ \bibnamefont {Bhat}}, \bibinfo {author}
  {\bibfnamefont {J.~A.}\ \bibnamefont {Sheikh}}, \bibinfo {author}
  {\bibfnamefont {N.}~\bibnamefont {Rather}},\ and\ \bibinfo {author}
  {\bibfnamefont {S.}~\bibnamefont {Frauendorf}},\ }\href
  {https://doi.org/10.1103/PhysRevC.107.L021303} {\bibfield  {journal}
  {\bibinfo  {journal} {Phys. Rev. C}\ }\textbf {\bibinfo {volume} {107}},\
  \bibinfo {pages} {L021303} (\bibinfo {year}
  {2023}{\natexlab{a}})}\BibitemShut {NoStop}%
\bibitem [{\citenamefont {Nazir}\ \emph
  {et~al.}(2023{\natexlab{b}})\citenamefont {Nazir}, \citenamefont {Jehangir},
  \citenamefont {Rouoof}, \citenamefont {Bhat}, \citenamefont {Sheikh},
  \citenamefont {Rather},\ and\ \citenamefont {Malik}}]{TPSM10}%
  \BibitemOpen
  \bibfield  {author} {\bibinfo {author} {\bibfnamefont {N.}~\bibnamefont
  {Nazir}}, \bibinfo {author} {\bibfnamefont {S.}~\bibnamefont {Jehangir}},
  \bibinfo {author} {\bibfnamefont {S.~P.}\ \bibnamefont {Rouoof}}, \bibinfo
  {author} {\bibfnamefont {G.~H.}\ \bibnamefont {Bhat}}, \bibinfo {author}
  {\bibfnamefont {J.~A.}\ \bibnamefont {Sheikh}}, \bibinfo {author}
  {\bibfnamefont {N.}~\bibnamefont {Rather}},\ and\ \bibinfo {author}
  {\bibfnamefont {M.~A.}\ \bibnamefont {Malik}},\ }\href
  {https://doi.org/10.1103/PhysRevC.108.044308} {\bibfield  {journal} {\bibinfo
   {journal} {Phys. Rev. C}\ }\textbf {\bibinfo {volume} {108}},\ \bibinfo
  {pages} {044308} (\bibinfo {year} {2023}{\natexlab{b}})}\BibitemShut
  {NoStop}%
\bibitem [{\citenamefont {Anguiano}\ \emph {et~al.}(2001)\citenamefont
  {Anguiano}, \citenamefont {Egido},\ and\ \citenamefont
  {Robledo}}]{ANGUIANO2001467}%
  \BibitemOpen
  \bibfield  {author} {\bibinfo {author} {\bibfnamefont {M.}~\bibnamefont
  {Anguiano}}, \bibinfo {author} {\bibfnamefont {J.}~\bibnamefont {Egido}},\
  and\ \bibinfo {author} {\bibfnamefont {L.}~\bibnamefont {Robledo}},\ }\href
  {https://doi.org/https://doi.org/10.1016/S0375-9474(01)01219-2} {\bibfield
  {journal} {\bibinfo  {journal} {Nuclear Physics A}\ }\textbf {\bibinfo
  {volume} {696}},\ \bibinfo {pages} {467} (\bibinfo {year}
  {2001})}\BibitemShut {NoStop}%
\bibitem [{\citenamefont {Bender}\ and\ \citenamefont
  {Heenen}(2008)}]{Bender2008}%
  \BibitemOpen
  \bibfield  {author} {\bibinfo {author} {\bibfnamefont {M.}~\bibnamefont
  {Bender}}\ and\ \bibinfo {author} {\bibfnamefont {P.-H.}\ \bibnamefont
  {Heenen}},\ }\href {https://doi.org/10.1103/PhysRevC.78.024309} {\bibfield
  {journal} {\bibinfo  {journal} {Phys. Rev. C}\ }\textbf {\bibinfo {volume}
  {78}},\ \bibinfo {pages} {024309} (\bibinfo {year} {2008})}\BibitemShut
  {NoStop}%
\bibitem [{\citenamefont {Lacroix}\ \emph {et~al.}(2009)\citenamefont
  {Lacroix}, \citenamefont {Duguet},\ and\ \citenamefont {Bender}}]{dl09}%
  \BibitemOpen
  \bibfield  {author} {\bibinfo {author} {\bibfnamefont {D.}~\bibnamefont
  {Lacroix}}, \bibinfo {author} {\bibfnamefont {T.}~\bibnamefont {Duguet}},\
  and\ \bibinfo {author} {\bibfnamefont {M.}~\bibnamefont {Bender}},\ }\href
  {https://doi.org/10.1103/PhysRevC.79.044318} {\bibfield  {journal} {\bibinfo
  {journal} {Phys. Rev. C}\ }\textbf {\bibinfo {volume} {79}},\ \bibinfo
  {pages} {044318} (\bibinfo {year} {2009})}\BibitemShut {NoStop}%
\bibitem [{\citenamefont {Bender}\ \emph {et~al.}(2009)\citenamefont {Bender},
  \citenamefont {Duguet},\ and\ \citenamefont {Lacroix}}]{mb09}%
  \BibitemOpen
  \bibfield  {author} {\bibinfo {author} {\bibfnamefont {M.}~\bibnamefont
  {Bender}}, \bibinfo {author} {\bibfnamefont {T.}~\bibnamefont {Duguet}},\
  and\ \bibinfo {author} {\bibfnamefont {D.}~\bibnamefont {Lacroix}},\ }\href
  {https://doi.org/10.1103/PhysRevC.79.044319} {\bibfield  {journal} {\bibinfo
  {journal} {Phys. Rev. C}\ }\textbf {\bibinfo {volume} {79}},\ \bibinfo
  {pages} {044319} (\bibinfo {year} {2009})}\BibitemShut {NoStop}%
\bibitem [{\citenamefont {Yao}\ \emph {et~al.}(2009)\citenamefont {Yao},
  \citenamefont {Meng}, \citenamefont {Ring},\ and\ \citenamefont
  {Arteaga}}]{Yao2009}%
  \BibitemOpen
  \bibfield  {author} {\bibinfo {author} {\bibfnamefont {J.~M.}\ \bibnamefont
  {Yao}}, \bibinfo {author} {\bibfnamefont {J.}~\bibnamefont {Meng}}, \bibinfo
  {author} {\bibfnamefont {P.}~\bibnamefont {Ring}},\ and\ \bibinfo {author}
  {\bibfnamefont {D.~P.}\ \bibnamefont {Arteaga}},\ }\href
  {https://doi.org/10.1103/PhysRevC.79.044312} {\bibfield  {journal} {\bibinfo
  {journal} {Phys. Rev. C}\ }\textbf {\bibinfo {volume} {79}},\ \bibinfo
  {pages} {044312} (\bibinfo {year} {2009})}\BibitemShut {NoStop}%
\bibitem [{\citenamefont {Meng}\ and\ \citenamefont {Zhang}(2010)}]{jm10}%
  \BibitemOpen
  \bibfield  {author} {\bibinfo {author} {\bibfnamefont {J.}~\bibnamefont
  {Meng}}\ and\ \bibinfo {author} {\bibfnamefont {S.~Q.}\ \bibnamefont
  {Zhang}},\ }\href {https://doi.org/10.1088/0954-3899/37/6/064025} {\bibfield
  {journal} {\bibinfo  {journal} {Journal of Physics G: Nuclear and Particle
  Physics}\ }\textbf {\bibinfo {volume} {37}},\ \bibinfo {pages} {064025}
  (\bibinfo {year} {2010})}\BibitemShut {NoStop}%
\bibitem [{\citenamefont {Yao}\ \emph {et~al.}(2010)\citenamefont {Yao},
  \citenamefont {Meng}, \citenamefont {Ring},\ and\ \citenamefont
  {Vretenar}}]{Yao2010}%
  \BibitemOpen
  \bibfield  {author} {\bibinfo {author} {\bibfnamefont {J.~M.}\ \bibnamefont
  {Yao}}, \bibinfo {author} {\bibfnamefont {J.}~\bibnamefont {Meng}}, \bibinfo
  {author} {\bibfnamefont {P.}~\bibnamefont {Ring}},\ and\ \bibinfo {author}
  {\bibfnamefont {D.}~\bibnamefont {Vretenar}},\ }\href
  {https://doi.org/10.1103/PhysRevC.81.044311} {\bibfield  {journal} {\bibinfo
  {journal} {Phys. Rev. C}\ }\textbf {\bibinfo {volume} {81}},\ \bibinfo
  {pages} {044311} (\bibinfo {year} {2010})}\BibitemShut {NoStop}%
\bibitem [{\citenamefont {Sun}(2016)}]{Sun2016}%
  \BibitemOpen
  \bibfield  {author} {\bibinfo {author} {\bibfnamefont {Y.}~\bibnamefont
  {Sun}},\ }\href {https://doi.org/10.1088/0031-8949/91/4/043005} {\bibfield
  {journal} {\bibinfo  {journal} {Physica Scripta}\ }\textbf {\bibinfo {volume}
  {91}},\ \bibinfo {pages} {043005} (\bibinfo {year} {2016})}\BibitemShut
  {NoStop}%
\bibitem [{\citenamefont {Egido}(2016)}]{Egido2016}%
  \BibitemOpen
  \bibfield  {author} {\bibinfo {author} {\bibfnamefont {J.~L.}\ \bibnamefont
  {Egido}},\ }\href {https://doi.org/10.1088/0031-8949/91/7/073003} {\bibfield
  {journal} {\bibinfo  {journal} {Physica Scripta}\ }\textbf {\bibinfo {volume}
  {91}},\ \bibinfo {pages} {073003} (\bibinfo {year} {2016})}\BibitemShut
  {NoStop}%
\bibitem [{\citenamefont {Robledo}\ \emph {et~al.}(2018)\citenamefont
  {Robledo}, \citenamefont {Rodríguez},\ and\ \citenamefont
  {Rodríguez-Guzmán}}]{Robledo2019}%
  \BibitemOpen
  \bibfield  {author} {\bibinfo {author} {\bibfnamefont {L.~M.}\ \bibnamefont
  {Robledo}}, \bibinfo {author} {\bibfnamefont {T.~R.}\ \bibnamefont
  {Rodríguez}},\ and\ \bibinfo {author} {\bibfnamefont {R.~R.}\ \bibnamefont
  {Rodríguez-Guzmán}},\ }\href {https://doi.org/10.1088/1361-6471/aadebd}
  {\bibfield  {journal} {\bibinfo  {journal} {Journal of Physics G: Nuclear and
  Particle Physics}\ }\textbf {\bibinfo {volume} {46}},\ \bibinfo {pages}
  {013001} (\bibinfo {year} {2018})}\BibitemShut {NoStop}%
\bibitem [{\citenamefont {Otsuka}\ \emph {et~al.}(2020)\citenamefont {Otsuka},
  \citenamefont {Gade}, \citenamefont {Sorlin}, \citenamefont {Suzuki},\ and\
  \citenamefont {Utsuno}}]{TO20}%
  \BibitemOpen
  \bibfield  {author} {\bibinfo {author} {\bibfnamefont {T.}~\bibnamefont
  {Otsuka}}, \bibinfo {author} {\bibfnamefont {A.}~\bibnamefont {Gade}},
  \bibinfo {author} {\bibfnamefont {O.}~\bibnamefont {Sorlin}}, \bibinfo
  {author} {\bibfnamefont {T.}~\bibnamefont {Suzuki}},\ and\ \bibinfo {author}
  {\bibfnamefont {Y.}~\bibnamefont {Utsuno}},\ }\href
  {https://doi.org/10.1103/RevModPhys.92.015002} {\bibfield  {journal}
  {\bibinfo  {journal} {Rev. Mod. Phys.}\ }\textbf {\bibinfo {volume} {92}},\
  \bibinfo {pages} {015002} (\bibinfo {year} {2020})}\BibitemShut {NoStop}%
\bibitem [{\citenamefont {Bally}\ and\ \citenamefont {Bender}(2021)}]{bb21}%
  \BibitemOpen
  \bibfield  {author} {\bibinfo {author} {\bibfnamefont {B.}~\bibnamefont
  {Bally}}\ and\ \bibinfo {author} {\bibfnamefont {M.}~\bibnamefont {Bender}},\
  }\href {https://doi.org/10.1103/PhysRevC.103.024315} {\bibfield  {journal}
  {\bibinfo  {journal} {Phys. Rev. C}\ }\textbf {\bibinfo {volume} {103}},\
  \bibinfo {pages} {024315} (\bibinfo {year} {2021})}\BibitemShut {NoStop}%
\bibitem [{\citenamefont {Frycz}\ \emph {et~al.}(2024)\citenamefont {Frycz},
  \citenamefont {Men\'endez}, \citenamefont {Rios}, \citenamefont {Bally},
  \citenamefont {Rodr\'{\i}guez},\ and\ \citenamefont {Romero}}]{Dorian24}%
  \BibitemOpen
  \bibfield  {author} {\bibinfo {author} {\bibfnamefont {D.}~\bibnamefont
  {Frycz}}, \bibinfo {author} {\bibfnamefont {J.}~\bibnamefont {Men\'endez}},
  \bibinfo {author} {\bibfnamefont {A.}~\bibnamefont {Rios}}, \bibinfo {author}
  {\bibfnamefont {B.}~\bibnamefont {Bally}}, \bibinfo {author} {\bibfnamefont
  {T.~R.}\ \bibnamefont {Rodr\'{\i}guez}},\ and\ \bibinfo {author}
  {\bibfnamefont {A.~M.}\ \bibnamefont {Romero}},\ }\href
  {https://doi.org/10.1103/PhysRevC.110.054326} {\bibfield  {journal} {\bibinfo
   {journal} {Phys. Rev. C}\ }\textbf {\bibinfo {volume} {110}},\ \bibinfo
  {pages} {054326} (\bibinfo {year} {2024})}\BibitemShut {NoStop}%
\bibitem [{\citenamefont {Vretenar}\ \emph {et~al.}(2005)\citenamefont
  {Vretenar}, \citenamefont {Afanasjev}, \citenamefont {Lalazissis},\ and\
  \citenamefont {Ring}}]{DV05}%
  \BibitemOpen
  \bibfield  {author} {\bibinfo {author} {\bibfnamefont {D.}~\bibnamefont
  {Vretenar}}, \bibinfo {author} {\bibfnamefont {A.}~\bibnamefont {Afanasjev}},
  \bibinfo {author} {\bibfnamefont {G.}~\bibnamefont {Lalazissis}},\ and\
  \bibinfo {author} {\bibfnamefont {P.}~\bibnamefont {Ring}},\ }\href
  {https://doi.org/https://doi.org/10.1016/j.physrep.2004.10.001} {\bibfield
  {journal} {\bibinfo  {journal} {Physics Reports}\ }\textbf {\bibinfo {volume}
  {409}},\ \bibinfo {pages} {101} (\bibinfo {year} {2005})}\BibitemShut
  {NoStop}%
\bibitem [{\citenamefont {Otsuka}\ \emph {et~al.}(2019)\citenamefont {Otsuka},
  \citenamefont {Tsunoda}, \citenamefont {Abe}, \citenamefont {Shimizu},\ and\
  \citenamefont {Van~Duppen}}]{TO19}%
  \BibitemOpen
  \bibfield  {author} {\bibinfo {author} {\bibfnamefont {T.}~\bibnamefont
  {Otsuka}}, \bibinfo {author} {\bibfnamefont {Y.}~\bibnamefont {Tsunoda}},
  \bibinfo {author} {\bibfnamefont {T.}~\bibnamefont {Abe}}, \bibinfo {author}
  {\bibfnamefont {N.}~\bibnamefont {Shimizu}},\ and\ \bibinfo {author}
  {\bibfnamefont {P.}~\bibnamefont {Van~Duppen}},\ }\href
  {https://doi.org/10.1103/PhysRevLett.123.222502} {\bibfield  {journal}
  {\bibinfo  {journal} {Phys. Rev. Lett.}\ }\textbf {\bibinfo {volume} {123}},\
  \bibinfo {pages} {222502} (\bibinfo {year} {2019})}\BibitemShut {NoStop}%
\bibitem [{\citenamefont {Tsunoda}\ and\ \citenamefont {Otsuka}(2021)}]{yt21}%
  \BibitemOpen
  \bibfield  {author} {\bibinfo {author} {\bibfnamefont {Y.}~\bibnamefont
  {Tsunoda}}\ and\ \bibinfo {author} {\bibfnamefont {T.}~\bibnamefont
  {Otsuka}},\ }\href {https://doi.org/10.1103/PhysRevC.103.L021303} {\bibfield
  {journal} {\bibinfo  {journal} {Phys. Rev. C}\ }\textbf {\bibinfo {volume}
  {103}},\ \bibinfo {pages} {L021303} (\bibinfo {year} {2021})}\BibitemShut
  {NoStop}%
\bibitem [{\citenamefont {Wang}\ \emph {et~al.}(2022)\citenamefont {Wang},
  \citenamefont {Zhao},\ and\ \citenamefont {Meng}}]{Wang2022}%
  \BibitemOpen
  \bibfield  {author} {\bibinfo {author} {\bibfnamefont {Y.~K.}\ \bibnamefont
  {Wang}}, \bibinfo {author} {\bibfnamefont {P.~W.}\ \bibnamefont {Zhao}},\
  and\ \bibinfo {author} {\bibfnamefont {J.}~\bibnamefont {Meng}},\ }\href
  {https://doi.org/10.1103/PhysRevC.105.054311} {\bibfield  {journal} {\bibinfo
   {journal} {Phys. Rev. C}\ }\textbf {\bibinfo {volume} {105}},\ \bibinfo
  {pages} {054311} (\bibinfo {year} {2022})}\BibitemShut {NoStop}%
\bibitem [{\citenamefont {Sheikh}\ \emph {et~al.}(2021)\citenamefont {Sheikh},
  \citenamefont {Dobaczewski}, \citenamefont {Ring}, \citenamefont {Robledo},\
  and\ \citenamefont {Yannouleas}}]{JD21}%
  \BibitemOpen
  \bibfield  {author} {\bibinfo {author} {\bibfnamefont {J.~A.}\ \bibnamefont
  {Sheikh}}, \bibinfo {author} {\bibfnamefont {J.}~\bibnamefont {Dobaczewski}},
  \bibinfo {author} {\bibfnamefont {P.}~\bibnamefont {Ring}}, \bibinfo {author}
  {\bibfnamefont {L.~M.}\ \bibnamefont {Robledo}},\ and\ \bibinfo {author}
  {\bibfnamefont {C.}~\bibnamefont {Yannouleas}},\ }\href
  {https://doi.org/10.1088/1361-6471/ac288a} {\bibfield  {journal} {\bibinfo
  {journal} {Journal of Physics G: Nuclear and Particle Physics}\ }\textbf
  {\bibinfo {volume} {48}},\ \bibinfo {pages} {123001} (\bibinfo {year}
  {2021})}\BibitemShut {NoStop}%
\bibitem [{\citenamefont {Sheikh}\ \emph {et~al.}(2016)\citenamefont {Sheikh},
  \citenamefont {Bhat}, \citenamefont {Dar}, \citenamefont {Jehangir},\ and\
  \citenamefont {Ganai}}]{WM39}%
  \BibitemOpen
  \bibfield  {author} {\bibinfo {author} {\bibfnamefont {J.~A.}\ \bibnamefont
  {Sheikh}}, \bibinfo {author} {\bibfnamefont {G.~H.}\ \bibnamefont {Bhat}},
  \bibinfo {author} {\bibfnamefont {W.~A.}\ \bibnamefont {Dar}}, \bibinfo
  {author} {\bibfnamefont {S.}~\bibnamefont {Jehangir}},\ and\ \bibinfo
  {author} {\bibfnamefont {P.~A.}\ \bibnamefont {Ganai}},\ }\href
  {https://doi.org/10.1088/0031-8949/91/6/063015} {\bibfield  {journal}
  {\bibinfo  {journal} {Physica Scripta}\ }\textbf {\bibinfo {volume} {91}},\
  \bibinfo {pages} {063015} (\bibinfo {year} {2016})}\BibitemShut {NoStop}%
\bibitem [{\citenamefont {Sheikh}\ \emph {et~al.}(2024)\citenamefont {Sheikh},
  \citenamefont {Jehangir},\ and\ \citenamefont {Bhat}}]{SJ24ch}%
  \BibitemOpen
  \bibfield  {author} {\bibinfo {author} {\bibfnamefont {J.~A.}\ \bibnamefont
  {Sheikh}}, \bibinfo {author} {\bibfnamefont {S.}~\bibnamefont {Jehangir}},\
  and\ \bibinfo {author} {\bibfnamefont {G.~H.}\ \bibnamefont {Bhat}},\ }in\
  \href {https://doi.org/10.1201/9781032691633} {\emph {\bibinfo {booktitle}
  {Chirality and Wobbling in Atomic Nuclei}}}\ (\bibinfo  {publisher} {Taylor
  and Francis},\ \bibinfo {address} {London, England},\ \bibinfo {year}
  {2024})\ pp.\ \bibinfo {pages} {237--254}\BibitemShut {NoStop}%
\bibitem [{\citenamefont {Nilsson}\ \emph {et~al.}(1969)\citenamefont
  {Nilsson}, \citenamefont {Tsang}, \citenamefont {Sobiczewski}, \citenamefont
  {Szymański}, \citenamefont {Wycech}, \citenamefont {Gustafson},
  \citenamefont {Lamm}, \citenamefont {Möller},\ and\ \citenamefont
  {Nilsson}}]{Ni69}%
  \BibitemOpen
  \bibfield  {author} {\bibinfo {author} {\bibfnamefont {S.~G.}\ \bibnamefont
  {Nilsson}}, \bibinfo {author} {\bibfnamefont {C.~F.}\ \bibnamefont {Tsang}},
  \bibinfo {author} {\bibfnamefont {A.}~\bibnamefont {Sobiczewski}}, \bibinfo
  {author} {\bibfnamefont {Z.}~\bibnamefont {Szymański}}, \bibinfo {author}
  {\bibfnamefont {S.}~\bibnamefont {Wycech}}, \bibinfo {author} {\bibfnamefont
  {C.}~\bibnamefont {Gustafson}}, \bibinfo {author} {\bibfnamefont {I.-L.}\
  \bibnamefont {Lamm}}, \bibinfo {author} {\bibfnamefont {P.}~\bibnamefont
  {Möller}},\ and\ \bibinfo {author} {\bibfnamefont {B.}~\bibnamefont
  {Nilsson}},\ }\href
  {https://doi.org/https://doi.org/10.1016/0375-9474(69)90809-4} {\bibfield
  {journal} {\bibinfo  {journal} {Nuclear Physics A}\ }\textbf {\bibinfo
  {volume} {131}},\ \bibinfo {pages} {1} (\bibinfo {year} {1969})}\BibitemShut
  {NoStop}%
\bibitem [{\citenamefont {Hara}\ and\ \citenamefont {Sun}(1995)}]{KY95}%
  \BibitemOpen
  \bibfield  {author} {\bibinfo {author} {\bibfnamefont {K.}~\bibnamefont
  {Hara}}\ and\ \bibinfo {author} {\bibfnamefont {Y.}~\bibnamefont {Sun}},\
  }\href {https://doi.org/10.1142/S0218301395000250} {\bibfield  {journal}
  {\bibinfo  {journal} {International Journal of Modern Physics E}\ }\textbf
  {\bibinfo {volume} {04}},\ \bibinfo {pages} {637} (\bibinfo {year}
  {1995})}\BibitemShut {NoStop}%
\bibitem [{\citenamefont {Ring}\ and\ \citenamefont {Schuck}(1980)}]{RS80}%
  \BibitemOpen
  \bibfield  {author} {\bibinfo {author} {\bibfnamefont {P.}~\bibnamefont
  {Ring}}\ and\ \bibinfo {author} {\bibfnamefont {P.}~\bibnamefont {Schuck}},\
  }\href {https://link.springer.com/book/9783540212065} {\emph {\bibinfo
  {title} {The Nuclear Many-Body Problem}}}\ (\bibinfo  {publisher} {Springer
  Berlin Heidelberg},\ \bibinfo {year} {1980})\BibitemShut {NoStop}%
\bibitem [{\citenamefont {Raman}\ \emph {et~al.}(1987)\citenamefont {Raman},
  \citenamefont {Malarkey}, \citenamefont {Milner}, \citenamefont {Nestor},\
  and\ \citenamefont {Stelson}}]{Raman}%
  \BibitemOpen
  \bibfield  {author} {\bibinfo {author} {\bibfnamefont {S.}~\bibnamefont
  {Raman}}, \bibinfo {author} {\bibfnamefont {C.~H.}\ \bibnamefont {Malarkey}},
  \bibinfo {author} {\bibfnamefont {W.~T.}\ \bibnamefont {Milner}}, \bibinfo
  {author} {\bibfnamefont {C.~W.}\ \bibnamefont {Nestor}},\ and\ \bibinfo
  {author} {\bibfnamefont {P.~H.}\ \bibnamefont {Stelson}},\ }\href
  {https://doi.org/https://doi.org/10.1016/0092-640X(87)90016-7} {\bibfield
  {journal} {\bibinfo  {journal} {Atomic Data and Nuclear Data Tables}\
  }\textbf {\bibinfo {volume} {36}},\ \bibinfo {pages} {1} (\bibinfo {year}
  {1987})}\BibitemShut {NoStop}%
\bibitem [{\citenamefont {Ayangeakaa}\ \emph {et~al.}(2019)\citenamefont
  {Ayangeakaa}, \citenamefont {Janssens}, \citenamefont {Zhu}, \citenamefont
  {Little}, \citenamefont {Henderson}, \citenamefont {Wu}, \citenamefont
  {Hartley}, \citenamefont {Albers}, \citenamefont {Auranen}, \citenamefont
  {Bucher}, \citenamefont {Carpenter}, \citenamefont {Chowdhury}, \citenamefont
  {Cline}, \citenamefont {Crawford}, \citenamefont {Fallon}, \citenamefont
  {Forney}, \citenamefont {Gade}, \citenamefont {Hayes}, \citenamefont
  {Kondev}, \citenamefont {Krishichayan}, \citenamefont {Lauritsen},
  \citenamefont {Li}, \citenamefont {Macchiavelli}, \citenamefont {Rhodes},
  \citenamefont {Seweryniak}, \citenamefont {Stolze}, \citenamefont {Walters},\
  and\ \citenamefont {Wu}}]{ge76}%
  \BibitemOpen
  \bibfield  {author} {\bibinfo {author} {\bibfnamefont {A.~D.}\ \bibnamefont
  {Ayangeakaa}}, \bibinfo {author} {\bibfnamefont {R.~V.~F.}\ \bibnamefont
  {Janssens}}, \bibinfo {author} {\bibfnamefont {S.}~\bibnamefont {Zhu}},
  \bibinfo {author} {\bibfnamefont {D.}~\bibnamefont {Little}}, \bibinfo
  {author} {\bibfnamefont {J.}~\bibnamefont {Henderson}}, \bibinfo {author}
  {\bibfnamefont {C.~Y.}\ \bibnamefont {Wu}}, \bibinfo {author} {\bibfnamefont
  {D.~J.}\ \bibnamefont {Hartley}}, \bibinfo {author} {\bibfnamefont
  {M.}~\bibnamefont {Albers}}, \bibinfo {author} {\bibfnamefont
  {K.}~\bibnamefont {Auranen}}, \bibinfo {author} {\bibfnamefont
  {B.}~\bibnamefont {Bucher}}, \bibinfo {author} {\bibfnamefont {M.~P.}\
  \bibnamefont {Carpenter}}, \bibinfo {author} {\bibfnamefont {P.}~\bibnamefont
  {Chowdhury}}, \bibinfo {author} {\bibfnamefont {D.}~\bibnamefont {Cline}},
  \bibinfo {author} {\bibfnamefont {H.~L.}\ \bibnamefont {Crawford}}, \bibinfo
  {author} {\bibfnamefont {P.}~\bibnamefont {Fallon}}, \bibinfo {author}
  {\bibfnamefont {A.~M.}\ \bibnamefont {Forney}}, \bibinfo {author}
  {\bibfnamefont {A.}~\bibnamefont {Gade}}, \bibinfo {author} {\bibfnamefont
  {A.~B.}\ \bibnamefont {Hayes}}, \bibinfo {author} {\bibfnamefont {F.~G.}\
  \bibnamefont {Kondev}}, \bibinfo {author} {\bibnamefont {Krishichayan}},
  \bibinfo {author} {\bibfnamefont {T.}~\bibnamefont {Lauritsen}}, \bibinfo
  {author} {\bibfnamefont {J.}~\bibnamefont {Li}}, \bibinfo {author}
  {\bibfnamefont {A.~O.}\ \bibnamefont {Macchiavelli}}, \bibinfo {author}
  {\bibfnamefont {D.}~\bibnamefont {Rhodes}}, \bibinfo {author} {\bibfnamefont
  {D.}~\bibnamefont {Seweryniak}}, \bibinfo {author} {\bibfnamefont {S.~M.}\
  \bibnamefont {Stolze}}, \bibinfo {author} {\bibfnamefont {W.~B.}\
  \bibnamefont {Walters}},\ and\ \bibinfo {author} {\bibfnamefont
  {J.}~\bibnamefont {Wu}},\ }\href
  {https://doi.org/10.1103/PhysRevLett.123.102501} {\bibfield  {journal}
  {\bibinfo  {journal} {Phys. Rev. Lett.}\ }\textbf {\bibinfo {volume} {123}},\
  \bibinfo {pages} {102501} (\bibinfo {year} {2019})}\BibitemShut {NoStop}%
\bibitem [{\citenamefont {Zajac}\ \emph {et~al.}(1999)\citenamefont {Zajac},
  \citenamefont {Prochniak}, \citenamefont {Pomorski}, \citenamefont
  {Rohozinski},\ and\ \citenamefont {Srebrny}}]{ZAJAC}%
  \BibitemOpen
  \bibfield  {author} {\bibinfo {author} {\bibfnamefont {K.}~\bibnamefont
  {Zajac}}, \bibinfo {author} {\bibfnamefont {L.}~\bibnamefont {Prochniak}},
  \bibinfo {author} {\bibfnamefont {K.}~\bibnamefont {Pomorski}}, \bibinfo
  {author} {\bibfnamefont {S.}~\bibnamefont {Rohozinski}},\ and\ \bibinfo
  {author} {\bibfnamefont {J.}~\bibnamefont {Srebrny}},\ }\href
  {https://doi.org/https://doi.org/10.1016/S0375-9474(99)00161-X} {\bibfield
  {journal} {\bibinfo  {journal} {Nuclear Physics A}\ }\textbf {\bibinfo
  {volume} {653}},\ \bibinfo {pages} {71} (\bibinfo {year} {1999})}\BibitemShut
  {NoStop}%
\bibitem [{\citenamefont {Wu}\ \emph {et~al.}(1996)\citenamefont {Wu},
  \citenamefont {Cline}, \citenamefont {Czosnyka}, \citenamefont {Backlin},
  \citenamefont {Baktash}, \citenamefont {Diamond}, \citenamefont {Dracoulis},
  \citenamefont {Hasselgren}, \citenamefont {Kluge}, \citenamefont {Kotlinski},
  \citenamefont {Leigh}, \citenamefont {Newton}, \citenamefont {Phillips},
  \citenamefont {Sie}, \citenamefont {Srebrny},\ and\ \citenamefont
  {Stephens}}]{WU}%
  \BibitemOpen
  \bibfield  {author} {\bibinfo {author} {\bibfnamefont {C.}~\bibnamefont
  {Wu}}, \bibinfo {author} {\bibfnamefont {D.}~\bibnamefont {Cline}}, \bibinfo
  {author} {\bibfnamefont {T.}~\bibnamefont {Czosnyka}}, \bibinfo {author}
  {\bibfnamefont {A.}~\bibnamefont {Backlin}}, \bibinfo {author} {\bibfnamefont
  {C.}~\bibnamefont {Baktash}}, \bibinfo {author} {\bibfnamefont
  {R.}~\bibnamefont {Diamond}}, \bibinfo {author} {\bibfnamefont
  {G.}~\bibnamefont {Dracoulis}}, \bibinfo {author} {\bibfnamefont
  {L.}~\bibnamefont {Hasselgren}}, \bibinfo {author} {\bibfnamefont
  {H.}~\bibnamefont {Kluge}}, \bibinfo {author} {\bibfnamefont
  {B.}~\bibnamefont {Kotlinski}}, \bibinfo {author} {\bibfnamefont
  {J.}~\bibnamefont {Leigh}}, \bibinfo {author} {\bibfnamefont
  {J.}~\bibnamefont {Newton}}, \bibinfo {author} {\bibfnamefont
  {W.}~\bibnamefont {Phillips}}, \bibinfo {author} {\bibfnamefont
  {S.}~\bibnamefont {Sie}}, \bibinfo {author} {\bibfnamefont {J.}~\bibnamefont
  {Srebrny}},\ and\ \bibinfo {author} {\bibfnamefont {F.}~\bibnamefont
  {Stephens}},\ }\href
  {https://doi.org/https://doi.org/10.1016/0375-9474(96)00181-9} {\bibfield
  {journal} {\bibinfo  {journal} {Nuclear Physics A}\ }\textbf {\bibinfo
  {volume} {607}},\ \bibinfo {pages} {178} (\bibinfo {year}
  {1996})}\BibitemShut {NoStop}%
\bibitem [{\citenamefont {Korten}\ \emph {et~al.}(1993)\citenamefont {Korten},
  \citenamefont {Härtlein}, \citenamefont {Gerl}, \citenamefont {Habs},\ and\
  \citenamefont {Schwalm}}]{KORTEN}%
  \BibitemOpen
  \bibfield  {author} {\bibinfo {author} {\bibfnamefont {W.}~\bibnamefont
  {Korten}}, \bibinfo {author} {\bibfnamefont {T.}~\bibnamefont {Härtlein}},
  \bibinfo {author} {\bibfnamefont {J.}~\bibnamefont {Gerl}}, \bibinfo {author}
  {\bibfnamefont {D.}~\bibnamefont {Habs}},\ and\ \bibinfo {author}
  {\bibfnamefont {D.}~\bibnamefont {Schwalm}},\ }\href
  {https://doi.org/https://doi.org/10.1016/0370-2693(93)91563-3} {\bibfield
  {journal} {\bibinfo  {journal} {Physics Letters B}\ }\textbf {\bibinfo
  {volume} {317}},\ \bibinfo {pages} {19} (\bibinfo {year} {1993})}\BibitemShut
  {NoStop}%
\bibitem [{\citenamefont {Martin}\ \emph {et~al.}(2000)\citenamefont {Martin},
  \citenamefont {Garrett}, \citenamefont {Kadi}, \citenamefont {Warr},
  \citenamefont {McEllistrem},\ and\ \citenamefont {Yates}}]{martin}%
  \BibitemOpen
  \bibfield  {author} {\bibinfo {author} {\bibfnamefont {A.}~\bibnamefont
  {Martin}}, \bibinfo {author} {\bibfnamefont {P.~E.}\ \bibnamefont {Garrett}},
  \bibinfo {author} {\bibfnamefont {M.}~\bibnamefont {Kadi}}, \bibinfo {author}
  {\bibfnamefont {N.}~\bibnamefont {Warr}}, \bibinfo {author} {\bibfnamefont
  {M.~T.}\ \bibnamefont {McEllistrem}},\ and\ \bibinfo {author} {\bibfnamefont
  {S.~W.}\ \bibnamefont {Yates}},\ }\href
  {https://doi.org/10.1103/PhysRevC.62.067302} {\bibfield  {journal} {\bibinfo
  {journal} {Phys. Rev. C}\ }\textbf {\bibinfo {volume} {62}},\ \bibinfo
  {pages} {067302} (\bibinfo {year} {2000})}\BibitemShut {NoStop}%
\bibitem [{\citenamefont {Bruce}\ \emph {et~al.}(2000)\citenamefont {Bruce},
  \citenamefont {Simpson}, \citenamefont {Warner}, \citenamefont {Baktash},
  \citenamefont {Barton}, \citenamefont {Bentley}, \citenamefont {Brinkman},
  \citenamefont {Cunningham}, \citenamefont {Dragulescu}, \citenamefont
  {Frankland}, \citenamefont {Ginter}, \citenamefont {Gross}, \citenamefont
  {Lemmon}, \citenamefont {MacDonald}, \citenamefont {O'Leary}, \citenamefont
  {Vincent}, \citenamefont {Wyss}, \citenamefont {Yu},\ and\ \citenamefont
  {Zamfir}}]{j1}%
  \BibitemOpen
  \bibfield  {author} {\bibinfo {author} {\bibfnamefont {A.~M.}\ \bibnamefont
  {Bruce}}, \bibinfo {author} {\bibfnamefont {J.}~\bibnamefont {Simpson}},
  \bibinfo {author} {\bibfnamefont {D.~D.}\ \bibnamefont {Warner}}, \bibinfo
  {author} {\bibfnamefont {C.}~\bibnamefont {Baktash}}, \bibinfo {author}
  {\bibfnamefont {C.~J.}\ \bibnamefont {Barton}}, \bibinfo {author}
  {\bibfnamefont {M.~A.}\ \bibnamefont {Bentley}}, \bibinfo {author}
  {\bibfnamefont {M.~J.}\ \bibnamefont {Brinkman}}, \bibinfo {author}
  {\bibfnamefont {R.~A.}\ \bibnamefont {Cunningham}}, \bibinfo {author}
  {\bibfnamefont {E.}~\bibnamefont {Dragulescu}}, \bibinfo {author}
  {\bibfnamefont {L.}~\bibnamefont {Frankland}}, \bibinfo {author}
  {\bibfnamefont {T.~N.}\ \bibnamefont {Ginter}}, \bibinfo {author}
  {\bibfnamefont {C.~J.}\ \bibnamefont {Gross}}, \bibinfo {author}
  {\bibfnamefont {R.~C.}\ \bibnamefont {Lemmon}}, \bibinfo {author}
  {\bibfnamefont {B.}~\bibnamefont {MacDonald}}, \bibinfo {author}
  {\bibfnamefont {C.~D.}\ \bibnamefont {O'Leary}}, \bibinfo {author}
  {\bibfnamefont {S.~M.}\ \bibnamefont {Vincent}}, \bibinfo {author}
  {\bibfnamefont {R.}~\bibnamefont {Wyss}}, \bibinfo {author} {\bibfnamefont
  {C.~H.}\ \bibnamefont {Yu}},\ and\ \bibinfo {author} {\bibfnamefont {N.~V.}\
  \bibnamefont {Zamfir}},\ }\href {https://doi.org/10.1103/PhysRevC.62.027303}
  {\bibfield  {journal} {\bibinfo  {journal} {Phys. Rev. C}\ }\textbf {\bibinfo
  {volume} {62}},\ \bibinfo {pages} {027303} (\bibinfo {year}
  {2000})}\BibitemShut {NoStop}%
\bibitem [{\citenamefont {Sohler}\ \emph {et~al.}(2005)\citenamefont {Sohler},
  \citenamefont {Tim\'ar}, \citenamefont {Rainovski}, \citenamefont {Joshi},
  \citenamefont {Starosta}, \citenamefont {Fossan}, \citenamefont {Moln\'ar},
  \citenamefont {Wadsworth}, \citenamefont {Algora}, \citenamefont
  {Bednarczyk}, \citenamefont {Curien}, \citenamefont {Dombr\'adi},
  \citenamefont {Duchene}, \citenamefont {Gizon}, \citenamefont {Gizon},
  \citenamefont {Jenkins}, \citenamefont {Koike}, \citenamefont
  {Krasznahorkay}, \citenamefont {Paul}, \citenamefont {Raddon}, \citenamefont
  {Scheurer}, \citenamefont {Simons}, \citenamefont {Vaman}, \citenamefont
  {Wilkinson},\ and\ \citenamefont {Zolnai}}]{Sohler2025}%
  \BibitemOpen
  \bibfield  {author} {\bibinfo {author} {\bibfnamefont {D.}~\bibnamefont
  {Sohler}}, \bibinfo {author} {\bibfnamefont {J.}~\bibnamefont {Tim\'ar}},
  \bibinfo {author} {\bibfnamefont {G.}~\bibnamefont {Rainovski}}, \bibinfo
  {author} {\bibfnamefont {P.}~\bibnamefont {Joshi}}, \bibinfo {author}
  {\bibfnamefont {K.}~\bibnamefont {Starosta}}, \bibinfo {author}
  {\bibfnamefont {D.~B.}\ \bibnamefont {Fossan}}, \bibinfo {author}
  {\bibfnamefont {J.}~\bibnamefont {Moln\'ar}}, \bibinfo {author}
  {\bibfnamefont {R.}~\bibnamefont {Wadsworth}}, \bibinfo {author}
  {\bibfnamefont {A.}~\bibnamefont {Algora}}, \bibinfo {author} {\bibfnamefont
  {P.}~\bibnamefont {Bednarczyk}}, \bibinfo {author} {\bibfnamefont
  {D.}~\bibnamefont {Curien}}, \bibinfo {author} {\bibfnamefont
  {Z.}~\bibnamefont {Dombr\'adi}}, \bibinfo {author} {\bibfnamefont
  {G.}~\bibnamefont {Duchene}}, \bibinfo {author} {\bibfnamefont
  {A.}~\bibnamefont {Gizon}}, \bibinfo {author} {\bibfnamefont
  {J.}~\bibnamefont {Gizon}}, \bibinfo {author} {\bibfnamefont {D.~G.}\
  \bibnamefont {Jenkins}}, \bibinfo {author} {\bibfnamefont {T.}~\bibnamefont
  {Koike}}, \bibinfo {author} {\bibfnamefont {A.}~\bibnamefont
  {Krasznahorkay}}, \bibinfo {author} {\bibfnamefont {E.~S.}\ \bibnamefont
  {Paul}}, \bibinfo {author} {\bibfnamefont {P.~M.}\ \bibnamefont {Raddon}},
  \bibinfo {author} {\bibfnamefont {J.~N.}\ \bibnamefont {Scheurer}}, \bibinfo
  {author} {\bibfnamefont {A.~J.}\ \bibnamefont {Simons}}, \bibinfo {author}
  {\bibfnamefont {C.}~\bibnamefont {Vaman}}, \bibinfo {author} {\bibfnamefont
  {A.~R.}\ \bibnamefont {Wilkinson}},\ and\ \bibinfo {author} {\bibfnamefont
  {L.}~\bibnamefont {Zolnai}},\ }\href
  {https://doi.org/10.1103/PhysRevC.71.064302} {\bibfield  {journal} {\bibinfo
  {journal} {Phys. Rev. C}\ }\textbf {\bibinfo {volume} {71}},\ \bibinfo
  {pages} {064302} (\bibinfo {year} {2005})}\BibitemShut {NoStop}%
\bibitem [{\citenamefont {Sohler}\ \emph {et~al.}(2012)\citenamefont {Sohler},
  \citenamefont {Kuti}, \citenamefont {Tim\'ar}, \citenamefont {Joshi},
  \citenamefont {Moln\'ar}, \citenamefont {Paul}, \citenamefont {Starosta},
  \citenamefont {Wadsworth}, \citenamefont {Algora}, \citenamefont
  {Bednarczyk}, \citenamefont {Curien}, \citenamefont {Dombr\'adi},
  \citenamefont {Duchene}, \citenamefont {Fossan}, \citenamefont {G\'al},
  \citenamefont {Gizon}, \citenamefont {Gizon}, \citenamefont {Jenkins},
  \citenamefont {Juh\'asz}, \citenamefont {Kalinka}, \citenamefont {Koike},
  \citenamefont {Krasznahorkay}, \citenamefont {Nyak\'o}, \citenamefont
  {Raddon}, \citenamefont {Rainovski}, \citenamefont {Scheurer}, \citenamefont
  {Simons}, \citenamefont {Vaman}, \citenamefont {Wilkinson},\ and\
  \citenamefont {Zolnai}}]{Sohler2012}%
  \BibitemOpen
  \bibfield  {author} {\bibinfo {author} {\bibfnamefont {D.}~\bibnamefont
  {Sohler}}, \bibinfo {author} {\bibfnamefont {I.}~\bibnamefont {Kuti}},
  \bibinfo {author} {\bibfnamefont {J.}~\bibnamefont {Tim\'ar}}, \bibinfo
  {author} {\bibfnamefont {P.}~\bibnamefont {Joshi}}, \bibinfo {author}
  {\bibfnamefont {J.}~\bibnamefont {Moln\'ar}}, \bibinfo {author}
  {\bibfnamefont {E.~S.}\ \bibnamefont {Paul}}, \bibinfo {author}
  {\bibfnamefont {K.}~\bibnamefont {Starosta}}, \bibinfo {author}
  {\bibfnamefont {R.}~\bibnamefont {Wadsworth}}, \bibinfo {author}
  {\bibfnamefont {A.}~\bibnamefont {Algora}}, \bibinfo {author} {\bibfnamefont
  {P.}~\bibnamefont {Bednarczyk}}, \bibinfo {author} {\bibfnamefont
  {D.}~\bibnamefont {Curien}}, \bibinfo {author} {\bibfnamefont
  {Z.}~\bibnamefont {Dombr\'adi}}, \bibinfo {author} {\bibfnamefont
  {G.}~\bibnamefont {Duchene}}, \bibinfo {author} {\bibfnamefont {D.~B.}\
  \bibnamefont {Fossan}}, \bibinfo {author} {\bibfnamefont {J.}~\bibnamefont
  {G\'al}}, \bibinfo {author} {\bibfnamefont {A.}~\bibnamefont {Gizon}},
  \bibinfo {author} {\bibfnamefont {J.}~\bibnamefont {Gizon}}, \bibinfo
  {author} {\bibfnamefont {D.~G.}\ \bibnamefont {Jenkins}}, \bibinfo {author}
  {\bibfnamefont {K.}~\bibnamefont {Juh\'asz}}, \bibinfo {author}
  {\bibfnamefont {G.}~\bibnamefont {Kalinka}}, \bibinfo {author} {\bibfnamefont
  {T.}~\bibnamefont {Koike}}, \bibinfo {author} {\bibfnamefont
  {A.}~\bibnamefont {Krasznahorkay}}, \bibinfo {author} {\bibfnamefont {B.~M.}\
  \bibnamefont {Nyak\'o}}, \bibinfo {author} {\bibfnamefont {P.~M.}\
  \bibnamefont {Raddon}}, \bibinfo {author} {\bibfnamefont {G.}~\bibnamefont
  {Rainovski}}, \bibinfo {author} {\bibfnamefont {J.~N.}\ \bibnamefont
  {Scheurer}}, \bibinfo {author} {\bibfnamefont {A.~J.}\ \bibnamefont
  {Simons}}, \bibinfo {author} {\bibfnamefont {C.}~\bibnamefont {Vaman}},
  \bibinfo {author} {\bibfnamefont {A.~R.}\ \bibnamefont {Wilkinson}},\ and\
  \bibinfo {author} {\bibfnamefont {L.}~\bibnamefont {Zolnai}},\ }\href
  {https://doi.org/10.1103/PhysRevC.85.044303} {\bibfield  {journal} {\bibinfo
  {journal} {Phys. Rev. C}\ }\textbf {\bibinfo {volume} {85}},\ \bibinfo
  {pages} {044303} (\bibinfo {year} {2012})}\BibitemShut {NoStop}%
\bibitem [{\citenamefont {Sithole}\ \emph {et~al.}(2021)\citenamefont
  {Sithole}, \citenamefont {Sharpey~Schafer}, \citenamefont {Lawrie},
  \citenamefont {Majola}, \citenamefont {Kardan}, \citenamefont {Bucher},
  \citenamefont {Lawrie}, \citenamefont {Mdletshe}, \citenamefont {Ntshangase},
  \citenamefont {Netshiya}, \citenamefont {Jones}, \citenamefont {Makhathini},
  \citenamefont {Malatji}, \citenamefont {Masiteng}, \citenamefont
  {Ragnarsson}, \citenamefont {Maqabuka}, \citenamefont {Ndayishimye},
  \citenamefont {Shirinda}, \citenamefont {Zikhali}, \citenamefont {Jongile},
  \citenamefont {O'Neill}, \citenamefont {Msebi}, \citenamefont {Someketa},
  \citenamefont {Kenfack}, \citenamefont {Mthembu}, \citenamefont {Khumalo},\
  and\ \citenamefont {Chisapi}}]{Sithole2021}%
  \BibitemOpen
  \bibfield  {author} {\bibinfo {author} {\bibfnamefont {M.~A.}\ \bibnamefont
  {Sithole}}, \bibinfo {author} {\bibfnamefont {J.~F.}\ \bibnamefont
  {Sharpey~Schafer}}, \bibinfo {author} {\bibfnamefont {E.~A.}\ \bibnamefont
  {Lawrie}}, \bibinfo {author} {\bibfnamefont {S.~N.~T.}\ \bibnamefont
  {Majola}}, \bibinfo {author} {\bibfnamefont {A.}~\bibnamefont {Kardan}},
  \bibinfo {author} {\bibfnamefont {T.~D.}\ \bibnamefont {Bucher}}, \bibinfo
  {author} {\bibfnamefont {J.~J.}\ \bibnamefont {Lawrie}}, \bibinfo {author}
  {\bibfnamefont {L.}~\bibnamefont {Mdletshe}}, \bibinfo {author}
  {\bibfnamefont {S.~S.}\ \bibnamefont {Ntshangase}}, \bibinfo {author}
  {\bibfnamefont {A.~A.}\ \bibnamefont {Netshiya}}, \bibinfo {author}
  {\bibfnamefont {P.}~\bibnamefont {Jones}}, \bibinfo {author} {\bibfnamefont
  {L.}~\bibnamefont {Makhathini}}, \bibinfo {author} {\bibfnamefont {K.~L.}\
  \bibnamefont {Malatji}}, \bibinfo {author} {\bibfnamefont {P.~L.}\
  \bibnamefont {Masiteng}}, \bibinfo {author} {\bibfnamefont {I.}~\bibnamefont
  {Ragnarsson}}, \bibinfo {author} {\bibfnamefont {B.}~\bibnamefont
  {Maqabuka}}, \bibinfo {author} {\bibfnamefont {J.}~\bibnamefont
  {Ndayishimye}}, \bibinfo {author} {\bibfnamefont {O.}~\bibnamefont
  {Shirinda}}, \bibinfo {author} {\bibfnamefont {B.~R.}\ \bibnamefont
  {Zikhali}}, \bibinfo {author} {\bibfnamefont {S.}~\bibnamefont {Jongile}},
  \bibinfo {author} {\bibfnamefont {G.}~\bibnamefont {O'Neill}}, \bibinfo
  {author} {\bibfnamefont {L.}~\bibnamefont {Msebi}}, \bibinfo {author}
  {\bibfnamefont {P.~M.}\ \bibnamefont {Someketa}}, \bibinfo {author}
  {\bibfnamefont {D.}~\bibnamefont {Kenfack}}, \bibinfo {author} {\bibfnamefont
  {S.~H.}\ \bibnamefont {Mthembu}}, \bibinfo {author} {\bibfnamefont {T.~C.}\
  \bibnamefont {Khumalo}},\ and\ \bibinfo {author} {\bibfnamefont {M.~V.}\
  \bibnamefont {Chisapi}},\ }\href
  {https://doi.org/10.1103/PhysRevC.103.024325} {\bibfield  {journal} {\bibinfo
   {journal} {Phys. Rev. C}\ }\textbf {\bibinfo {volume} {103}},\ \bibinfo
  {pages} {024325} (\bibinfo {year} {2021})}\BibitemShut {NoStop}%
\bibitem [{\citenamefont {D{\"o}nau}\ and\ \citenamefont
  {Frauendorf}(1982)}]{FD82}%
  \BibitemOpen
  \bibfield  {author} {\bibinfo {author} {\bibfnamefont {F.}~\bibnamefont
  {D{\"o}nau}}\ and\ \bibinfo {author} {\bibfnamefont {S.}~\bibnamefont
  {Frauendorf}},\ }in\ \href@noop {} {\emph {\bibinfo {booktitle} {High Angular
  Momentum Properties of Nuclei}}}\ (\bibinfo  {publisher} {Nucl. Sci.},\
  \bibinfo {address} {Oak Ridge},\ \bibinfo {year} {1982})\BibitemShut
  {NoStop}%
\bibitem [{\citenamefont {Wang}\ \emph
  {et~al.}(2020{\natexlab{b}})\citenamefont {Wang}, \citenamefont {Chen},\ and\
  \citenamefont {Sun}}]{WANG2020135676}%
  \BibitemOpen
  \bibfield  {author} {\bibinfo {author} {\bibfnamefont {L.-J.}\ \bibnamefont
  {Wang}}, \bibinfo {author} {\bibfnamefont {F.-Q.}\ \bibnamefont {Chen}},\
  and\ \bibinfo {author} {\bibfnamefont {Y.}~\bibnamefont {Sun}},\ }\href
  {https://doi.org/https://doi.org/10.1016/j.physletb.2020.135676} {\bibfield
  {journal} {\bibinfo  {journal} {Phys. Lett. B}\ }\textbf {\bibinfo {volume}
  {808}},\ \bibinfo {pages} {135676} (\bibinfo {year}
  {2020}{\natexlab{b}})}\BibitemShut {NoStop}%
\bibitem [{\citenamefont {Frauendorf}\ and\ \citenamefont
  {D\"onau}(2014)}]{sf14}%
  \BibitemOpen
  \bibfield  {author} {\bibinfo {author} {\bibfnamefont {S.}~\bibnamefont
  {Frauendorf}}\ and\ \bibinfo {author} {\bibfnamefont {F.}~\bibnamefont
  {D\"onau}},\ }\href {https://doi.org/10.1103/PhysRevC.89.014322} {\bibfield
  {journal} {\bibinfo  {journal} {Phys. Rev. C}\ }\textbf {\bibinfo {volume}
  {89}},\ \bibinfo {pages} {014322} (\bibinfo {year} {2014})}\BibitemShut
  {NoStop}%
\bibitem [{\citenamefont {Qi}\ \emph {et~al.}(2009)\citenamefont {Qi},
  \citenamefont {Zhang}, \citenamefont {Meng}, \citenamefont {Wang},\ and\
  \citenamefont {Frauendorf}}]{BQ09}%
  \BibitemOpen
  \bibfield  {author} {\bibinfo {author} {\bibfnamefont {B.}~\bibnamefont
  {Qi}}, \bibinfo {author} {\bibfnamefont {S.~Q.}\ \bibnamefont {Zhang}},
  \bibinfo {author} {\bibfnamefont {J.}~\bibnamefont {Meng}}, \bibinfo {author}
  {\bibfnamefont {S.~Y.}\ \bibnamefont {Wang}},\ and\ \bibinfo {author}
  {\bibfnamefont {S.}~\bibnamefont {Frauendorf}},\ }\href
  {https://doi.org/https://doi.org/10.1016/j.physletb.2009.02.061} {\bibfield
  {journal} {\bibinfo  {journal} {Phys. Lett. B}\ }\textbf {\bibinfo {volume}
  {675}},\ \bibinfo {pages} {175} (\bibinfo {year} {2009})}\BibitemShut
  {NoStop}%
\end{thebibliography}%
\end{document}